\documentclass[twocolumn,tighten,times]{aastex63}

\date{June 17, 2020}
\received{Dec 3, 2019}
\submitjournal{the Astrophysical Journal}

\usepackage{amsmath}    % Advanced maths commands
\def\gtsim {>\kern-1.2em\lower1.1ex\hbox{$\sim$}~}   % Greater than sim
\def\ltsim {<\kern-1.2em\lower1.1ex\hbox{$\sim$}~}   % Less than sim
\newcommand{\Msun}{M_\odot}

\begin{document}

\title{The Origin of Elements from Carbon to Uranium}
\author{Chiaki Kobayashi}
%\thanks{E-mail: c.kobayashi@herts.ac.uk}
\affiliation{Centre for Astrophysics Research, Department of Physics, Astronomy and Mathematics, University of Hertfordshire, Hatfield, AL10 9AB, UK}
\author{Amanda I. Karakas}
\affiliation{School of Physics \& Astronomy, Monash University, Clayton VIC 3800, Australia}
\affiliation{ARC Centre of Excellence for All Sky Astrophysics in 3 Dimensions (ASTRO 3D), Australia}
\author{Maria Lugaro}
\affiliation{Konkoly Observatory, Research Centre for Astronomy and Earth Sciences, Hungarian Academy of Sciences, Konkoly Thege Miklos ut 15-17, H-1121 Budapest, Hungary}
\affiliation{ELTE E\"{o}tv\"{o}s Lor\'and University, Institute of Physics, Budapest 1117, P\'azm\'any P\'eter s\'et\'any 1/A, Hungary}
\affiliation{School of Physics \& Astronomy, Monash University, Clayton VIC 3800, Australia}

\begin{abstract}
To reach a deeper understanding of the origin of elements in the periodic table, we construct Galactic chemical evolution (GCE) models for all stable elements from C ($A\!=\!12$) to U ($A\!=\!238$) from first principles,
i.e., using theoretical nucleosynthesis yields and event rates of all chemical enrichment sources.
This enables us to predict the origin of elements as a function of time and environment.
In the solar neighborhood, we find that stars with initial masses of $M>30M_\odot$ can become failed supernovae if there is a significant contribution from hypernovae (HNe) at $M\sim20-50M_\odot$.
The contribution to GCE from super asymptotic giant branch (AGB) stars (with $M\sim8-10M_\odot$ at solar metallicity) is negligible, unless hybrid white dwarfs from low-mass super-AGB stars explode as so-called Type Iax supernovae, or high-mass super-AGB stars explode as electron-capture supernovae (ECSNe).
Among neutron-capture elements, the observed abundances of the second (Ba) and third (Pb) peak elements are well reproduced with our updated yields of the slow neutron-capture process (s-process) from AGB stars.
The first peak elements, Sr, Y, and Zr, are sufficiently produced by ECSNe together with AGB stars.
Neutron star mergers can produce rapid neutron-capture process (r-process) elements up to Th and U, but the timescales are too long to explain observations at low metallicities.
The observed evolutionary trends, such as for Eu, can well be explained if $\sim3$\% of $25-50M_\odot$ hypernovae are magneto-rotational supernovae producing r-process elements.
Along with the solar neighborhood, we also predict the evolutionary trends in the halo, bulge, and thick disk for future comparison with galactic archaeology surveys.
\end{abstract}

\keywords{Galaxy: abundances --- Galaxy: evolution --- stars: abundances --- stars: AGB and post-AGB --- stars: supernovae}

\section{Introduction}

Since the time of \citet{bbfh} the question of the origin of the elements is one that has been studied at the interface between nuclear physics and astrophysics.
We now know that different elements are produced by different astronomical sources, i.e., different masses of stars, supernovae, and binary systems.
The relative contribution of each source depends on time and environment (i.e., mass and type of galaxies), and hence it is necessary to use galactic chemical evolution (GCE) models to understand this question.
Observationally, elemental abundances have been estimated the best in the Sun and in the stars in the Local Group, as well as meteorites, planetary nebulae, and globular clusters. For a limited number of elements, these are also some estimates for damped Ly $\alpha$ systems \citep[e.g.,][]{pet94,wol05}, the intracluster medium \citep[e.g.,][]{mus96,hitomi17}, stellar populations in early-type galaxies \citep[e.g.,][]{tho03,con14}, and star forming galaxies \citep[e.g.,][]{gar90,pil10}.

Elemental abundances in the Milky Way Galaxy provide stringent constraints not only on stellar astrophysics but also on the formation and evolutionary history of the Milky Way Galaxy itself.
Elements heavier than helium are synthesized inside and then ejected by dying stars.
The next generation of stars form from gas clouds that include heavy elements from the previous stellar generations. Therefore, stars in the present-day galaxy are fossils that retain the information on the properties of stars from the past.
From the elemental abundances of the present-day stars, it is possible to disentangle the star formation history of the host galaxy.
This approach is called Galactic archaeology and
can be applied not only to our Milky Way Galaxy but also to other galaxies \citep[e.g.,][]{kob16,vin18a}.
For constraining the star formation histories of galaxies, the most important uncertainty is the set of nucleosynthesis yields.

A vast amount of observational elemental abundance data are being or will be taken by Galactic archaeology surveys, together with data from space astrometry missions (e.g., Gaia) and medium-resolution multi-object spectroscopy\footnote{There are also surveys with lower-resolution multi-object spectroscopy, such as SDSS (the Sloan Digital Sky Survey), RAVE (the Radial Velocity Experiment), LAMOST (the Large Sky Area Multi-Object Fibre Spectroscopic Telescope), and PFS (Prime Focus Spectrograph) on Subaru Telescope.} such as APOGEE (the Apache Point Observatory Galactic Evolution Experiment), HERMES (the High Efficiency and Resolution Multi-Element Spectrograph) on the Anglo-Australian Telescope, 4MOST (4-metre Multi-Object Spectroscopic Telescope) on the VISTA telescope, WEAVE on the William Herschel Telescope, and MSE (Maunakea Spectroscopic Explorer).
These data are revealing the chemodynamical structure of the Milky Way Galaxy \citep{hay15,bud18} and the Local Group by mapping the elemental abundance patterns of millions of stars.
In contrast, for a smaller number of stars, more detailed spectral analysis is made with non-local thermodynamic equilibrium (NLTE) and/or three-dimensional (3D) stellar atmosphere modelling, which increases the accuracy for estimating elemental abundances from high-resolution spectra.
This was done for the solar abundances \citep[][]{asp09}, for some metal-poor stars \citep{nor17,pra17}, and recently for a wide range of metallicities \citep{and07,zha16,ama19b}, and should be used for constraining stellar nucleosynthesis.

Because of the nature of the triple $\alpha$ reactions, elements with $A\ge12$ are produced not during the Big Bang but are instead formed inside stars. 
Roughly half of the light elements such as C, N and F are produced by low- and intermediate-mass stars during their asymptotic giant branch (AGB) phase (\citealt[][hereafter K11]{kar10,kob11agb}; see also \citealt{van97,mar01,kar07,cri11,ven13}).
Isotopes such as $^{13}$C, $^{17}$O, and $^{25,26}$Mg are also enhanced by AGB stars and thus these isotopic ratios can also be used for Galactic archaeology \citep[e.g.,][]{spi06,car18}.
The $\alpha$-elements (O, Mg, Si, S, and Ca) are mostly produced in massive stars before being ejected by core-collapse (Type II, Ib, and Ic) supernovae \citep[e.g.,][hereafter K06]{tim95,kob06}.
The production of some elements such as F, K, Sc, and V can be increased by neutrino processes in core-collapse supernovae \citep{kob11f}.
Conversely, half of the iron-peak elements (Cr, Mn, Fe, Ni, Co, Cu, and Zn) are produced by Type Ia Supernovae (SNe Ia), which are the explosions of C+O white dwarfs (WDs) in binary systems \citep[e.g.,][]{kob09,kob19ia}.
The production of odd-Z elements (Na, Al, P, ... and Cu) depends on the metallicity of the progenitor, as their production depends on the surplus of neutrons from $^{22}$Ne, which is made during He-burning from $^{14}$N produced in the CNO cycle.
The production of minor isotopes ($^{13}$C, $^{17,18}$O, $^{25,26}$Mg, ...) also depends on the metallicity (K11).

GCE models have been used to test the production sources and the nucleosynthesis yields \citep[e.g.,][hereafter K00]{tin80,pra93,tim95,pagel1997,chi97,matteucci2001,kob00}.
For example, the [$\alpha$/Fe]--[Fe/H] relation in the Milky Way Galaxy is explained by the delayed enrichment of Fe from SNe Ia, which have a longer timescale than core-collapse supernovae. Therefore, the [$\alpha$/Fe] ratios can be used to constrain star formation timescales in other galaxies \citep{tay15,kob16,vin18c}.
The average evolutionary tracks of most of the elements from C to Zn (except for Ti) are well reproduced by GCE models (\citealt{kob11agb}, see also \citealt{rom10}).

The elements beyond Fe ($A\gtsim64$) are synthesized mostly by the two extreme cases of neutron-capture processes: the slow (s, $N_{\rm n}\sim10^7$ cm$^3$) and rapid (r, $N_{\rm n}$ $\gtsim$ $10^{20}$ cm$^3$) processes depending on the neutron density\footnote{Intermediate process (i-process) has also been discussed \citep[e.g.,][]{cow77,herwig11}, although the contribution to GCE may be small \citep{cot18}, depending however on the currently unknown stellar site.}.
The traditional main and strong s-process components (producing elements from Sr to Pb) are produced in the He-rich intershell of low-mass AGB stars \citep{busso99,herwig05,kar14} where the neutron source is mainly $^{13}$C($\alpha$,n)$^{16}$O.
The weak s-process component (from Fe to Sr) is produced instead in massive stars near solar metallicity \citep{pig10}, as well as in low-metallicity stars if high rotational rates are assumed \citep{fri16,lim18,cho18}; here the neutrons are mostly provided by the $^{22}$Ne($\alpha$,n)$^{25}$Mg reaction.

For the r-process, the astrophysical sites have been debated. Detailed simulations have shown that electron capture supernovae \citep[ECSNe,][]{hof08,wan11,wan13b} and $\nu$-driven winds \citep{arc07,fis10,arc11,wan13a} cannot produce the elements heavier than $A\sim110$.
Neutron star mergers (NSMs) provide suitable conditions for the r-process \citep[][and references therein]{lat74,ros99,wan14}, and recently, the existence of such an event was confirmed by the gravitational wave source GW170817 \citep{abb17a}, associated with an astronomical transient AT2017gfo \citep{sma17,val17} and a short $\gamma$-ray burst GRB170817A \citep{abb17b}.
In GCE models however, the timescale of NSMs seems to be too long to explain the observations \citep{arg04}, and magneto-rotational supernovae \citep[MRSNe,][]{win12,mos14,nis15} is also invoked as a main site of the r-process in the Galaxy \citep{ces15,weh15,hay19,cot19}.

In this paper, in order to reach a deeper understanding of the origin of elements, we construct GCE models for all stable elements from C ($A\!=\!12$) to U ($\!A=\!238$), using the latest results of stellar astrophysics and the observations of elemental abundances in the Milky Way Galaxy.
We include theoretical nucleosynthesis yields and event rates, avoiding empirical relations, so that our models are calculated from the first principles.
Our novel and comprehensive approach of addressing the origin of all the elements within the same framework allows us to discover consistencies, and inconsistencies, that may arise only by considering all the elements together.
This approach is fundamentally different from that in \citet{pra20}, where the r-process is assumed to be primary and follows the evolution of $\alpha$ elements.
In \S 2, we describe our chemical evolution models summarizing the enrichment sources.
In \S 3, after addressing the impact of failed supernovae and super-AGB stars for GCE, we show the time/metallicity evolution of neutron capture elements for the solar neighborhood, halo, bulge, and thick disk.
Since we aim to discuss elemental abundances on 0.1 dex accuracy, we adopt the latest solar abundances throughout this paper, and shift observational data if necessary.
We focus on the average evolution of abundances in the systems, excluding the carbon-enhanced metal-poor stars \citep[CEMP,][]{bee05}, which are explained with other effects such as faint supernovae and binary mass transfer, but including so-called r-II stars \citep[][{\rm [Eu/Fe]} $>+1$ and {\rm [Ba/Eu]} $<0$]{bee05}.
We then summarize the origin of the elements in \S 4 and end with conclusions in \S 5.

\begin{table*} 
\begin{tabular}{l|cccccc}
%\hline 
%AGB star models with s-process
\hline
$Z$ & 0.0001 & 0.001 & 0.0028 & 0.007 & 0.014 & 0.03\\
\hline
$M_{\rm mix}=0$ & - & & $1 M_\odot$ & $1-1.25 M_\odot$ & $1-1.25M_\odot$ & $1-2.25M_\odot$ \\
$M_{\rm mix}=2\times10^{-3}$ & $0.9-2.25 M_\odot$ & $1-2.5 M_\odot$ & - & - & - & - \\
$M_{\rm mix}=1\times10^{-3}$ & $2.5-3M_\odot$ & $2.75 M_\odot$ & $1.15-2.75M_\odot$ & $1.5-3.75M_\odot$ & $1.5-4M_\odot$ & $2.5-4M_\odot$ \\
$M_{\rm mix}=5\times10^{-4}$ & - & $3M_\odot$ & $3-3.5 M_\odot$ & - & - & - \\
$M_{\rm mix}=1\times10^{-4}$ & - & - & $3.75-4 M_\odot$ & $4-4.25M_\odot$ & $4.25-5M_\odot$ & $4.25-5M_\odot$ \\
$M_{\rm mix}=0$ & $3.5-6M_\odot$ & $3.25-7M_\odot$ & $4.5-7 M_\odot$ & $4.5-7.5M_\odot$ & $5.5-8M_\odot$ & $5.5-8M_\odot$ \\
\hline
%super-AGB star models
%\hline
$Z$ & 0.0001 & 0.001 & 0.004 & 0.008 & 0.02 & - \\
\hline
$M_{\rm mix}=0$ & $6.5-7.5M_\odot$ & $7.5M_\odot$ & $7.5-8M_\odot$ & $8-8.5M_\odot$ & $8.5-9M_\odot$ & \\
\hline 
\end{tabular} 
\label{tab:mix} 
\caption{The mass of partial mixing zones, $M_{\rm mix}$, adopted for the AGB and super-AGB models as a function of initial mass and metallicity.
}
\end{table*}

\section{The Model}

\subsection{Chemical Enrichment Sources}
\label{sec:sn}

Often GCE model predictions directly come from the input stellar physics and nucleosynthesis yields.
Based on recent developments in stellar astrophysics, we summarize the chemical enrichment sources that are chosen to be included in this section.

\subsubsection{AGB stars and core-collapse supernovae}

{\bf Stellar winds} ---
All dying stars return a fraction or all of their envelope mass to the interstellar medium (ISM) by stellar winds. 
These winds (for massive stars occurring before the final supernova explosions) carry newly processed metals and the unprocessed metals that were trapped inside the star at its formation and is returned to the ISM.
Usually, both the processed and unprocessed components are included in the nucleosynthesis yield table of AGB stars, while only the former is included for supernova yields (see Eq.~9 in K00) and 
the latter is added in the GCE models using the abundance pattern of the ISM at the time when the stars formed (Eq.~8 in K00).
The wind mass is given by $M_{\rm wind}=M_{\rm init}-M_{\rm remnant}-\Sigma_i\,p_{z_im}$, where the initial mass $M_{\rm init}$, the mass of remnant $M_{\rm remnant}$, i.e., black hole (BH), neutron star (NS), or white dwarf (WD) mass. Nucleosynthesis yields, $p_{z_im}$, of an element/isotope $i$ are given in the yield tables (see below for more details).
For stars of initial masses $0.7$ and $0.9M_\odot$, the He core mass is set as $M_{\rm remnant}=0.459$ and $0.473M_\odot$, respectively, and $p_{z_im}=0$, as in K06/K11.

{\bf Asymptotic Giant Branch (AGB) stars} ---
Stars with initial masses between roughly $0.9-8 M_\odot$ (depending on
metallicity) pass through the thermally-pulsing AGB phase. The
He-burning shell is thermally unstable and can drive mixing of material from the core into the envelope, which has been processed by nuclear reactions. This mixing is known as 
third dredge-up (TDU), and is responsible for enriching
the surface in $^{12}$C and other products of He-burning,
as well as s-process elements.
In AGB stars with initial masses $\gtsim 4\Msun$, the base of the convective envelope becomes hot enough to sustain proton-capture nucleosynthesis 
(hot bottom burning, HBB). HBB can change the surface 
composition because the entire envelope is exposed to the hot 
burning region a few thousand times per interpulse period. 
The CNO cycles operate to convert the freshly synthesized $^{12}$C
into {\em primary} $^{14}$N, and the NeNa and MgAl chains may also
operate to produce $^{23}$Na and Al.

At the deepest extent of each TDU, it is assumed that the bottom of the H-rich convective envelope penetrates into the $^{12}$C-rich intershell layer resulting into a partial mixing zone (PMZ) leading to the formation of a $^{13}$C pocket via the $^{12}$C(p,$\gamma$)$^{13}$N($\beta^+$)$^{13}$C reaction chain. While many physical processes have been proposed, there is still not full agreement on which process(es) drives the mixing. The inclusion of $^{13}$C pockets in theoretical calculations of AGB stars is still one of the most significant uncertainties affecting predictions of the s process and in particular the absolute values of the yields \citep[][and references therein]{kar16,bun17}. Other major uncertainties come from the rates of the neutron source reactions $^{13}$C($\alpha$,n)$^{16}$O and $^{22}$Ne($\alpha$,n)$^{25}$Mg \citep{bisterzo15} and the neutron-capture cross sections of some key isotopes \citep{ces18}.

In this paper, we take the nucleosynthesis yields including s-process and WD masses primarily from \citet{lugaro12} for $Z=0.0001$, \citet{fis14} for $Z=0.001$, \citet{kar18} for $Z=0.0028$, and \citet{kar16} for $Z=0.007, 0.014$ and $0.03$. In these post-processing nucleosynthesis, protons are added to the top layers of the He-intershell at the deepest extent of each TDU episode by means of an artificial PMZ.
The mass of the PMZ, i.e., how deep it reaches below the base of the convective envelope, is given by a free parameter $M_{\rm mix}$ as a function of mass and metallicity, as discussed in detail by \citet{kar16}.
In addition, for this paper we calculated some selected low-mass star models with $Z = 0.014, 0.007$ and $0.0028$ using a smaller PMZ mass; namely these models set the PMZ mass to be $0.001 M_{\odot}$ compared to the standard size, 0.002$M_{\odot}$, used in \citet{kar16}.
The adopted PMZ mass of our fiducial model is summarized in Table \ref{tab:mix}. We also show a GCE model with the original \citet{kar16}'s yields for Ba in this paper (Fig.\,\ref{fig:ba}).
Non time-dependent overshoot, which essentially only affects the depth of the TDU and not the formation of the PMZ, is also included in some models with the parameter $N_{\rm ov}$ set to 1.0 for the $1.5$ and $1.75 M_\odot$ models of $Z=0.007$, to 3.0 and 2.0, respectively, for the $1.5$ and $1.75M_\odot$ models of $Z=0.014$, and to 2.5, 2.0, and 1.0, respectively, for the 2.5, 2.75, and $3.0 M_\odot$ models of $Z=0.03$.
For the $Z=0.0028$ models overshoot was included in the 1.15 and 1.25 $M_\odot$ models where $N_{\rm ov}$ was set to 1.0 for both cases.

In these yield tables, the mass of each element expelled over the stellar lifetime is listed, which contains the unprocessed metals.
The newly produced metal yields, $p_{z_im}$, are calculated as the difference between the amount of the species in the winds and the initial amount in the envelope of the progenitor star.
The initial abundances are set as the scaled proto-solar abundances, which are calculated from Tables 1, 3, and 5 from \citet[][]{asp09}; the meteoritic values are chosen if the errors are smaller than the photospheric values, and the proto-solar abundances are used for C, N, O, Ne, Mg, Si, S, Ar, and Fe.
Therefore, $p_{z_im}$ can have negative values especially for H, but after adding the unprocessed metals in the GCE, the mass of each element becomes positive.
For $Z=0$, the models of $Z=0.0001$ are used, although the yields from \citet{cam08} were adopted in K11.
The upper and lower mass ranges of the AGB models can also be found in Table \ref{tab:mix} as a function of initial metallicity.
% ML
It is important to note that these AGB models successfully reproduce both the trend with metallicity observed in a large sample of Ba stars \citep{cseh18}, and the heavy element composition of meteoritic stardust silicon carbide (SiC) grains that formed around C-rich AGB stars \citep{lugaro18}.

Note that, however, when we discuss the model dependence for the elements up to Zn (namely, Figs.~\ref{fig:xfe}, \ref{fig:xfe2}, \ref{fig:coo}, and \ref{fig:noo}), in order to make a fair comparison, we use the same AGB yields as in K11, i.e., \citet{kar10}'s yields. There are no differences between this model with K11's yields and the fiducial model with s-process yields, except for C and N.

{\bf Super AGB stars} ---
The fate of stars with initial masses between about $8-10 M_\odot$ (at $Z=0.02$) is uncertain \citep[][for a review]{doh17} and their contributions were not included in K11.
The upper limit of AGB stars, $M_{\rm up,C}$, is defined as the minimum mass for carbon ignition, and is 
estimated to be larger at high metallicity and also for  metallicities lower than $Z\sim10^{-4}$ \citep{gil07,sie07}.
Just above $M_{\rm up,C}$, neutrino cooling and contraction leads to the off-centre ignition of a C flame, which moves inward but does not propagate to the centre. This may form a hybrid C+O+Ne WD \citep[see \S2.2 of][for more details]{kob15}.
These hybrid WDs can be progenitors of a sub-class of SNe Ia, called SNe Iax \citep{fol13}, which are expected preferably in dwarf galaxies \citep{kob15,ces17}.
We take the nucleosynthesis yields of an SN Iax from \citet{fin14}.

Above this mass range, the off-centre C ignition moves inward all the way to the centre ($\ltsim 9M_\odot$), or stars undergo central carbon ignition ($\gtsim 9M_\odot$).
For both cases, a strongly degenerate O+Ne+Mg core is formed (O+Ne dominant, but Mg is essential for electron capture).
If the stellar envelope is lost by winds or binary interaction, an O+Ne+Mg WD may be formed.
This upper mass limit is defined as the minimum mass for the Ne ignition, $M_{\rm up,Ne}\sim9\pm1 M_\odot$, and is smaller for lower metallicities \citep[e.g.,][]{sie07,doh15}.

Stars with $M_{\rm up,Ne}<M<10M_\odot$ may have cores as massive as $\gtsim 1.35M_\odot$ and ignite Ne off-centre. 
If Ne burning is not ignited at the centre \citep[for a core mass $<1.37M_\odot$,][]{nom84}, or if off-centre Ne burning does not propagate to the centre \citep[$M=8.8M_\odot$,][]{jon14}, it has been believed that such a core eventually undergoes an electron-capture-induced collapse.
Electron-capture supernova \citep[ECSN, see \S2.3.2 of][for more details]{nom13} are one of the candidate r-process sites (\S 1, see \S \ref{sec:r} for more details).
Note that recent 3D simulations showed that the fate of the O+Ne+Mg core may depend on the density, and the explosion may result in thermonuclear disruption leaving behind an O+Ne+Fe WD instead \citep{jon16}.

In this paper, the mass ranges of the C+O+Ne WDs, O+Ne+Mg WDs, and ECSNe are taken from \citet{doh15}.
Note that Ne burning is not followed in \citet{doh15}; the lower-limit of ECSNe is defined with the temperature $\sim1.2 \times 10^9$K, and the upper limit is defined with the core mass $=1.375M_\odot$ at the end of C burning. These may underestimate the ECSN rate.
We also note that these mass ranges are highly affected by convective overshooting, mass-loss, and reaction rates, as well as binary effects, and some of the important physics, such as the URCA process \citep{jon14}, are also not included.
There is no region where the core mass is larger than the Chandrasekhar mass limit in the models considered by \citet{doh15}; if there were, the stars could explode as so-called Type 1.5 SNe, although no signature of such supernovae has yet been observed (K06).
In previous GCE models \citep[e.g.,][]{ces14}, a much larger mass range was adopted; for example, if $8-10M_\odot$, the ECSN rate is $\sim 8-18$ times larger than in our models depending on the metallicity.

The nucleosynthesis yields (up to and including Ni) of super AGB stars are taken from \citet{doh14a,doh14b}; the available models are $6.5-7.5M_\odot$, $6.5-7.5M_\odot$, $6.5-8M_\odot$, $6.5-8M_\odot$, and $7-9M_\odot$ respectively for $Z=0.0001, 0.001, 0.004, 0.008$, and $0.02$, and we use these at the masses where Karakas's yields are not available (Table \ref{tab:mix}). The initial abundances are the scaled solar abundances from \citet{gre96}.

\begin{table} 
\begin{tabular}{l|ccc}
\hline
& stellar mass [$M_\odot$] & rotation & magnetic field \\
\hline
ECSN       & $\sim 8.8-9$ & no & no \\
SNII/Ibc   & $10-30$ & no & no \\
failed SN  & $30-50$ & no & no \\
HN         & $20-50$ & yes & weak? \\
MRSN       & $25-50$ & yes & strong \\
\hline 
\end{tabular} 
\caption{\label{tab:mrsn}
The mass ranges of core-collapse supernovae used in our fiducial GCE model, and necessary conditions for the explosions. See the text for the details.}
\end{table}

{\bf Core-collapse supernovae} --- 
Although a few groups have presented multi-dimensional simulations of exploding $10-25 M_\odot$ stars \citep{mar09,kot12,bru13,bur13},
the explosion mechanism of core-collapse (Type II, Ib, and Ic) supernovae is still uncertain;
the ejected iron mass in explosion simulations is not as large as observed \citep{bru16} and the formation of black holes is also not followed in most cases, except for \citet{kur18}.
Therefore, we use the nucleosynthesis yields from one-dimensional (1D) calculations of K06/K11\footnote{Three models of K06 were replaced in K11, which is important for isotopic ratios.
The K11 yield table is identical to that in \citet{nom13}.}.

Similar 1D nucleosynthesis yield of massive stars and supernovae have been provided by three different groups (\citealt{woo95,nom97,lim03}, see Fig.\,5 of \citealt{nom13} for the comparison) and are constantly being updated \citep[e.g., ][]{kob06,heg10,lim18}.
The uncertainties include the reaction rates (namely, of $^{12}$C($\alpha$,$\gamma$)$^{16}$O), mixing in stellar interiors, rotationally induced mixing processes, and mass loss via stellar winds, which affect the yields of elements/isotopes formed during hydrostatic burning.
Furthermore, the most important uncertainty in the yields is associated with the formation of remnants (i.e., neutron stars or blackholes) in massive stars, and different methods have been used to address this problem.
The iron mass of \citet{woo95} is known to be too large, and is usually reduced by a factor of 2 or 3 \citep[e.g.,][]{rom10}, but this modification causes an inconsistency for the other iron-peak elements, which are formed in the same layer as iron and should be reduced by the same amount.
In \citet{nom97}, the remnant masses were determined from one parameter, the mass cut, which self-consistently determined the yields of iron-peak elements as well.
As shown in multidimensional simulations \citep[e.g.,][]{jan12,bru16}, remnant formation is not well described with the mass cut, and the material around the boundary is mixed and partially ejected or falls back onto the remnant. To mimic these phenomena in 1D calculations, the mixing-fallback model was introduced by \citet{ume02}.

As in K06, the ejected explosion energy and $^{56}$Ni mass (which decays to $^{56}$Fe) are determined to meet an independent observational constraint: the light curves and spectral fitting of individual supernova \citep{nom01,nom13}.
As a result it is found that many core-collapse supernovae ($M \ge 20M_\odot$) have an explosion
energy that is more than 10 times that of a regular supernova ($E_{51}\equiv E/10^{51}$ erg $\gtsim10$), as well as production of more
iron and $\alpha$ elements. These are called hypernovae (HNe), while all other supernovae with $E_{51}=1$ are refereed to as SNe II.
The nucleosynthesis yields are provided separately for SNe II and HNe as a function of the 
progenitor mass ($M= 13$, 15, 18, 20, 25, 30, and $40 M_\odot$) and metallicity ($Z= 0$, 0.001, 0.004, 0.02, and 0.05).
As mentioned above, the yield tables provide the amount of processed metals ($p_{z_im}$) in the ejecta (in $M_\odot$), and the unprocessed metals are added in the GCE.
The fraction of HNe at any given time is uncertain and is set $\epsilon_{\rm HN}=0.5$ for masses $M\ge20M_\odot$ following previous works (K06/K11), while a metal-dependent fraction 
$\epsilon_{\rm HN}=0.5, 0.5, 0.4, 0.01$, and $0.01$ for $Z=0, 0.001, 0.004, 0.02$, and $0.05$ was introduced in \citet{kob11mw} in order to match the observed rate of broad-line SNe Ibc at the present day \citep[e.g.,][]{pod04}.
The metallicity-dependent HN fraction is also tested in this paper.

It is known that multi-dimensional effects are particularly important for some elements, e.g., Sc, V, Ti, and Co \citep{mae03,tom09}. We calculated the K15 GCE model, which is plotted in \citet{sne16}, \citet{zha16}, and \citet{reg17}, applying constant factors, $+1.0$, 0.45, 0.3, 0.2, and 0.2 dex for [(Sc, Ti, V, Co, and $^{64}$Zn)/Fe] yields, respectively, which takes the 2D jet effects of HNe into account.
We also show this K15 model for some elements in this paper.

Stellar rotation induce mixing of C into the H-burning shell, producing a large amount of 
primary nitrogen, which is mixed back into the He burning shell \citep{mey02,hir07}. For high initial rotational velocities at low metallicity (``spin stars''), this process results in the production of s-process elements, even at low metallicities \citep{fri16,lim18,cho18}.
\citet{chi06} showed that rotation is necessary to explain the observed N/O--O/H relations with a GCE model and the same result was shown in Fig.13 of K11.
However, using more self-consistent cosmological simulations, \citet{vin18b} reproduced the observed relation not with rotation but with inhomogeneous enrichment from AGB stars.
Therefore, we do not include yields from rotating massive stars in this paper.
\citet{pra18} showed a GCE model assuming a metallicity-dependent function of the rotational velocities, and concluded that because of the contribution from rotating massive stars a further light element primary process (LEPP) is not necessary to explain the elemental abundances with $A<100$.
In the following sections, we will show that we do not need to include fast rotating stars for these elements neither since other sources are present in our models.

{\bf Failed supernovae} ---
The upper limit of SNe II supernovae, $M_{\rm u,2}$, is not well known owing to uncertainties in 
the physics of blackhole formation,
and was set as $M_{\rm u,2}=50M_\odot$ in K06/K11, which is the same for HNe.
However, recently it has been questioned if massive SNe II can explode or not, both observationally and theoretically.
In searching for the progenitor stars at the locations of nearby SNe II-P, no progenitor stars have been found with initial masses $M>30M_\odot$ \citep{sma09}.
In multidimensional simulations of supernova explosions, it seems very difficult to explode stars $\gtsim 25M_\odot$ with the neutrino mechanism \citep[e.g.,][]{jan12}, and similar results are obtained with parametrized 1D models \citep{ugl12,mue16}.
At lower metallicities, since the stellar cores become more compact, it might even be harder to explode. However, the metallicity dependence is probably not very straight forward and may be non-monotonic \citep{pej15}.

Therefore, in this paper we include new nucleosynthesis yields of `failed' supernovae (Kobayashi \& Tominaga 2020, in prep.) at the massive end of SNe II, while keeping the contributions from HNe.
It is assumed that all CO cores fall onto black holes and is not ejected to the ISM, since the timescales of the multi-dimensional simulations are not long enough to follow this process.
The upper mass limit of SNe II, $M_{\rm u,2}$, is treated as a free parameter, while the upper mass limit of HNe is the same as the upper limit of initial mass function, $M_{\rm u}$.
In our fiducial model
$M_{\rm u,2} = 30M_\odot$ and $M_{\rm u} = 50M_\odot$ are adopted (at $>30M_\odot$, the yields are interpolated between the values at $30M_\odot$ and 0 at $40M_\odot$).
If we assume all ejecta collapses onto blackholes, the evolution of C and N is slightly different, but there is no significant difference in the evolution of heavier elements.

Confusingly, failed supernovae are not related to faint supernovae \citep{nom13,ish14,ish18}, which are suggested by completely different observational results.
At [Fe/H] $\ltsim -2.5$, a large fraction of stars are carbon enhanced relative to iron (CEMP stars, [C/Fe] $> 0.7$ in \citealt{aok07}, but see \citealt{bee05} for a different definition), with increasing the fraction toward lower metallicities \citep[e.g.,][]{pla14}.
CEMP stars with an s-process enrichment (CEMP-s, [Ba/Fe] $>1$) are well explained by the binary mass transfer from AGB stars, while CEMP with no s-process enhancement (CEMP-no stars, [Ba/Fe] $<0$) are observed to be both single and binaries \citep{han16binary} and several scenarios suggested (see \citealt{nom13} and the references therein).
Faint supernovae are core-collapse supernovae from massive ($\gtsim$  $13M_\odot$) stars possibly only at $Z=0$, with normal or large explosion energy (i.e., faint SNe or faint HNe) that leave relatively large black holes and eject C-rich envelope.
Because of the small ejecta mass, the contribution to GCE is negligible and thus we do not include the yields of faint supernovae and exclude CEMP stars from most of figures in this paper.

{\bf Pair-instability supernovae} ---
Stars with $100M_\odot\!\ltsim$ $M\ltsim 300M_\odot$ encounter the electron-positron pair instability 
and do not reach the temperature of iron photodisintegration.
Pair-instability supernovae (PISNe) are predicted to produce a large amount of metals such as S and Fe \citep{barkat,heg02,ume02}.
Despite searching for many years, no conclusive signature of PISNe has been detected in metal-poor stars in the solar neighborhood \citep{ume02,cay04,kel14}, the bulge \citep{how15}, or in metal-poor damped Lyman $\alpha$ system \citep{kob11dla}.
Therefore, we do not include PISNe in this paper.

\begin{table*}
\center
\caption{\label{tab:yield}
IMF-weighted return fractions and yields of core-collapse supernovae for the models in Fig.~\ref{fig:sfr} for different upper mass limits of the IMF ($M_{\rm u}$) and SNe II ($M_{\rm u,2}$).}
\begin{tabular}{l|ccccccc}
\hline
 & $M_{\rm u}$ [$M_\odot$] & $M_{\rm u,2}$ [$M_\odot$] & $R(Z=0)$ & $R(0.02)$ & $y(0)$ & $y(0.02)$ \\
\hline
K20           & 50 & 30 & 0.43 & 0.51 & 0.017 & 0.015\\
K20 no failed SNe & 40 & 40  & 0.43 & 0.50 & 0.018 & 0.015 \\
K20 metal-dependent $\epsilon_{\rm HN}$   & 50 & 30  & 0.43 & 0.50 & 0.017 & 0.012 \\
VK18          & 50 & 25, $Z\ge Z_\odot$ & 0.44 & 0.50 & 0.021 & 0.006 \\
K11 & 50 & 50  & 0.43 & 0.51 & 0.021 & 0.019 \\
\hline
\end{tabular}
\end{table*}

Table \ref{tab:mrsn} summarizes the possible necessary conditions for the different types of core-collapse supernovae discussed in this and the next subsection. Note that these are very uncertain, and should be investigated with 3D/GR/MHD simulations of supernova explosions.

\subsubsection{Sites for rapid neutron-capture processes}
\label{sec:r}

The solar abundances of neutron-capture elements require both the s-process and r-process \citep[e.g.,][]{cam73}.
Observations of neutron-capture elements in nearby metal-poor stars have revealed both cases of universality, where the elemental abundance patterns of r-process rich stars are almost identical to that in the Sun \citep{sne08}, and of diversity, where some stars show a deficiency of heavy r-process elements at $A \gtsim 130$, similar to the weak-r process pattern \citep{hon04}.
In this paper, the following sites are included for the r-process.

{\bf Electron-capture supernovae (ECSNe)} ---
After the super-AGB phase, because of electron captures $^{24}$Mg(e$^-, \nu$)$^{24}$Na(e$^-, \nu$)$^{24}$Ne and $^{20}$Ne(e$^-, \nu$)$^{20}$F(e$^-, \nu$)$^{20}$O, the electron fraction $Y_{\rm e}$ decreases, which can trigger collapse \citep{miy80,nom87}.
The collapsing O+Ne+Mg cores have a steep surface density gradient and loosely bound H/He envelope, which can cause prompt explosions. Indeed, \cite{kit06} obtained self-consistent explosions with a 1D hydrodynamical code with neutrino transport.
This is the case for SN 1054 that formed the Crab Nebula \citep{nom82}.
Although 1D nucleosynthesis calculations did not have low enough $Y_{\rm e}$ \citep{hof08,wan09} for heavy r-process elements, 2D calculations showed $Y_{\rm e}$ down to $0.40$ \citep{wan11}, which leads to a weak r-process up to $A \sim 110$.
We apply the nucleosynthesis yields from the 2D calculation of an ECSN from an $8.8M_\odot$ star \citep{wan13b} for all ECSNe.
Note that neutrino oscillations may affect the nucleosynthesis yields of ECSNe \citep{wu14,pll15}.

{\bf Neutrino-driven winds ($\nu$-winds)} ---
Neutron stars (NSs) are born as hot and dense environments from which neutrinos diffuse out leading to a process of mass loss known as $\nu$-driven winds.
1D hydrodynamical codes with neutrino transport showed that the conditions of these winds are not suitable for the occurrence of the r-process \citep{arc07,fis10}.
\citet{wan13a} confirmed this with semi-analytic nucleosynthesis calculations, and showed the dependence of proto-NS mass.
Although proto-NSs with masses $>2.0 M_\odot$ can eject heavy r-process elements, the others eject light trans-iron elements made by quasi nuclear statistical equilibrium (Sr, Y, and Zr) and by a weak r-process up to $A \sim 110$.
Based on the initial mass to NS mass relation from 1D hydrodynamical simulations by \citet{arc07}, we add the nucleosynthesis yields of $\nu$-driven winds from the proto-NS masses $1.4, 1.6, 1.8$, and $2.0M_\odot$ \citep{wan13a} to our SNe II yields of $13, 15, 20$, and $40M_\odot$ stars, respectively.
Similar results as \citet{arc07} are obtained with 2D simulations \citep{arc11jan}, although the impact of multi-dimensional modelling on nucleosynthesis yields needs to be studied further.

{\bf Neutron-star mergers (NSMs)} ---
Compact binary mergers, i.e., NS-NS and NS-BH mergers, have been considered as a possible site of the r-process \citep[e.g.,][]{lat74}.
Recently, the existence of such an event was confirmed by the gravitational wave source GW170817 \citep{abb17a}, associated with an astronomical transient AT2017gfo \citep{sma17,val17} and a short $\gamma$-ray burst GRB170817A \citep{abb17b}.
The spectra of AT2017gfo can be well explained with the emissions peaking in near-infrared from the dynamical ejecta with heavy r-process elements including lanthanides, and the emissions peaking at optical wavelengths from the outflow from BH discs \citep{met14,pia17,tan17}.

Newtonian \citep{ruf97,ros99} and approximate general-relativity (GR) \citep{bau13} 3D simulations showed unbound matter of $\sim 10^{-2}M_\odot$ after NSMs. The ejecta had extremely low $Y_{\rm e}<0.1$ \citep{fre99,bau13}, which can explain the ``universal'' r-process pattern \citep{sne08} at $A\gtsim 130$ but not at $A\ltsim 130$.
However, in a full-GR 3D simulation with approximate neutrino transport, the dynamical ejecta
exhibit a wide range of $Y_{\rm e} \sim 0.09-0.45$, which gives good agreement with the ``universal'' r-process pattern for $A\sim90-240$.
We use the nucleosynthesis yields from the 3D-GR calculation of a NS-NS merger ($1.3M_\odot+1.3M_\odot$, \citealt{wan13b}) both for NS-NS and NS-BH mergers.
Note that, however, double NS systems with a mass ratio $<1$ \citep{fer18} might lead tidal disruption and larger r-process production. Also,
a recent full-GR simulation of a NS-BH merger ($1.35M_\odot+5.4M_\odot$) shows a smaller outflow but the ejecta is very neutron rich \citep{kyu18}, so that the nucleosynthesis yields may be significantly different.
% ML
Furthermore, the overall outflow may be dominated by winds from the accretion disks formed after merger \citep{radice18}.

The rate of NS-NS mergers is estimated as $10^{-5}$ per year per galaxy from the Galactic pulsar population \citep[e.g.,][]{van96}.
The delay-time distributions of NSMs are predicted from binary population synthesis codes \citep[e.g.,][]{tut93,men14,bel18,kru18,vig18}, but the results depend on many parameters that describe uncertain physics such as Roche lobe overflow and common envelope evolution, as well as on the distribution of initial binary parameters.
We adopt the delay-time distributions of the standard model from \citet{men14} for $Z=0.002$ and $Z=0.02$, which are shown in Fig.\,3 of \citet{men16}, assuming a binary fraction of 100\%; we use the rates at $Z=0.002$ and $Z=0.02$ for $Z \le 0.002$ and $Z \ge 0.02$, respectively.
Supernova kick is also one of the most important assumptions for NSMs rates, and the average velocity 450 km s$^{-1}$ is adopted in these rates.
With 265 km s$^{-1}$ \citep{hob05}, the NS-NS and NS-BH merger rates are increased by a factor of 1.6 and 1.2, respectively, and with other parameters, these rates can be increased by a factor of $\sim 20$ and $\sim 30$, respectively \citep{men14}.

{\bf Magneto-rotational supernovae (MRSNe)} ---
While the explosions of normal core-collapse supernovae (referred to as SNe II in this paper) are likely to be triggered by a standing accretion shock instability \citep[e.g.,][]{jan12}, strong magnetic fields and/or fast rotation could also induce core-collapse supernovae.
Such magneto-rotational supernovae are also considered as an r-process site \citep{sym84,cam03}.
Followed by a few axisymmetric magneto-hydrodynamic (MHD) simulations \citep[e.g.,][]{tak09}, a full 3D MHD simulation is performed for $15M_\odot$ star with $5\times10^{12}$ G, which shows a clear jet-like explosion \citep{win12}. \citet{mos14} showed in a 3D MHD GR simulation for $25M_\odot$ star with $10^{12}$ G that the jet is disturbed and no runaway explosion is obtained during the simulation time. It is not sure whether they can explode and produce enough r-process elements or not.

\citet{nis15} calculated nucleosynthesis yields as a post-process based on 2D special relativistic MHD simulations for a $25M_\odot$ star \citep{tak09} depending on the strength of magnetic fields and the rotational energy. Actually, \citet{tak09} calculated $1.69M_\odot$ iron core from a star with initial mass $25M_\odot$, solar metallicity, and equatorial rotational velocity of $\sim 200$ km s$^{-1}$ \citep{heg00}. The late phase evolution of jet propagation and shock-breakout is not followed. Therefore, it is unknown whether the envelope of the iron core is also ejected or falls back onto the remnant. The proto-NSs have $10^{15}$ G and could be the origin of the magnetars, but probably do not become long gamma-ray burst or HNe since the jet is only mildly relativistic.
Because of the necessary conditions of rotation and some magnetic fields, MRSNe and HNe may be related but not the same.

In this paper, we replace 3\% of $25-50M_\odot$ HNe with MRSNe. In the model with the metal-dependent HN fraction, the MRSN rate also depends on the metallicity:
$\epsilon_{\rm MRSN}(Z)=0.03\,\epsilon_{\rm HN}(Z)$.
This fraction of MRSNe relative to HNe is chosen from the observed [Eu/Fe]--[Fe/H] relation in the solar neighborhood (\S \ref{sec:sr}), 
but may differ in the Galactic bulge (\S \ref{sec:bulge}).
If we also allow $20M_\odot$ or $15M_\odot$ stars for MRSNe, the rate can be larger by a factor of $\sim 2$ or $\sim 4$, respectively, but the chemical enrichment timescale is not so different.

\subsubsection{Type Ia Supernovae}
\label{sec:ia}

{\bf Type Ia Supernovae (SNe Ia)} ---
The progenitor systems of SNe Ia are still a matter of extensive debate; plausible scenarios are (1) deflagrations or delayed detonations of Chandrasekhar (Ch) mass WDs from single degenerate systems, (2) sub-Ch-mass explosions from double degenerate systems, or (3) double detonation of sub-Ch-mass WDs in single or double degenerate systems \citep[e.g.,][for a review]{hil00}.
Observationally, the progenitors of the majority of `normal' SNe Ia are most likely to be 
Ch-mass WDs \citep{sca14}. 

From the nucleosynthetic point of view, \citet{kob19ia} showed that more than 75\% of SNe Ia should be Ch-mass explosions \citep[see also][]{sei13}, using the nucleosynthesis yields calculated with their 2D hydrodynamical code both for Ch and sub-Ch mass explosions.
Therefore, in this paper we adopt the same yield set but only for delayed detonations in Ch-mass C+O WDs, as a function of metallicity ($Z= 0, 0.002, 0.01, 0.02, 0.04, 0.06$, and $0.10$).
This new yield set solved the Ni overproduction problem in the yields of \citet{nom97ia} and \citet{iwa99}.

The adopted progenitor systems are the binaries of C+O WDs with main-sequence (MS) or red-giants (RG) secondary stars \citep[see][for more details, hereafter KN09]{kob09}, and the mass ranges of the secondary stars depend on the metallicity because the optically thick winds from WDs are essential for the evolution of these progenitor systems \citep[][hereafter K98]{kob98}.
In GCE, the lifetime distribution function of SNe Ia is calculated 
with Eq.[2] in KN09, with the metallicity dependence of the WD winds (K98) and 
the mass-stripping effect on the binary companion stars (KN09).
MS$+$WD systems have timescales of 
$\sim 0.1-1$ Gyr, which are dominant in star-forming galaxies (the so-called prompt population),
while RG$+$WD systems have lifetimes of $\sim 1-20$ Gyr, which are dominant in 
early-type galaxies.
The binary parameters of MS+WD and RG+WD systems, $b_{\rm MS}$ and $b_{\rm RG}$, are mainly\footnote{The total number of SNe Ia ($\sim b_{\rm MS}+b_{\rm RG}$) is determined from the [O/Fe] slope, and $b_{\rm RG}$ is determined from the metal-rich tail of the metallicity distribution function.} determined from the observed [O/Fe]--[Fe/H] relation at [Fe/H] $>-1$, and ($b_{\rm RG}, b_{\rm MS}$) are set to be (0.02, 0.04) for the new models in this paper, while (0.023, 0.023) were used for the K11 and K15 models.
Both sets give very similar [O/Fe]--[Fe/H] relations.
Note that these parameters do not only account for the binary fractions, but include a suitable range of binary separations and any other conditions that are necessary for the systems to explode as SNe Ia \citep[see][for more details]{kob09}.
The resultant delay-time/lifetime distribution is very similar to that observed at $Z \sim 0.02$ \citep[see Fig.\,12 of][]{kob19ia}.

{\bf Type Iax Supernovae (SNe Iax)} ---
There is also a significant number of faint or super-luminous SNe Ia \citep[e.g.,][]{gal17} that are likely to be from sub-Ch or super-Ch WDs \citep{sca19}.
The rate of super-luminous SNe Ia is so small that we do not include them in this paper.
Possibly the subset of faint SNe Ia are included as SNe Iax in the following sections. Since the secondary star of an SN Iax is observed \citep{mcc14}, we adopt the single degenerate model from \citet{kob15}. It is assumed that the progenitors are hybrid C+O+Ne WDs, and we take the mass ranges from the results of super AGB calculations \citep{doh15} depending on the metallicity.

\begin{table}
\center
\caption{\label{tab:param}
Parameters of the GCE models: timescales for infall ($\tau_{\rm i}$), star formation ($\tau_{\rm s}$), and outflow ($\tau_{\rm o}$) , and the galactic wind epoch $\tau_{\rm w}$, all in Gyr.}
\begin{tabular}{l|cccc}
\hline
 & $\tau_{\rm i}$ & $\tau_{\rm s}$ & $\tau_{\rm o}$ & $\tau_{\rm w}$ \\
\hline
solar neighborhood  & 5 & 4.7 & - & - \\
halo                & - & 15  & 1 & - \\
bulge               & 5 & 0.2 & - & 3 \\
thick disk          & 5 & 2.2 & - & 3 \\
halo, outflow       & - & 5   & 0.3 & - \\
bulge, outflow      & 5 & 0.1 & 0.3 & - \\
\hline
\end{tabular}
\end{table}

\subsection{Galactic Chemical Evolution Model}
\label{sec:gce}

\subsubsection{Basic Equations and Constants}

The basic equations of chemical evolution are described in K00. The code follows the time evolution of elemental and isotopic abundances in a system where the ISM is instantaneously well mixed (and thus it is called a one-zone model). No instantaneous recycling approximation is adopted and chemical enrichment sources with long time-delays (\S \ref{sec:sn}) are properly included.

We adopt the initial mass function (IMF) from \citet{kro08}, which is a power-law mass spectrum
$\phi(m) \propto m^{-x}$ with three slopes at different mass ranges:
$x= 1.3$ for $0.5M_\odot \le m \le 50M_\odot$,
$x= 0.3$ for $0.08M_\odot \le m \le 0.5M_\odot$, and
$x=-0.7$ for $0.01M_\odot \le m \le 0.08M_\odot$, which are the same as the canonical stellar IMF in \citet{kro01} and very similar to \citet{chab03} IMF\footnote{The \citet{sal55} IMF ($x=1.35$) is adopted in K06. K11 also used the early star formation Chabrier IMF for comparison.}.
The IMF is normalized to unity at $0.01M_\odot \le m \le M_{\rm u}$, and $M_{\rm u}=50M_\odot$ is adopted in the fiducial model of this paper.
Table \ref{tab:yield} shows the IMF weighted return fractions and yields of core-collapse supernovae for the models we show in the following sections.
The net yields are defined as $1/(1-R)\,y = 1-2 Z_\odot$ \citep{tin80}.

The metallicity dependent main-sequence lifetimes are taken from \citet{kod97} for $0.6-80M_\odot$, which are calculated with the stellar evolution code described in \citet{iwan99}. These are in excellent agreement with the lifetimes in \citet{kar10} for low- and intermediate-mass stars.

We use similar models as K06 for the star formation histories of the solar neighborhood, halo, bulge, and thick disk, but with the Kroupa IMF.
The formation of these components are not simple and affected by dynamical effect such as radial migration and satellite accretion (see \citealt{ath02,bar18,zoc18} for the bulge, and \citealt{min13,gri17,spito19} for the thick disk).
In this paper, in order to give rough evolutionary tracks, we adopt one-zone models.

The gas fraction and the metallicity of the system evolve as a function of time as a consequence of star formation, as well as inflow and outflow of matter to/from the outside of the system.
The star formation rate (SFR) is assumed to be proportional to the gas fraction; $\frac{1}{\tau_{\rm s}} f_{\rm g}$.
The infall of primordial gas from the outside of the system is given by the rate 
$\propto t\exp[-\frac{t}{\tau_{\rm i}}]$ for the solar neighborhood \citep{pagel1997}, and 
$\frac{1}{\tau_{\rm i}}\exp(-\frac{t}{\tau_{\rm i}})$ for the other systems.
In addition, during the star burst at early stages of galaxy formation (such as in the bulge) or the star formation in the shallow gravitational potential well (such as in the halo), outflow is an important process.
The driving source of the outflow is the feedback from supernovae, and hence the outflow rate is also proportional to the gas fraction; $\frac{1}{\tau_{\rm o}}f_{\rm g}$.
The outflow also removes some metals with the composition of the average ISM in the system at the time.
The outflow gas could later fall onto the disk (so-called `fountain'), but this process is not included in the model.
For the bulge and thick disk, star formation may be quenched suddenly by a galactic wind at a given epoch ($t=t_{\rm w}$), which is driven by a large number of supernovae or by a central super-massive blackhole.
The adopted GCE model parameters are summarized in Table \ref{tab:param}.

Compared to K06 and K11, one of the major revision here is the adopted solar abundances, which are now taken from Tables 1 and 3 from \citet[][hereafter AGSS09]{asp09} for all elements except for O, Th and U. For most elements (except for Li and Pb) we use the photospheric value, when that is not available we adopt the meteoritic values.
For O we adopt the oxygen abundance, $A_\odot({\rm O})=8.76 \pm 0.02$, from \citet{ste15}, which is higher than $A_\odot({\rm O})=8.69 \pm 0.05$ in AGSS09.
Note that in K06 and K11, the solar abundances were taken from \citet[][hereafter AG89]{and89}, where the oxygen abundance was $A_\odot({\rm O})=8.93$.
There is no difference in the solar Fe abundance ($A_\odot({\rm Fe})=7.51$ in AG89 and 7.50 in AGSS09).
Thus the [O/Fe] ratios in this paper appear to be 1.5 times larger than in K06 and K11. This difference appears only in the comparison to observational data, and affects the choice of GCE model parameters.
For Th and U, we adopt the initial solar system values $A_\odot({\rm Th})=0.22$ and $A_\odot({\rm U})=-0.02$ from \citet{lod20}, which took into account the radioactive decay of the Th and U isotopes over the past 4.567 Gyr, and we show the model predictions after the long-term decay at each time.

These solar abundances are also applied to the observational data plotted in the figures of the following sections, if necessary. When compared with theoretical predictions, it is better to compare the relative abundances for some cases such as NLTE analysis \citep[e.g.,][]{zha16} and LTE differential analysis \citep{reg17}. If the solar abundances are not measured by the same analysis and are taken from the literature (e.g., AG98, AGS09), then constant shifts are applied to re-normalise with our solar abundances (including \citet{ste15}'s O). For Fe, most of the observational papers used $A_\odot({\rm Fe})=7.50$ or $7.51$, and re-normalisation is applied only for the other cases such as with $7.45$.

The primordial abundances are also updated from K06/K11, which does not affect the figures showing [X/Fe]. The adopted values are D/H $=2.527 \times 10^{-5}$ \citep[][metal-poor damped Ly$\alpha$ systems]{coo18}, $^3$He/H $=1.1 \times 10^{-5}$ \citep[][Milky Way HII regions]{ban02}, $Y=0.2449$ (\citealt{ave15}, low-metallicity H II regions, but see \citealt{izo14}), and theoretical values of $^6$Li/H $=1.27 \times 10^{-14}$ and $^7$Li/H $=5.623 \times 10^{-10}$ \citep{pit18}, which is higher than in \citet{sbo10} but is comparable to the value in \citet{mel10} with a correction of stellar depletion. For $^9$Be and $^{10,11}$B, theoretical values are taken from \citet{coc12}.

\begin{figure*}\center
\includegraphics[width=15cm]{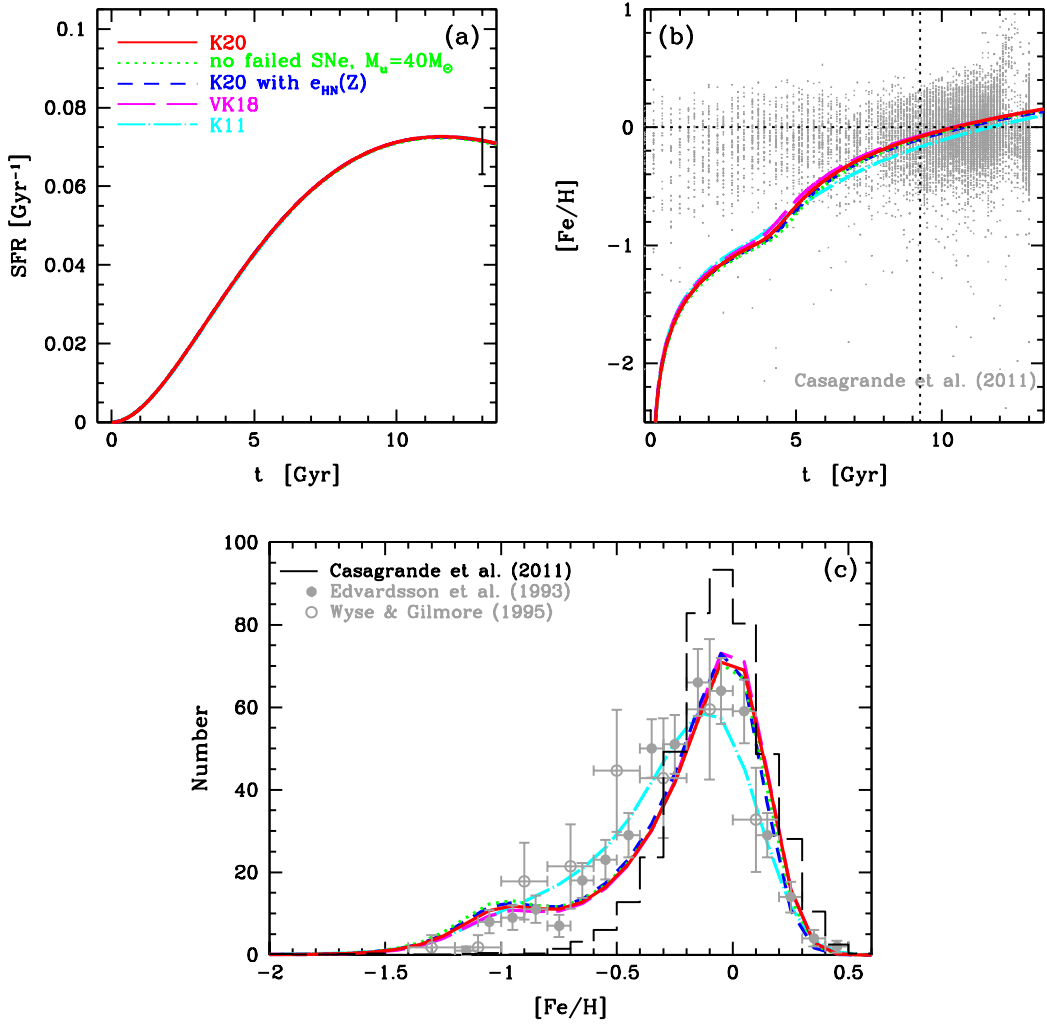}
\caption{\label{fig:sfr}
Star formation histories (panel a), age-metallicity relations (panel b), and metallicity distribution functions (panel c) for the models in the solar neighborhood, where K20 is the fiducial presented in this paper (see discussion in the text).
The observational data sources are:
an error estimate from \citet{mat97} in panel (a);
gray points in panel (b) and 
histogram in panel(c) from \citet{cas11};
gray filled circles, \citet{edv93};
gray open circles, \citet{wys95}.
}
\end{figure*}

\begin{figure*}\center
\includegraphics[width=15cm]{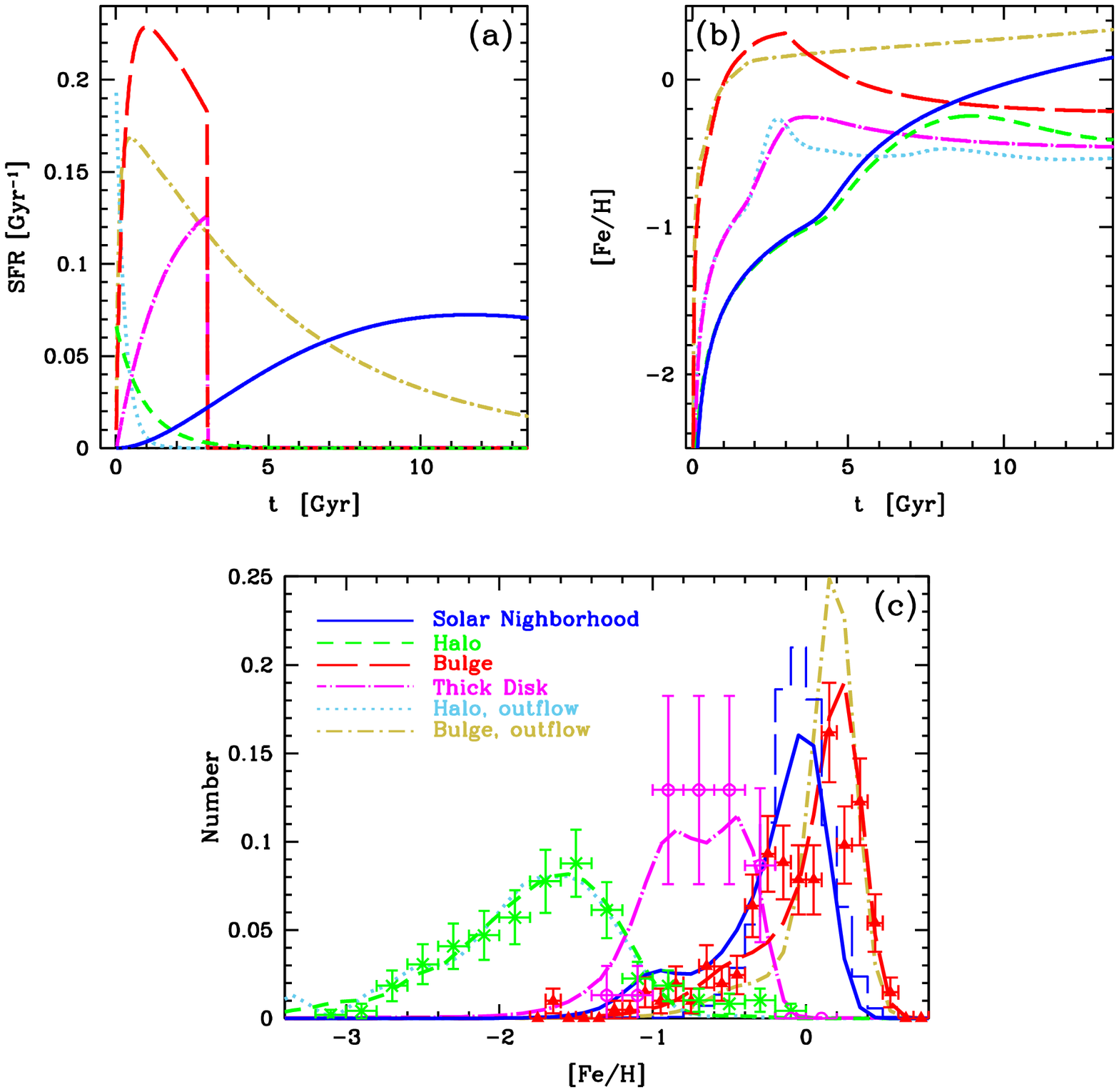}
\caption{\label{fig:sfr2}
Star formation histories (panel a), age-metallicity relations (panel b), and metallicity 
distribution functions (panel c) for the 
solar neighborhood (blue solid lines),
halo (green short-dashed lines),
halo with stronger outflow (light-blue dotted lines),
bulge (red long-dashed lines),
bulge with outflow (olive dot-short-dashed lines)
and thick disk (magenta dot-long-dashed lines).
The observational data sources are:
histogram, \citet{cas11};
crosses, \citet{chi98};
filled triangles, \citet{zoc08};
open circles, \citet{wys95}.
}
\end{figure*}

\subsubsection{Metallicity Distribution Functions and Star Formation Histories}

The parameters for stellar physics (e.g., the IMF) can be determined from independent observations, while the galactic parameters (e.g., $\tau_{\rm i}$, $\tau_{\rm s}$, and $\tau_{\rm o}$) that describe the formation history of the system have to be determined by comparing GCE model predictions to observations. The metallicity distribution function (MDF) is the most important constraint for this purpose.
The GCE model parameters chosen to match the observed MDF of each system are summarized in Table \ref{tab:param}.

The resultant SFR histories (panel a), age-metallicity relations (panel b), and MDF (panel c) of our solar neighborhood models are shown in Figure \ref{fig:sfr}.
In the solar neighborhood, star formation takes place over 13 Gyr, the SFR peaked $\sim 3$ Gyr ago, and declined at the age $\gtsim 5$ Gyr, which is consistent with WD observations \citep{tre14}.
In the recent observational data, there is no tight relation between stellar ages and metallicities \citep{hol07,cas11}.
Our model value is slightly lower than the solar ratio ([Fe/H] $=0$) at the time of the Sun's formation (4.6 Gyr ago), which implies that the Sun is slightly more metal-rich than the average ISM of the solar neighborhood.
The recent MDF \citep{cas11} is narrower than in previous works \citep{edv93,wys95}, where thick disk stars were also included.
The peak is almost solar but is slightly sub-solar, which also means that the Sun is slightly more metal-rich than the average of low-mass stars at present in the solar neighborhood.
K11 model (cyan dot-dashed lines) was constructed to meet the previous MDFs, while in this paper the models are updated in order to match the recent MDF as much as possible.
Since we do not assume pre-enrichment or unreasonably slow infall, it is difficult to perfectly reproduce the narrow MDF as observed.

Figure \ref{fig:sfr2} shows the same as in Figure \ref{fig:sfr} but for the bulge, halo, and thick disk models.
The GCE model parameters are chosen to match the observed MDFs and are summarized in Table \ref{tab:param}; the first four models are the same as in K11, and the second halo model (with stronger outflow) is very similar to the model used in \citet{car18}.

In the bulge, the MDF is peaked at super solar metallicity and has a sharp-cut at the metal-rich end.
This is well reproduced with a rapid star formation (with a short star-formation timescale) truncated with a strong outflow or galactic wind.
Infall is also required to explain the lack of metal-poor stars.
Then the metallicity increases rapidly and reaches solar metallicity only after 1 Gyr, which results in the high [$\alpha$/Fe] ratios at [Fe/H] $\gtsim -1$ \citep{mat90}.
The first bulge model in this paper (red long-dashed lines) includes infall and winds at 3 Gyr after formation,
which results in a peak metallicity of [Fe/H] $\sim +0.3$ at 3 Gyr.
A much higher efficiency of chemical enrichment, e.g., a flatter IMF 
is not required, unless the duration is much shorter than 3 Gyr.
Note that the 3 Gyr duration is consistent with chemodynamical simulations of Milky Way-type galaxies from cold dark matter initial conditions (KN11).
A similar MDF can be produced with the outflow model (olive dot-short-dashed lines) where the star formation is more gradually suppressed by outflows.
In this second bulge model, young and metal-rich stars can be formed, and [Fe/H] increases steadily to $+0.3$ by the present.

Also for the thick disk (magenta dot-long-dashed lines), we use the infall$+$wind model, which gives 
good agreement with the observed age-metallicity relation \citep{ben04b}.
The formation timescale is as short as in the bulge ($\sim 3$ Gyr), and the star formation 
efficiency is smaller than for the bulge but is larger than for the solar neighborhood.
Not only the short duration of star formation but also the intense star formation is necessary to reproduce the observed [$\alpha$/Fe]--[Fe/H] relations in the thick disk stars \citep{ben04}.

In the halo, the MDF has a peak at a much lower metallicity ([Fe/H] $\sim -1.6$, e.g., \citealt{chi98}) and distributes over a wider range of metallicities.
This can be well reproduced with an outflow model without infall.
In the first halo model (green short-dashed lines), the age-metallicity relation is similar to that in the solar neighborhood.
However,  \citet{car18} suggested a faster star formation in the halo from the observed Mg isotopic ratios.
With a shorter star formation timescale, the metallicity would become too high; for this reason a stronger outflow is adopted in the second halo model (light-blue dotted lines).
This model gives the age-metallicity relation similar to that in the thick disk.

\begin{figure}\center
\includegraphics[width=8.5cm]{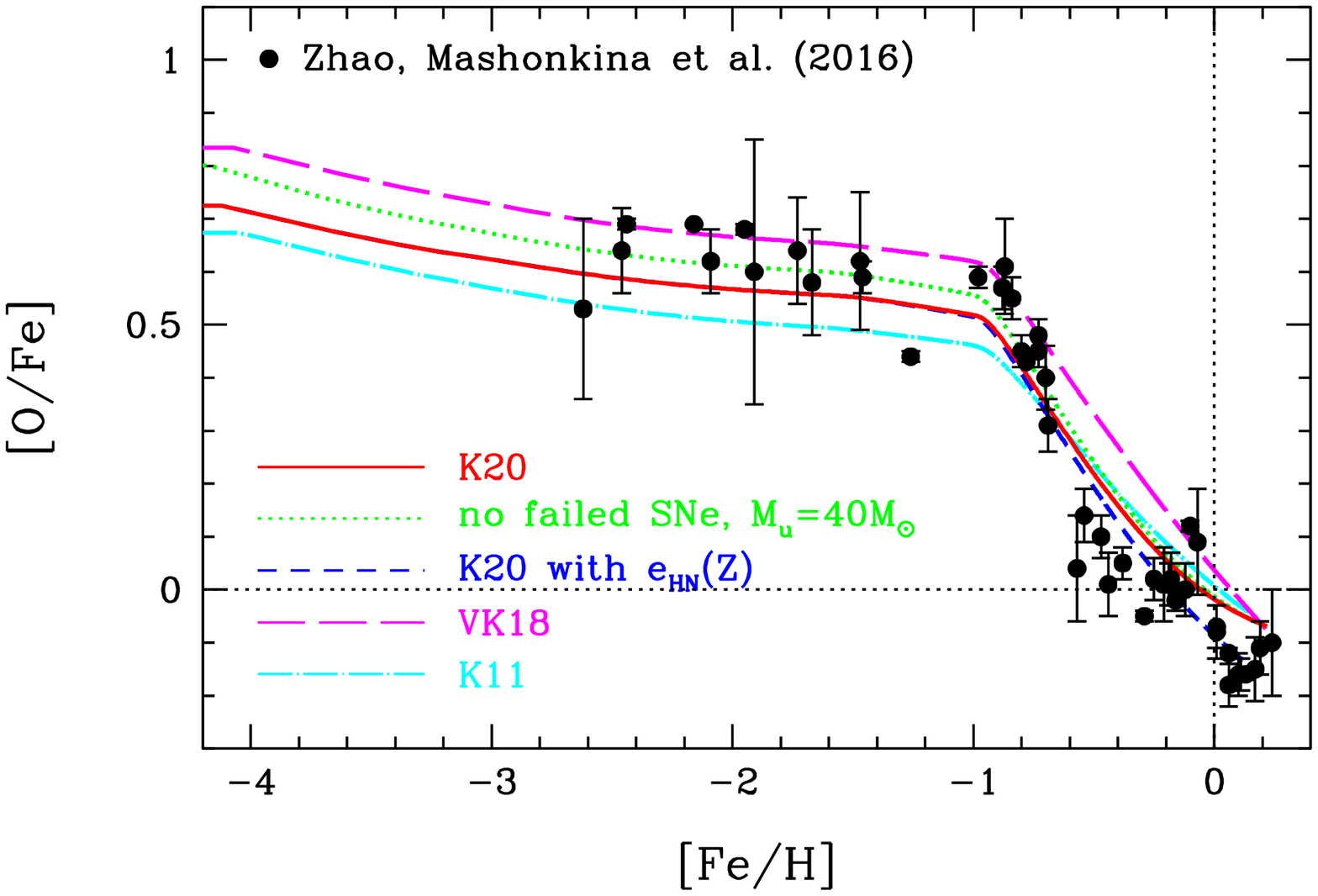}
\caption{\label{fig:ofe}
[O/Fe]--[Fe/H] relations for the models in the solar neighborhood, 
with failed SNe (red solid line), 
without failed SNe but with the IMF upper limit of $40M_\odot$ (green dotted line), 
with failed SNe but with the metal-dependent HN fraction (blue short-dashed line), 
with failed SNe as in \citet[][magenta long-dashed line]{vin18a}
and the model in \citet[][cyan dot-dashed lines]{kob11agb}.
The observational data are obtained with NLTE analysis \citep{zha16}.
}
\end{figure}

\begin{figure}\center
\includegraphics[width=8.5cm]{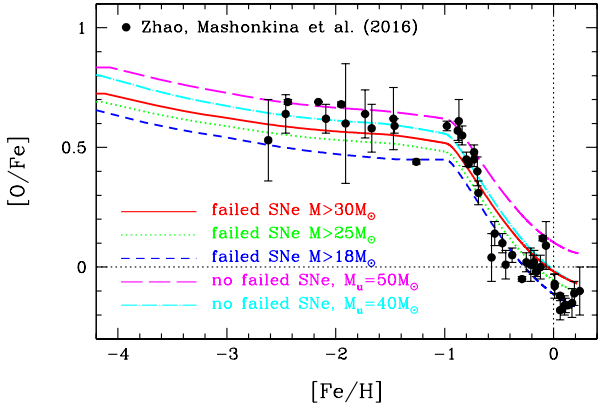}
\caption{\label{fig:ofe2}
[O/Fe]--[Fe/H] relations for the models in the solar neighborhood, 
with failed SNe at $>30M_\odot$ (red solid line, the fiducial model indicated as K20),
at $>25M_\odot$ (green dotted line),
at $>18M_\odot$ (blue short-dashed line),
and without failed SNe
with the same IMF upper limit, $M_{\rm u}=50M_\odot$ (magenta long-dashed line),
and a different IMF upper limit, $M_{\rm u}=40M_\odot$ (cyan dot-dashed line).
The observational data are obtained with NLTE analysis \citep{zha16}.
}
\end{figure}

\begin{figure*}\center
\includegraphics[width=17.5cm]{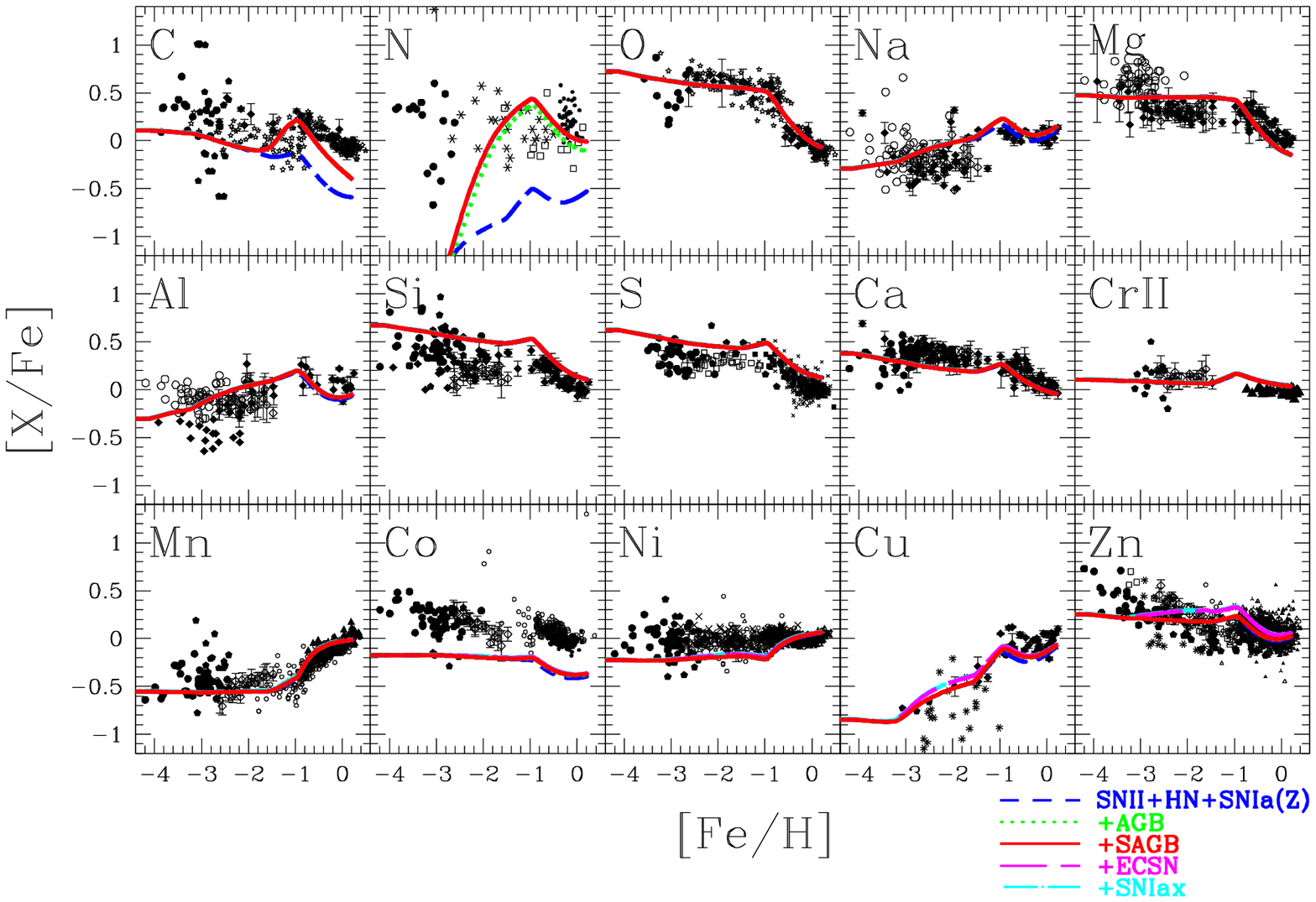}
\caption{\label{fig:xfe}
Evolution of the elemental abundances [X/Fe] from C to Zn against [Fe/H]
for the models in the solar neighborhood,
with only supernovae (without AGB and super-AGB stars, blue short-dashed line),
with AGB without super-AGB stars (green dotted lines),
with AGB and super-AGB stars (red solid line, fiducial model),
with ECSNe (magenta long-dashed lines),
and with SNe Iax (cyan dot-dashed lines).
The observational data sources are:
filled diamonds with error bars, \citet{zha16,mas17,mas19} with NLTE for C, O, Na, Mg, Al, Si, Ca, and Cu;
open diamonds with error bars, \citet{reg17} with differential analysis for Na, Mg, Al, Si, Ca, Cr II, Mn, Co, Ni and Zn;
stars, \citet{ama19b} with 3D/NLTE C, O, and 3D Fe;
filled circles, \citet{spi06} for C, N, and O (unmixed stars only);
asterisks, \citet{isr04} for N;
diamonds, \citet{car00} for N (unmixed stars only);
small filled and open circles, \citet{red03,red06,red08} for N, Mn, Co, Ni, Zn of thin and thick/halo stars, respectively;
filled circles, \citet{spi11} for S;
filled pentagons, \citet{tak05} for S;
filled squares, \citet{che02} for S;
crosses, \citet{cos20} for S;
open diamonds, \citet{nis07} for S and Zn;
filled pentagons, \citet{hon04} for C, Cr II, Mn, Co, Ni, Cu, Zn;
filled triangles, \citet{ben04} for Cr II;
\citet{cay04} for Mn, Co, Ni, Zn;
\citet{fel07} for Mn;
crosses, \citet{ful00} for Ni;
filled and opened triangles, and three-pointed stars, \citet{ben14} for Ni, Zn of thin, thick disk, and intermediate stars, respectively;
asterisks, \citet{pri00} for Cu and Zn;
filled pentagons, \citet{sai09} for Zn.
All observational data are shifted for our adopted solar abundances if necessary.
}
\end{figure*}

\section{Results}

\subsection{Constraining failed supernovae from the [O/Fe]--[Fe/H] relation}

Figure \ref{fig:ofe} shows the evolution of [O/Fe] against [Fe/H] for the solar neighborhood.
In the early stages of galaxy formation, only SNe II/HNe contribute and the [O/Fe] ratios form a plateau at a wide range of [Fe/H]
([O/Fe] $= 0.62, 0.57, 0.52$ at [Fe/H] $=-3, -2, -1.1$).
The small slope at the low-metallicity end is caused by the mass dependence of the SN II/HN yields.
Around [Fe/H] $\sim -1$, SNe Ia start to occur, which produce more iron than $\alpha$ elements such as oxygen.
This delayed enrichment of SNe Ia causes the decrease in [O/Fe] with increasing [Fe/H] \citep{mat86}.
The observational data are the NLTE abundances obtained from the homogeneous analysis of a relatively large sample of high-resolution spectra of nearby stars and of the Sun \citep{zha16}, which reveal the following three features.
First, the [O/Fe] plateau value obtained here is $\sim 0.6$, slightly higher than in K06 and K11 (cyan dot-dashed line).
Second, the [O/Fe] plateau continues to [Fe/H] $\sim-0.8$ and then the [O/Fe] ratio sharply decreases.
The [Fe/H] at which the [$\alpha$/Fe] starts to decrease depends on the adopted SN Ia progenitor model, and is determined not by the lifetime but by the metallicity dependence of SN Ia progenitors in our models.
It is very difficult to reproduce this rapid evolutionary change without the metallicity effect of SNe Ia (\citealt{kob98,kob09}; Fig.15 of \citealt{kob19ia}).
Third, the abundance ratios approach the solar ratios (i.e., [O/Fe] $=$ [Fe/H] $=0$).
Our fiducial model (red solid line) can reproduce all of these features very well. 

With the metal-dependent HN fraction (blue short-dashed line), at $Z\gtsim Z_\odot$, the metal production from core-collapse supernovae is assumed to be very small compared to SNe Ia; the present-day HN fraction is only 1\% and the rest of massive supernovae are failed supernovae (i.e., no O and Fe production).
This results in lower [O/Fe] at [Fe/H] $\sim 0$, which may be more consistent with these observational data.
If we simply exclude failed SNe, the predicted [O/Fe] plateau value would become higher than observed. Therefore, in the model without failed SNe, we reduce the upper mass limit of the IMF from $50M_\odot$ to $40M_\odot$ (green dotted line), so that the model has a [O/Fe] plateau value consistent with the observations.
In VK18 (magenta long-dashed line), we assumed failed SNe at mass $\ge 25M_\odot$ and metallicity $\ge 0.02$, where all synthesised O and heavier elements fall back onto a blackhole, except for H, He, C, N, and F that are synthesised in the outermost layers of the SN ejecta.

More parameter studies are shown in Figure \ref{fig:ofe2} for the [O/Fe]--[Fe/H] relation.
In the adopted nucleosynthesis yields, at any given mass, [O/Fe] is  larger for SNe II than for HNe.
Therefore the models with failed SNe at lower progenitor masses give systematically lower [O/Fe] ratios (red solid, green dotted, and blue short-dashed lines).
Adopting $30M_\odot$ as the upper limit of SNe II provides the best fit to the observations.
As mentioned above, without failed SNe, i.e., if SNe II occur up to $50M_\odot$ as HNe (magenta long-dashed line), the [O/Fe] ratio becomes too high. 
With changing the IMF upper limit from $50M_\odot$ to $40M_\odot$ (cyan dot-dashed line), both SNe II and HNe occur only up to $40M_\odot$, then the [O/Fe] ratio is consistent with observations.
This conclusion depends on the updated solar oxygen abundance, and was not drawn in K06 or K11.

Note that in the NLTE analysis of \citet{zha16} the solar abundances are obtained for each line and the oxygen solar abundances vary from $A_\odot({\rm O})=8.74$ to 8.82 (and are not shifted, see \S \ref{sec:gce}). Our value 8.76 (that applied to other observations) lies in the range, but not the AGSS09 value. Since there is a 0.05 dex uncertainty depending on the choice of the solar abundance, the model without failed SNe but with $M_{\rm u}=40M_\odot$ (green dotted line) and the model with failed SNe at $> 25 M_\odot$ (blue short-dashed line) are also acceptable.

\begin{figure*}\center
\includegraphics[width=17.5cm]{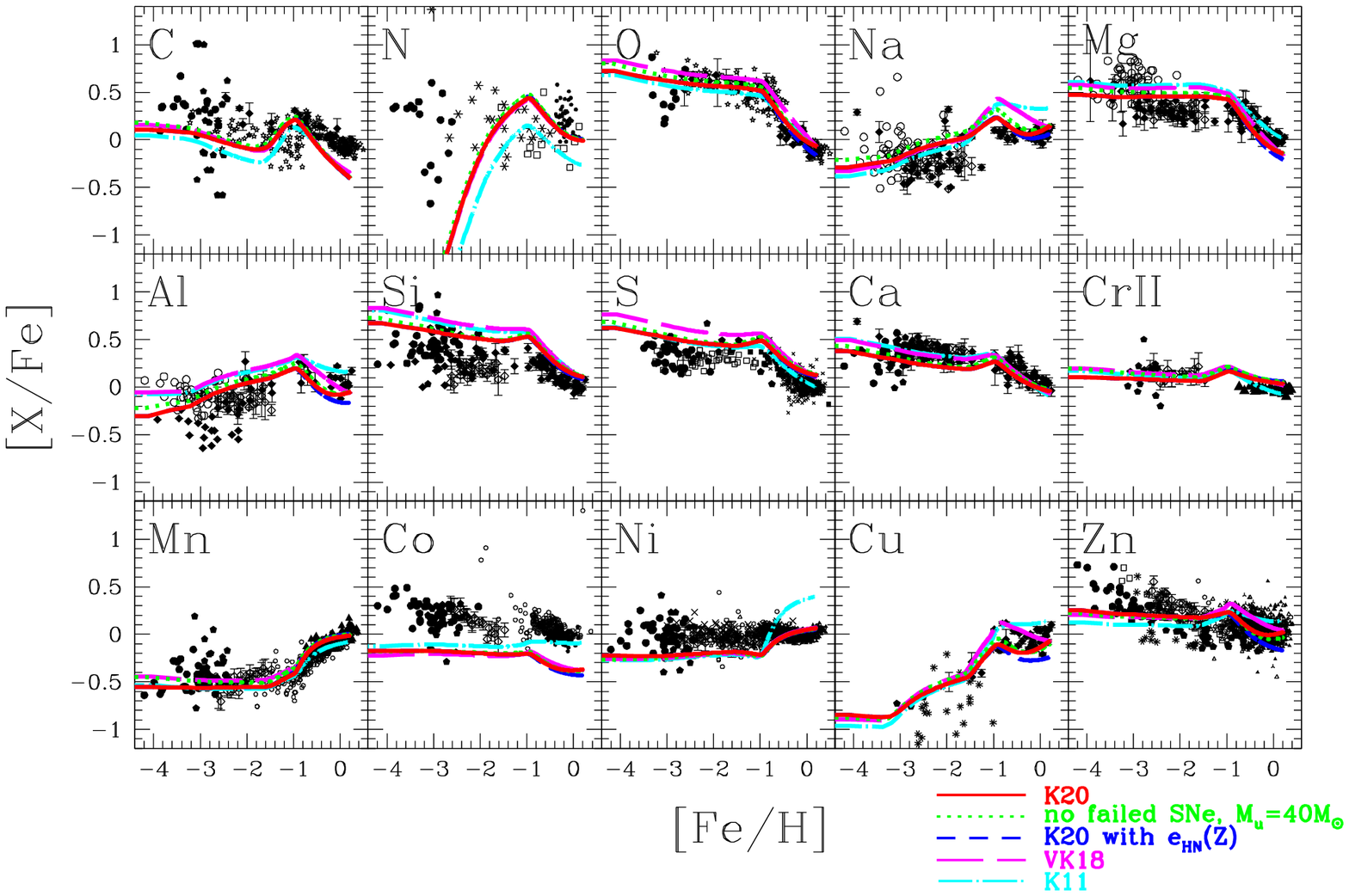}
\caption{\label{fig:xfe2}
Same as Fig.~\ref{fig:ofe} but the elemental abundances [X/Fe] from C to Zn.
See Fig.~\ref{fig:xfe} for the observational data sources.
}
\end{figure*}

\subsection{Elemental abundances from C to Zn}

\begin{figure}\center
\includegraphics[width=8.5cm]{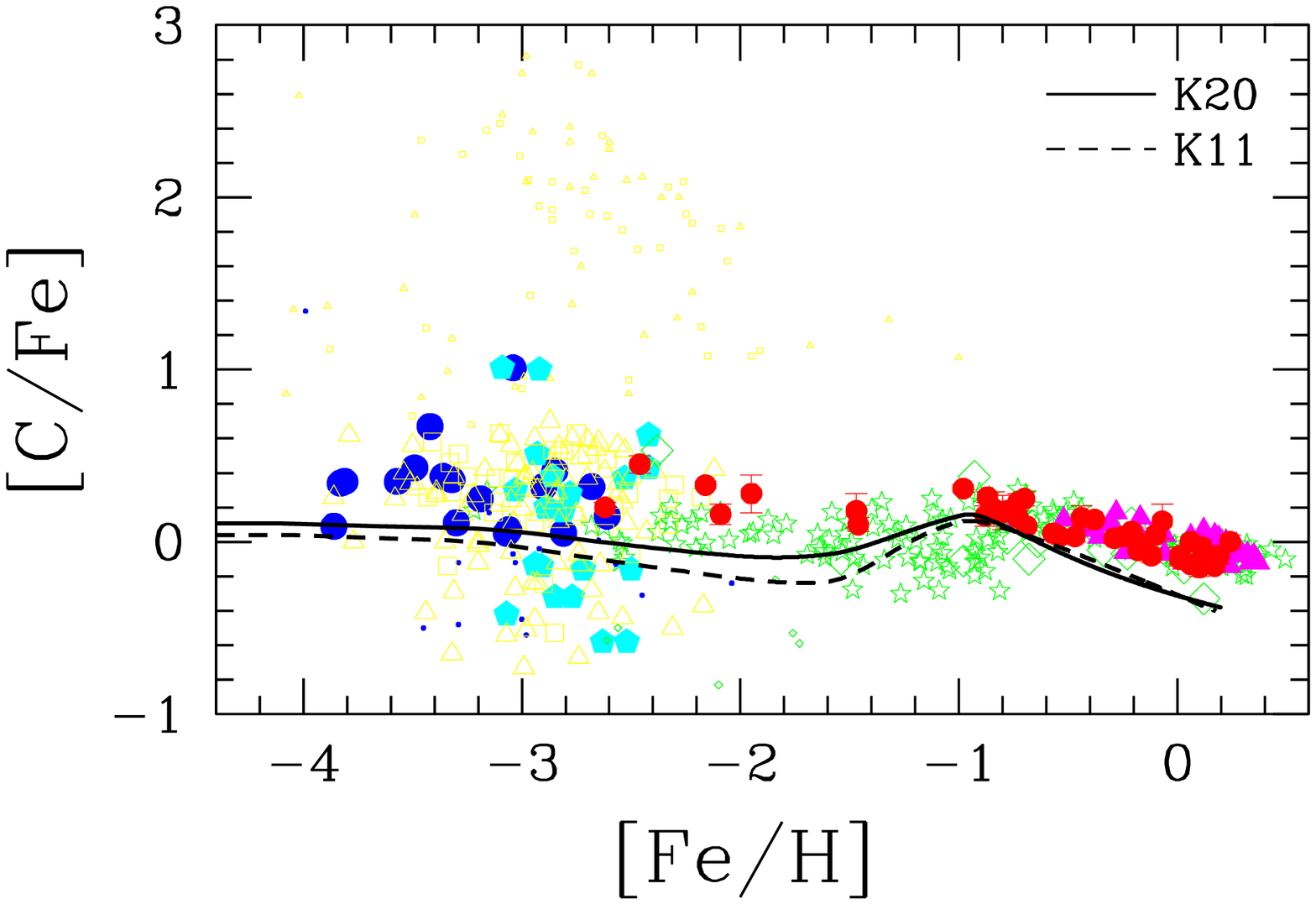}
\caption{\label{fig:c}
[C/Fe]--[Fe/H] relation for the solar neighborhood models in this paper (solid line) and in K11 (dashed line).
Observational data sources (mostly for CH) are:
blue filled circles, \citet{spi06} with smaller symbols denoting mixed stars; %4224 (G band) 
cyan filled pentagons, \citet{hon04}; %CH 4323
yellow open squares, \citet{coh13} with smaller symbols denoting CEMP stars;
yellow open triangles, \citet{yon13} with smaller symbols denoting CEMP stars;
green stars, \citet{ama19b} with 3D/NLTE C and 3D Fe;
green open diamonds, \citet{car00} with smaller symbols denoting mixed stars; %with CH 4207-4225
magenta filled triangle, \citet{ben06} for [CI]; %872.7 nm
red filled circles with error bars, \citet{zha16} with NLTE.
}
\end{figure}

\begin{figure}\center
\includegraphics[width=8.5cm]{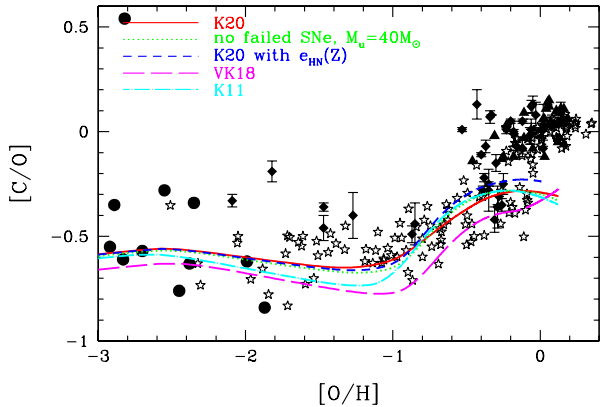}
\caption{\label{fig:coo}
Same as Fig.\ref{fig:ofe} but for [C/O] plotted against [O/H].
Observational data sources are:
filled circles, \citet{spi05} for unmixed stars with LTE O and 3D correction $-0.23$;
stars, \citet{ama19b} with 3D/NLTE;
filled diamonds with error bars, \citet{zha16,mas17,mas19} with NLTE;
filled triangles, \citet{ben06}.
}
\end{figure}

\begin{figure}\center
\includegraphics[width=8.5cm]{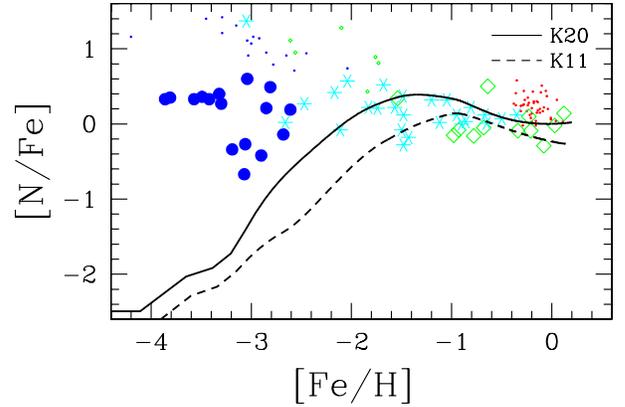}
\caption{\label{fig:n}
Same as Fig.\ref{fig:c}, but for [N/Fe]--[Fe/H] relation.
Observational data sources are:
blue filled circles, \citet{spi06} for UV NH, with 3D correction $-0.40$, with smaller symbols denoting mixed stars; %at 336nm
cyan asterisks, \citet{isr04} for UV NH;
green open diamonds, \citet{car00} for CN, with smaller symbols denoting mixed stars;
red small filled circles, \citet{red03} for NI.
}
\end{figure}

\begin{figure}\center
\includegraphics[width=8.5cm]{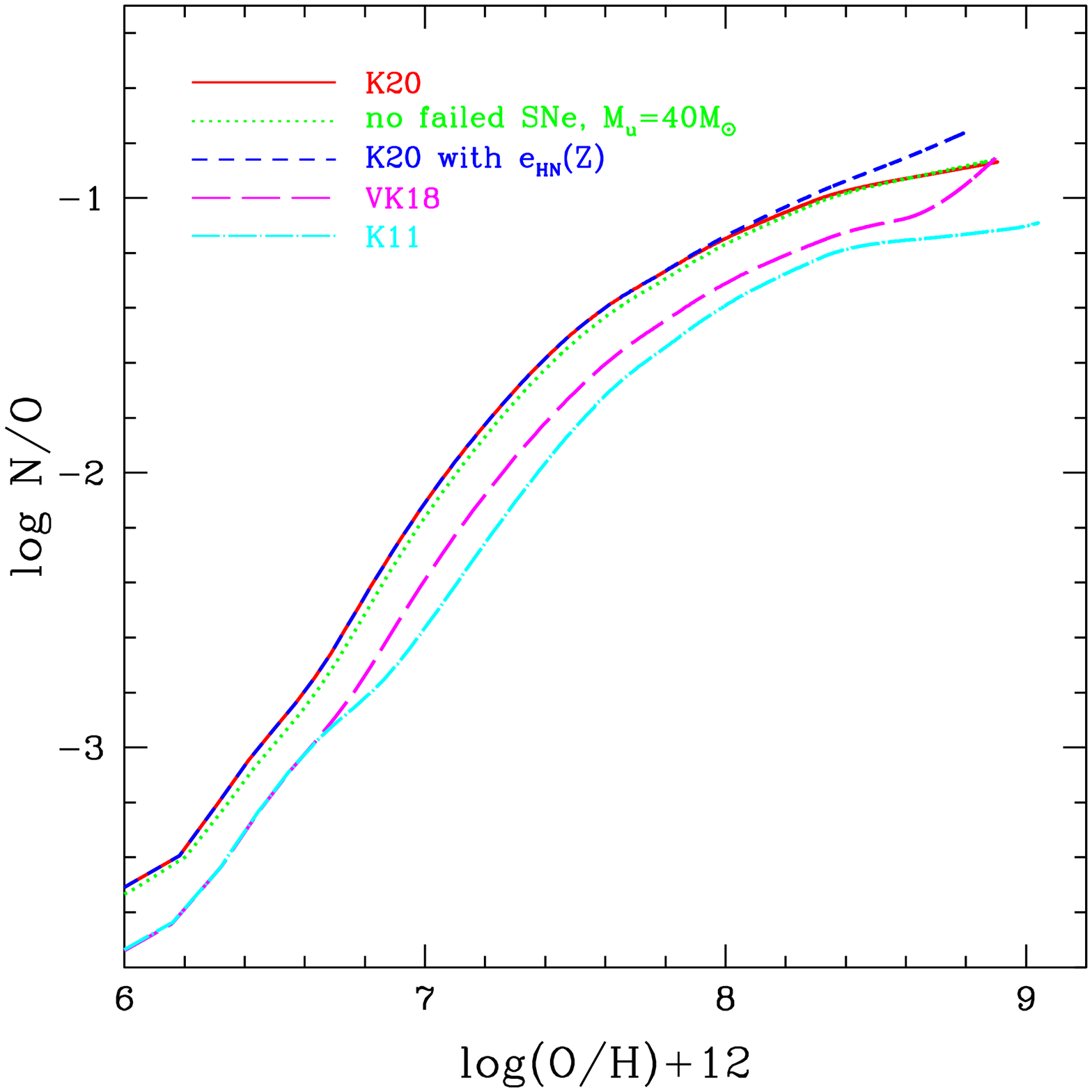}
\caption{\label{fig:noo}
Same as Fig.\ref{fig:ofe} but for the N/O ratio plotted against the oxygen abundance.
}
\end{figure}

\begin{figure}\center
\includegraphics[width=8.5cm]{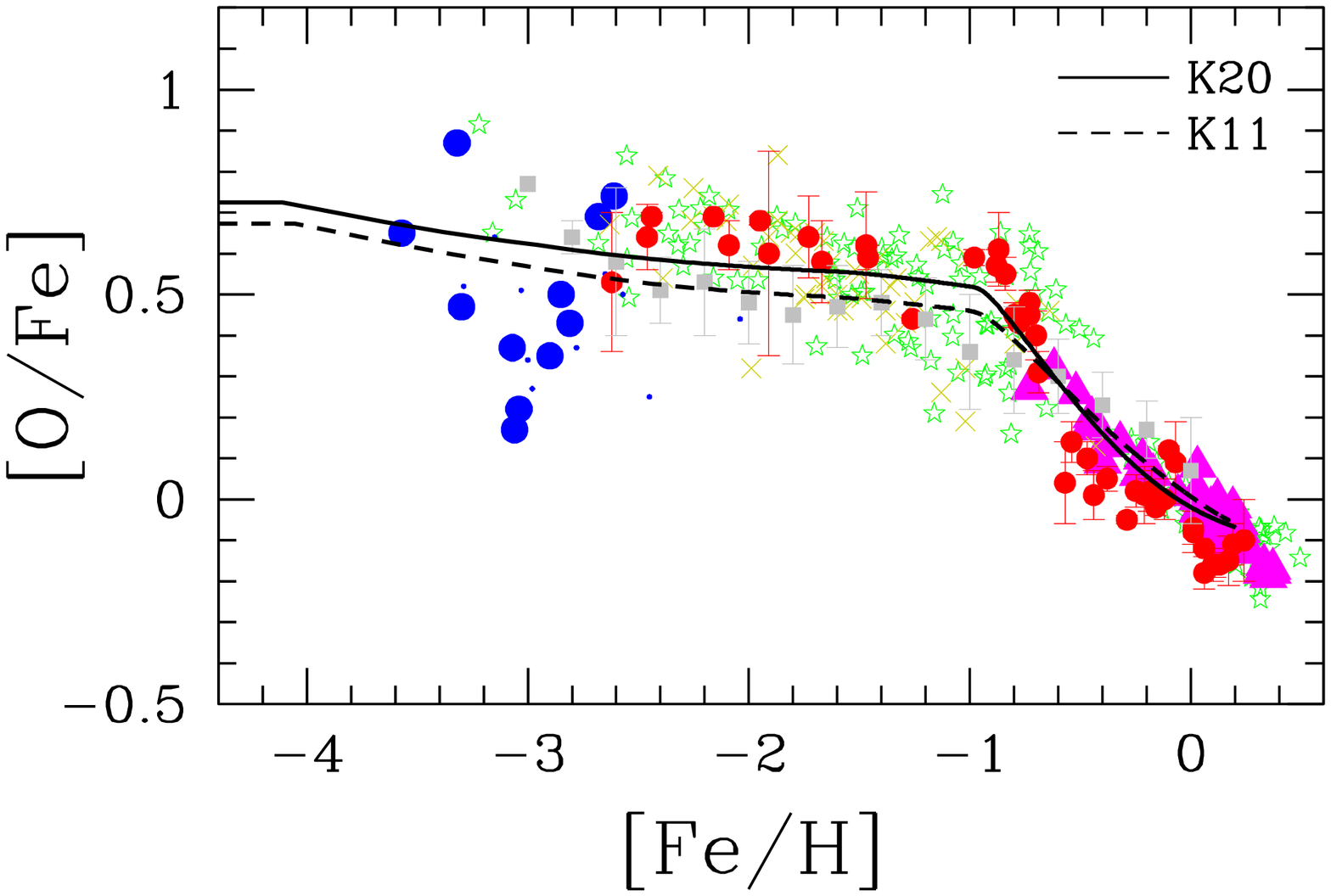}
\caption{\label{fig:o}
Same as Fig.~\ref{fig:c}, but for the [O/Fe]--[Fe/H] relation.
Observational data sources (mostly for [OI]) are:
blue filled circles, \citet{spi05} with 3D correction $-0.23$, with smaller symbols denoting mixed stars;
green filled squares with error bars, \citet{mel02} for IR OH;
olive crosses, \citet{ful03};
magenta filled triangles, \citet{ben04};
red filled circles with error bars, \citet{zha16} with NLTE;
green stars, \citet{ama19b} with 3D/NLTE O and 3D Fe.
}
\end{figure}

\begin{figure}\center
\includegraphics[width=8.5cm]{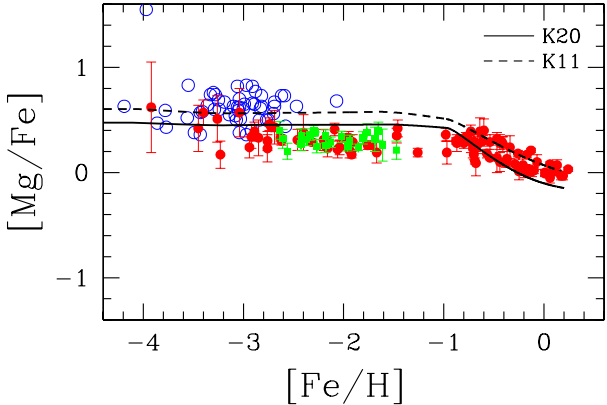}
\caption{\label{fig:mg}
Same as Fig.~\ref{fig:c}, but for the [Mg/Fe]--[Fe/H] relation.
Observational data sources are:
blue open squares, \citet{and10}, NLTE;
green filled squares with error bars, \citet{reg17};
red filled circles with error bars, \citet{zha16}, NLTE.
}
\end{figure}

\begin{figure}\center
\includegraphics[width=8.5cm]{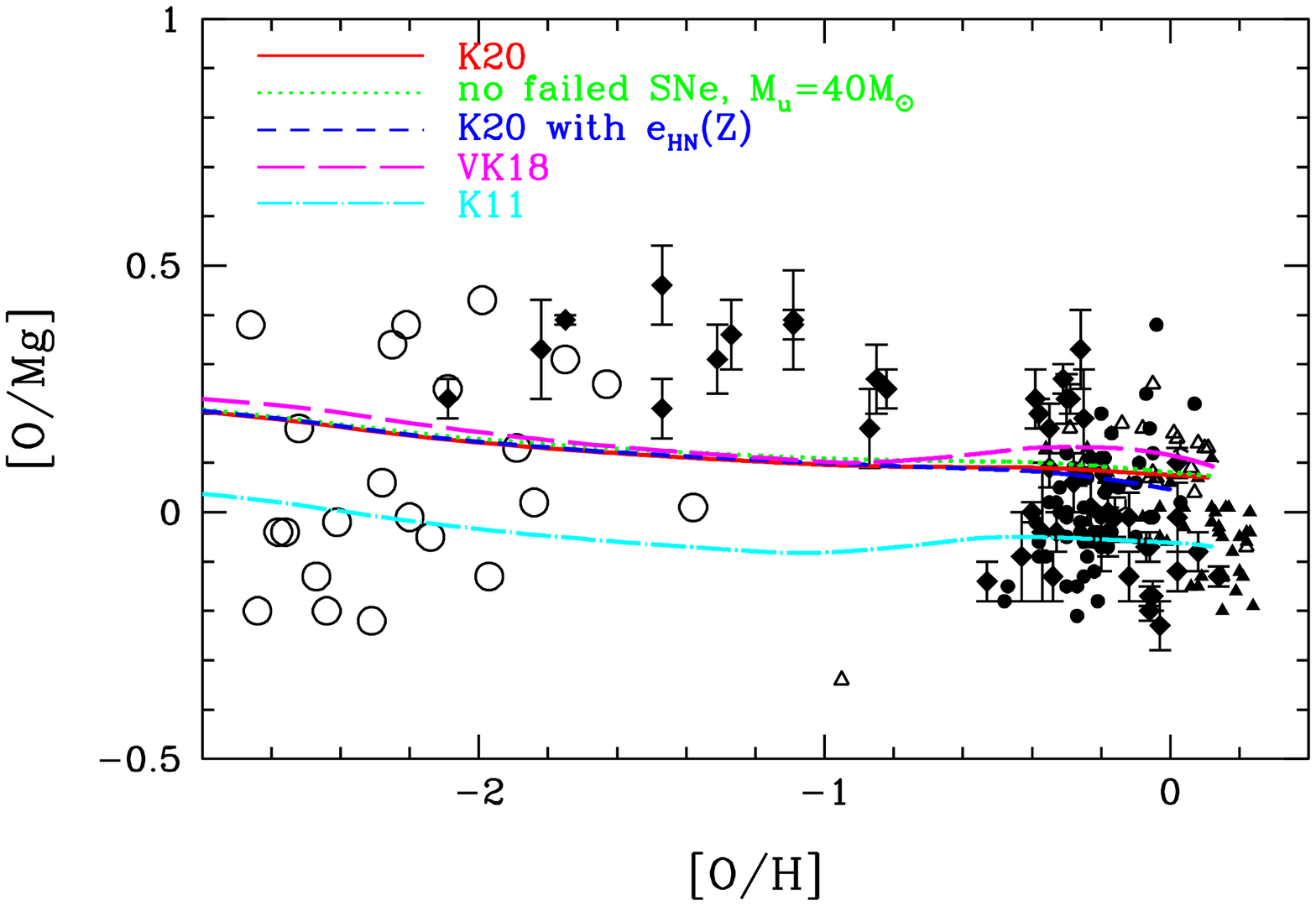}
\caption{\label{fig:mgoo}
Same as Fig.~\ref{fig:ofe} but for [Mg/O] against [O/H].
Observational data sources are:
open circles, \citet{and10} for NLTE Mg and LTE O with 3D correction $-0.23$;
filled diamonds with error bars, \citet{zha16}, NLTE;
filled triangles, opened triangles, and three-pointed stars, \citet{ben14} for thin, thick disk, and intermediate stars, respectively;
small filled circles, \citet{red03} for thin disk stars with [OI].
}
\end{figure}

\begin{figure}\center
\includegraphics[width=8.5cm]{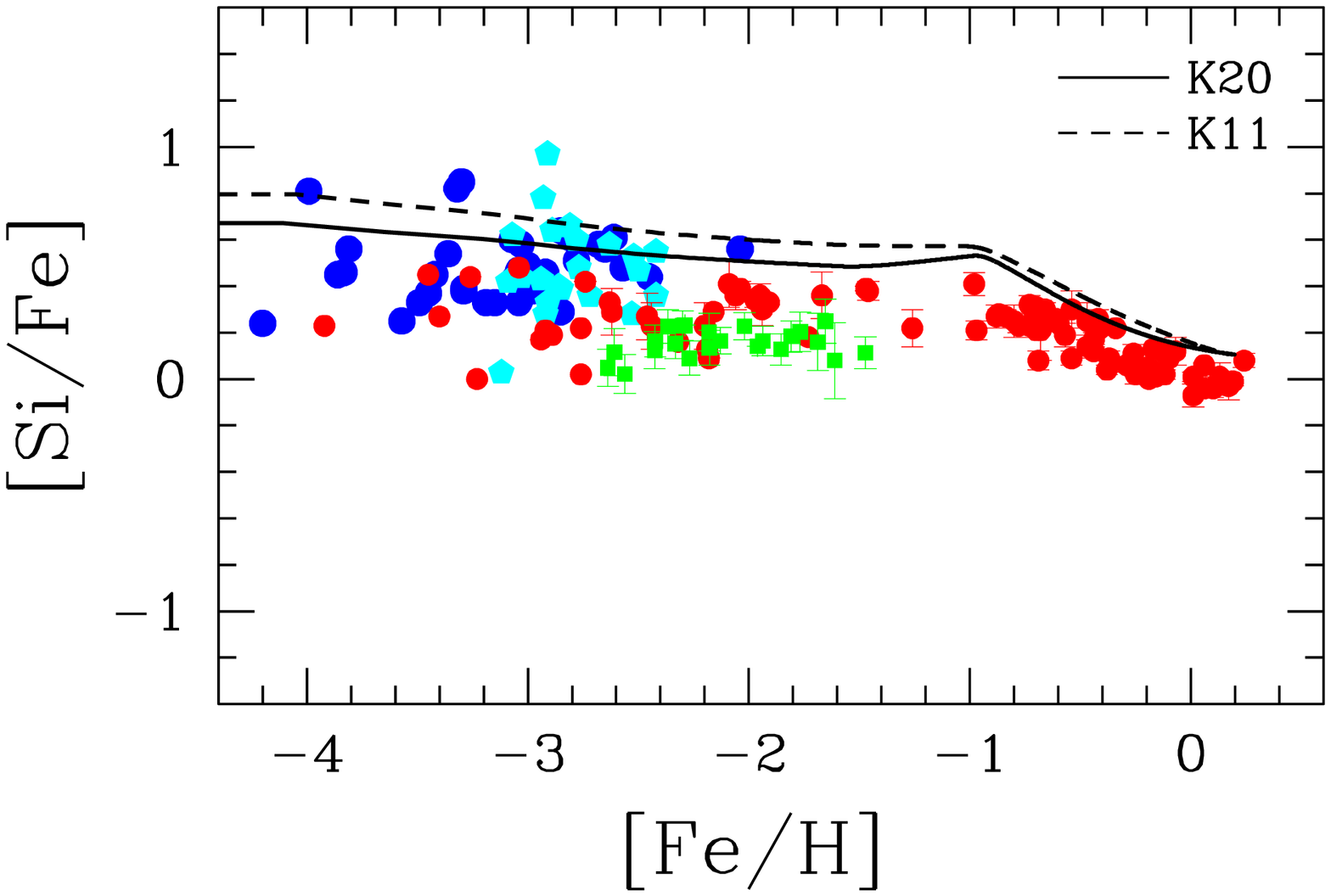}
\caption{\label{fig:si}
Same as Fig.~\ref{fig:c}, but for the [Si/Fe]--[Fe/H] relation.
Observational data sources are:
blue filled squares, \citet{cay04};
cyan filled pentagons, \citet{hon04};
green filled squares with error bars, \citet{reg17};
red filled circles with error bars, \citet{zha16,mas17,mas19}, NLTE.
}
\end{figure}

\begin{figure}\center
\includegraphics[width=8.5cm]{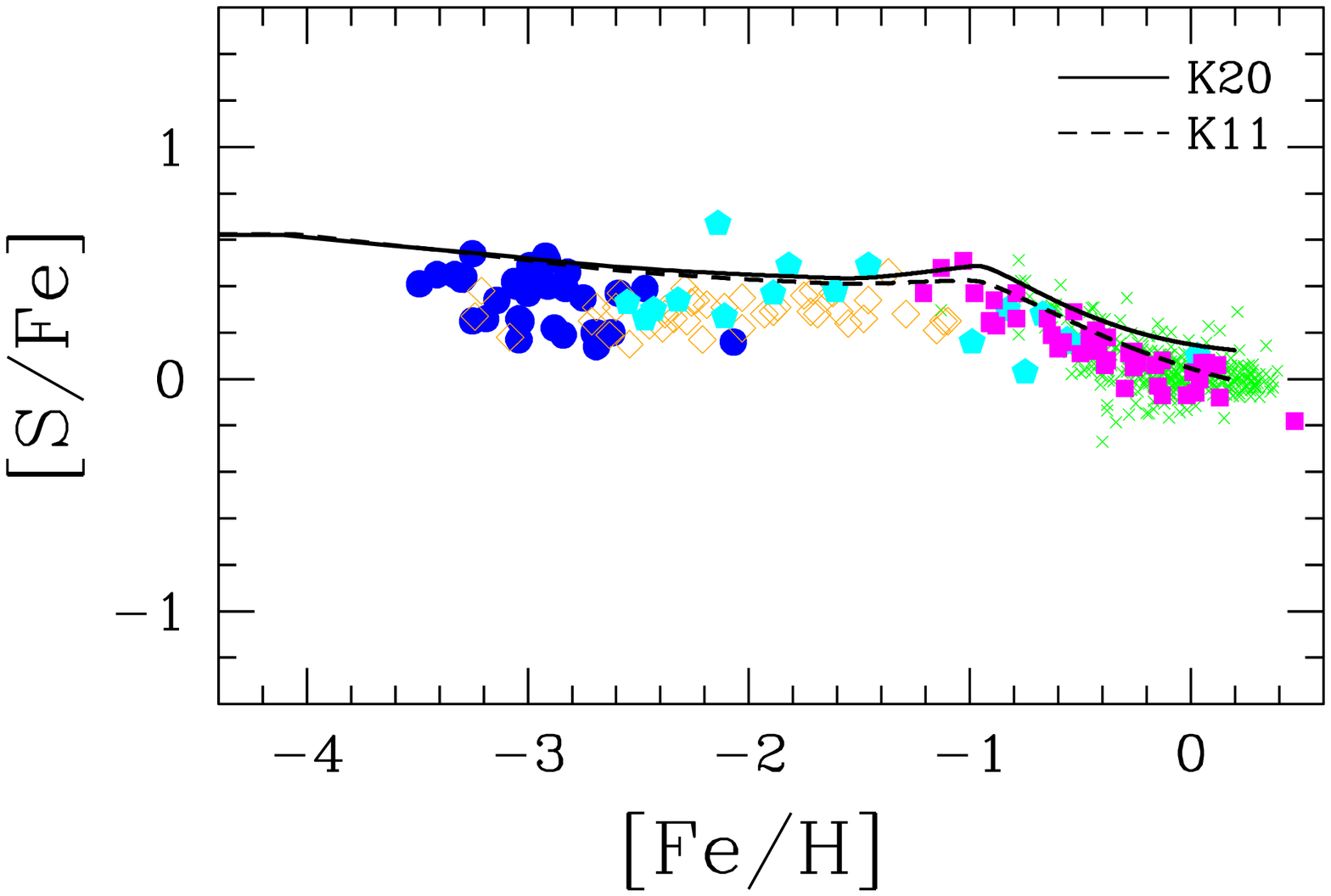}
\caption{\label{fig:s}
Same as Fig.\ref{fig:c}, but for the [S/Fe]--[Fe/H] relation.
Observational data sources are:
blue filled circles, \citet{spi11};
cyan filled pentagons, \citet{tak05};
orange open diamonds, \citet{nis07};
magenta filled squares, \citet{che02};
crosses, \citet{cos20}.
}
\end{figure}

\begin{figure}\center
\includegraphics[width=8.5cm]{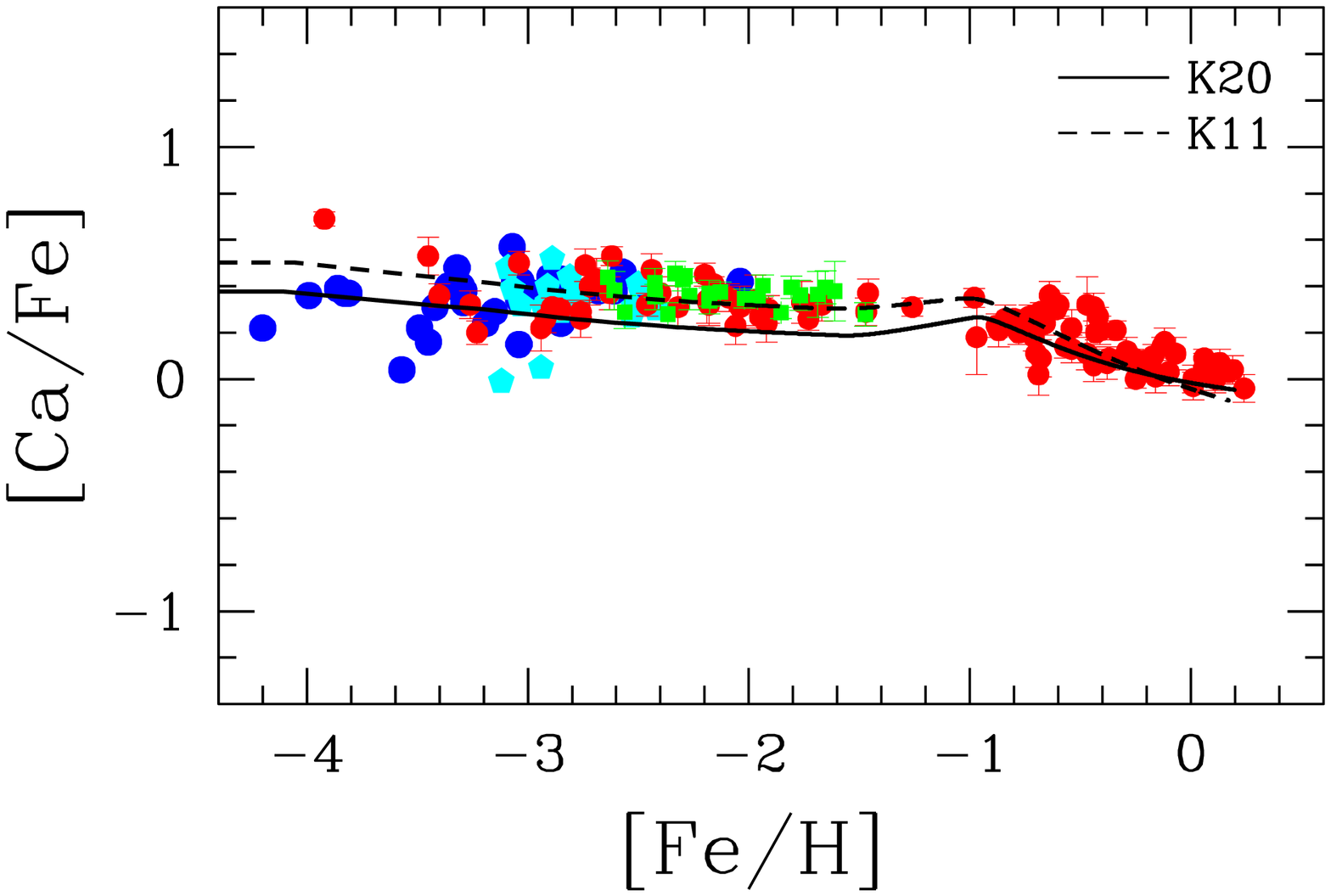}
\caption{\label{fig:ca}
Same as Fig.~\ref{fig:c}, but for the [Ca/Fe]--[Fe/H] relation.
See Fig.~\ref{fig:si} for the observational data sources.
}
\end{figure}

\begin{figure}\center
\includegraphics[width=8.5cm]{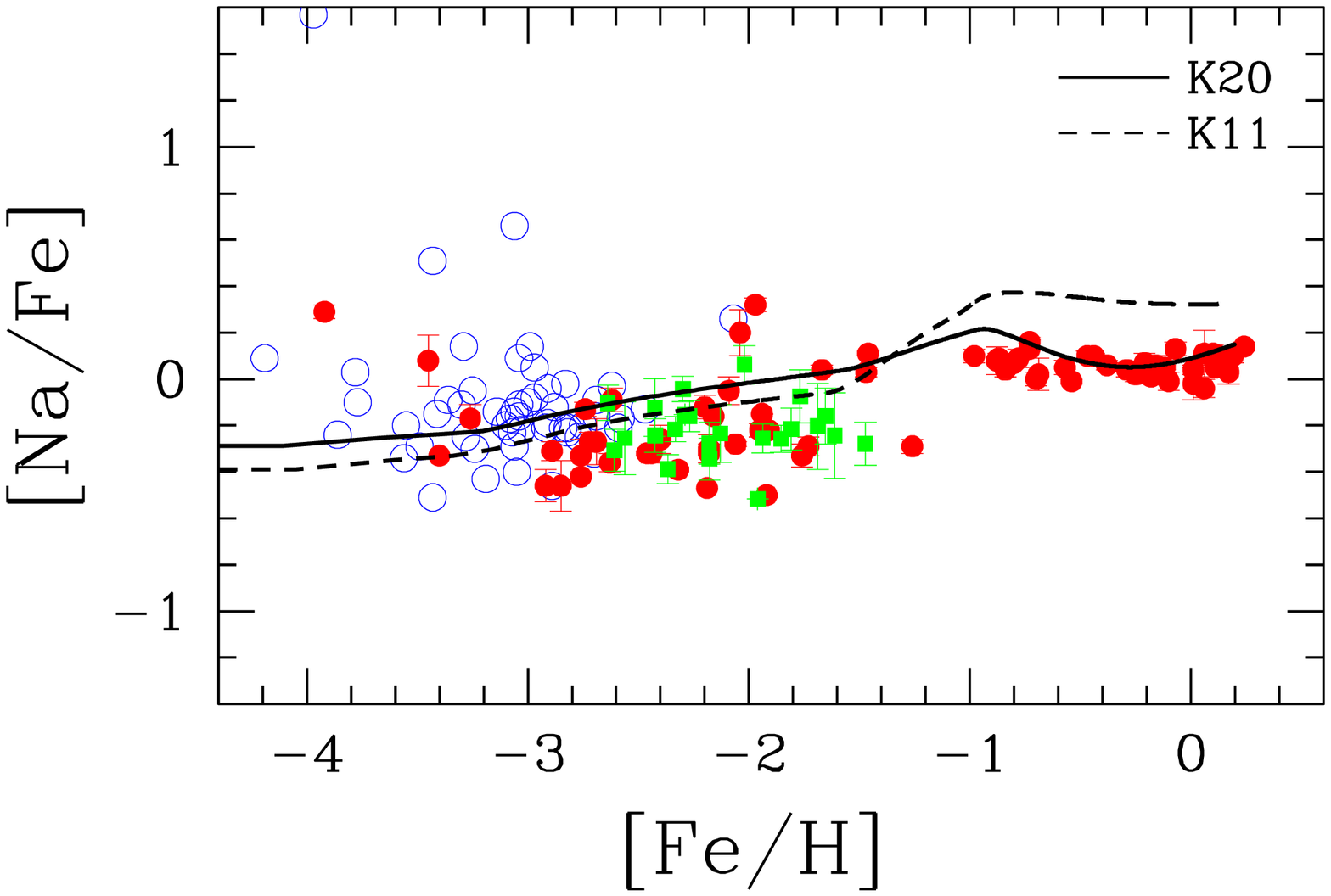}
\caption{\label{fig:na}
Same as Fig.~\ref{fig:c}, but for the [Na/Fe]--[Fe/H] relation.
Observational data sources are:
blue open squares, \citet{and07};
%red filled circles with error bars, \citet{zha16};
%green filled squares with error bars, \citet{reg17};
and as in Fig.~\ref{fig:si} for the other datapoints.
}
\end{figure}

\begin{figure}\center
\includegraphics[width=8.5cm]{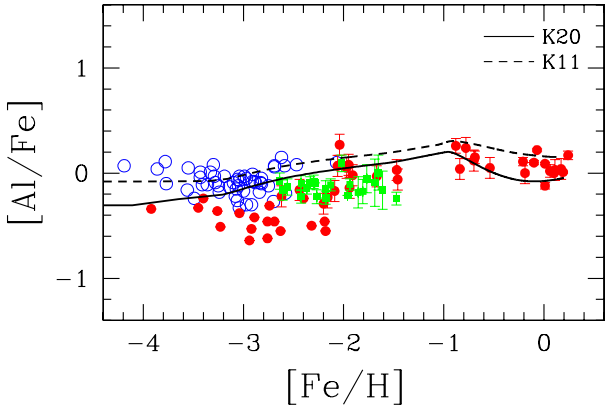}
\caption{\label{fig:al}
Same as Fig.~\ref{fig:c}, but for the [Al/Fe]--[Fe/H] relation.
Observational data sources are:
blue open squares, \citet{and08};
%red filled circles with error bars, \citet{zha16};
%green filled squares with error bars, \citet{reg17};
and as in Fig.~\ref{fig:si} for the other datapoints.
}
\end{figure}

\begin{figure}\center
\includegraphics[width=8.5cm]{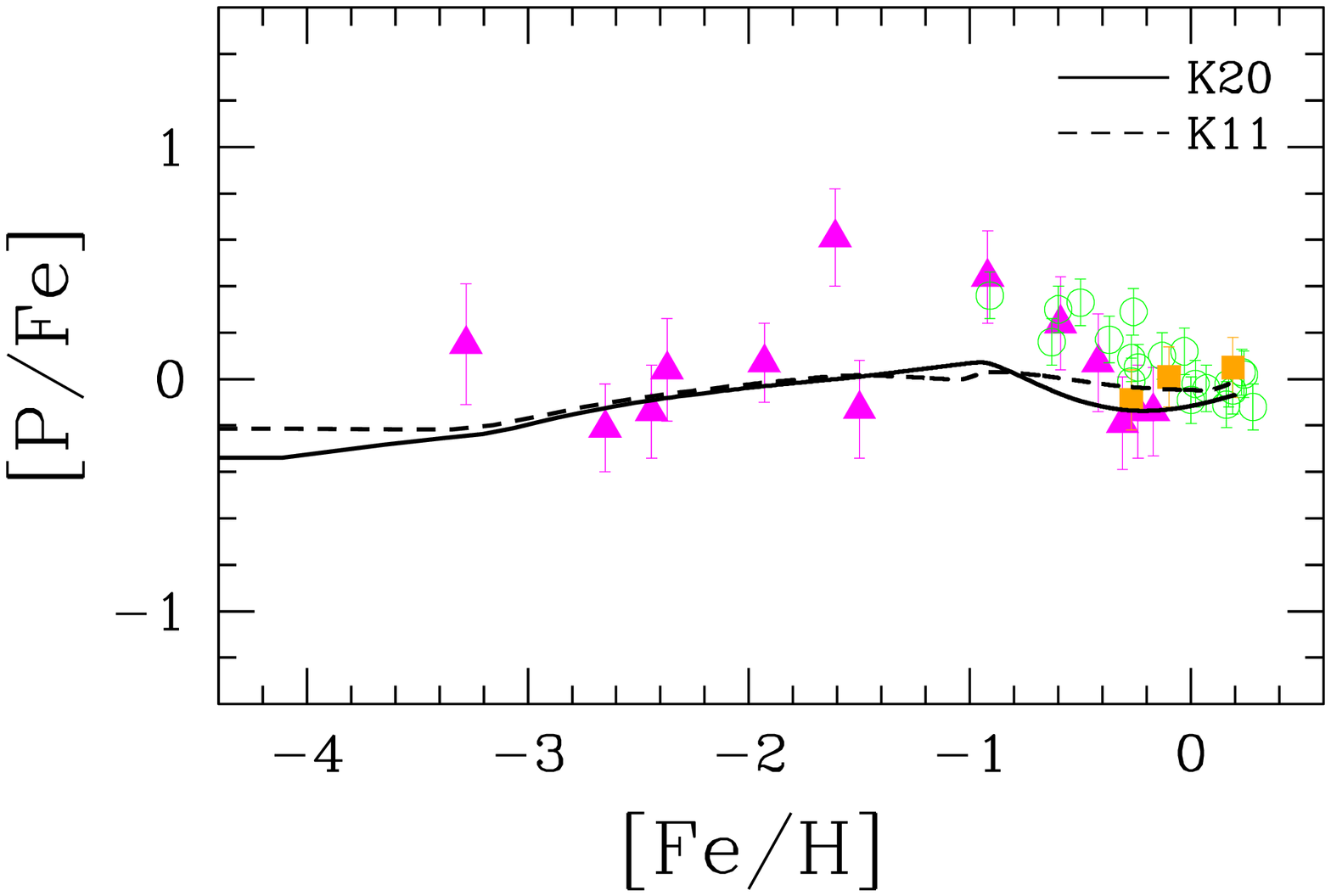}
\caption{\label{fig:p}
Same as Fig.~\ref{fig:c}, but for the [P/Fe]--[Fe/H] relation.
Observational data sources are:
magenta filled triangle, \citet{roe14p};
green open circles, \citet{caf11};
orange filled squares, \citet{caf16}.
}
\end{figure}

\begin{figure}\center
\includegraphics[width=8.5cm]{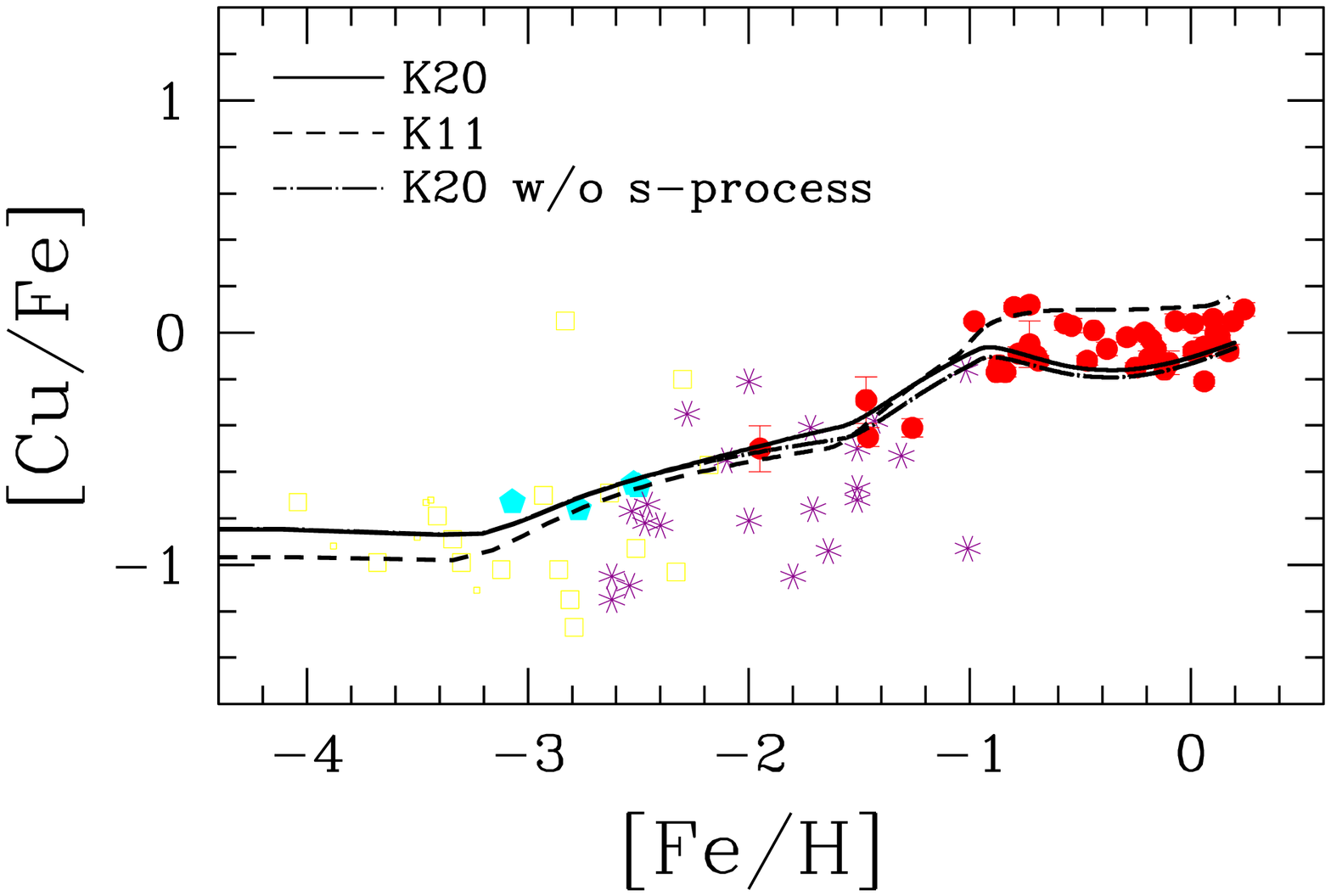}
\caption{\label{fig:cu}
The [Cu/Fe]--[Fe/H] relation for the solar neighborhood models from this paper (solid line), K11 (dashed line), and the model including the s-process (dot-dashed line).
Observational data sources are:
yellow open squares, \citet{coh13}
with smaller symbols denoting CEMP stars;
filled cyan pentagons, \citet{hon04};
purple asterisks, \citet{pri00};
red filled circles with error bars, \citet{zha16};
and as in Fig.\ref{fig:si} for the other datapoints.
See the text for Cu II and NLTE abundances.
}
\end{figure}

\begin{figure}\center
\includegraphics[width=8.5cm]{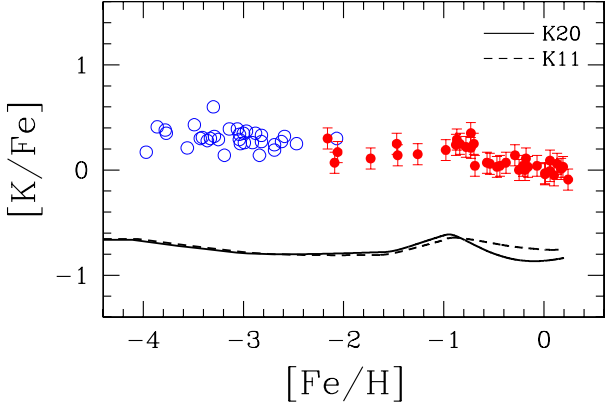}
\caption{\label{fig:k}
Same as Fig.~\ref{fig:c}, but for the [K/Fe]--[Fe/H] relation.
Observational data sources are:
blue open squares, \citet{and10};
red filled circles with error bars, \citet{zha16}.
}
\end{figure}

\begin{figure}\center
\includegraphics[width=8.5cm]{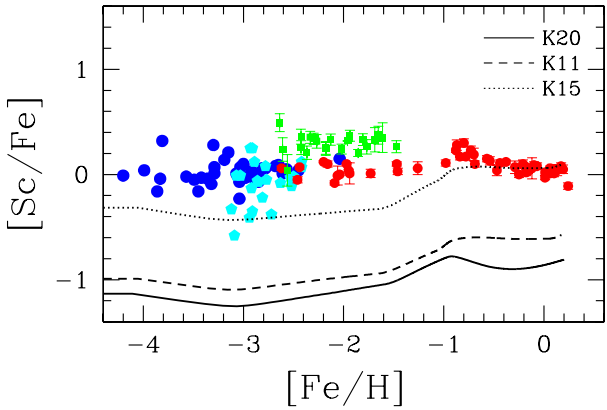}
\caption{\label{fig:sc}
The [Sc/Fe]--[Fe/H] relation for the solar neighborhood models from this paper (solid line), K11 (dashed line), and K15 including the HN jet effects (dotted line).
See Fig.~\ref{fig:si} for the observational data sources.
}
\end{figure}

\begin{figure}\center
\includegraphics[width=8.5cm]{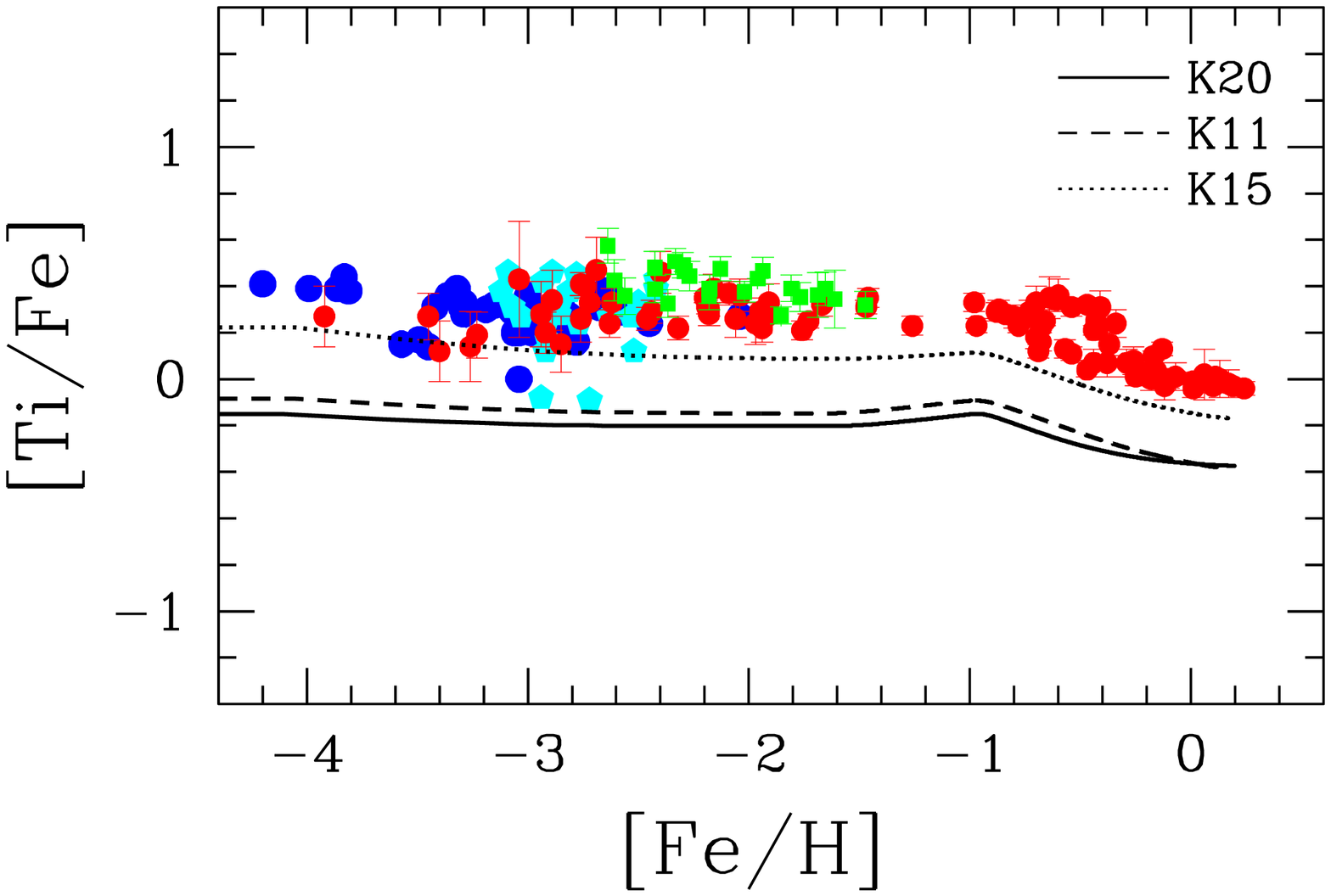}
\caption{\label{fig:ti}
Same as Fig.~\ref{fig:sc}, but for the [Ti/Fe]--[Fe/H] relation.
See Fig.~\ref{fig:si} for the observational data sources.
}
\end{figure}

\begin{figure}\center
\includegraphics[width=8.5cm]{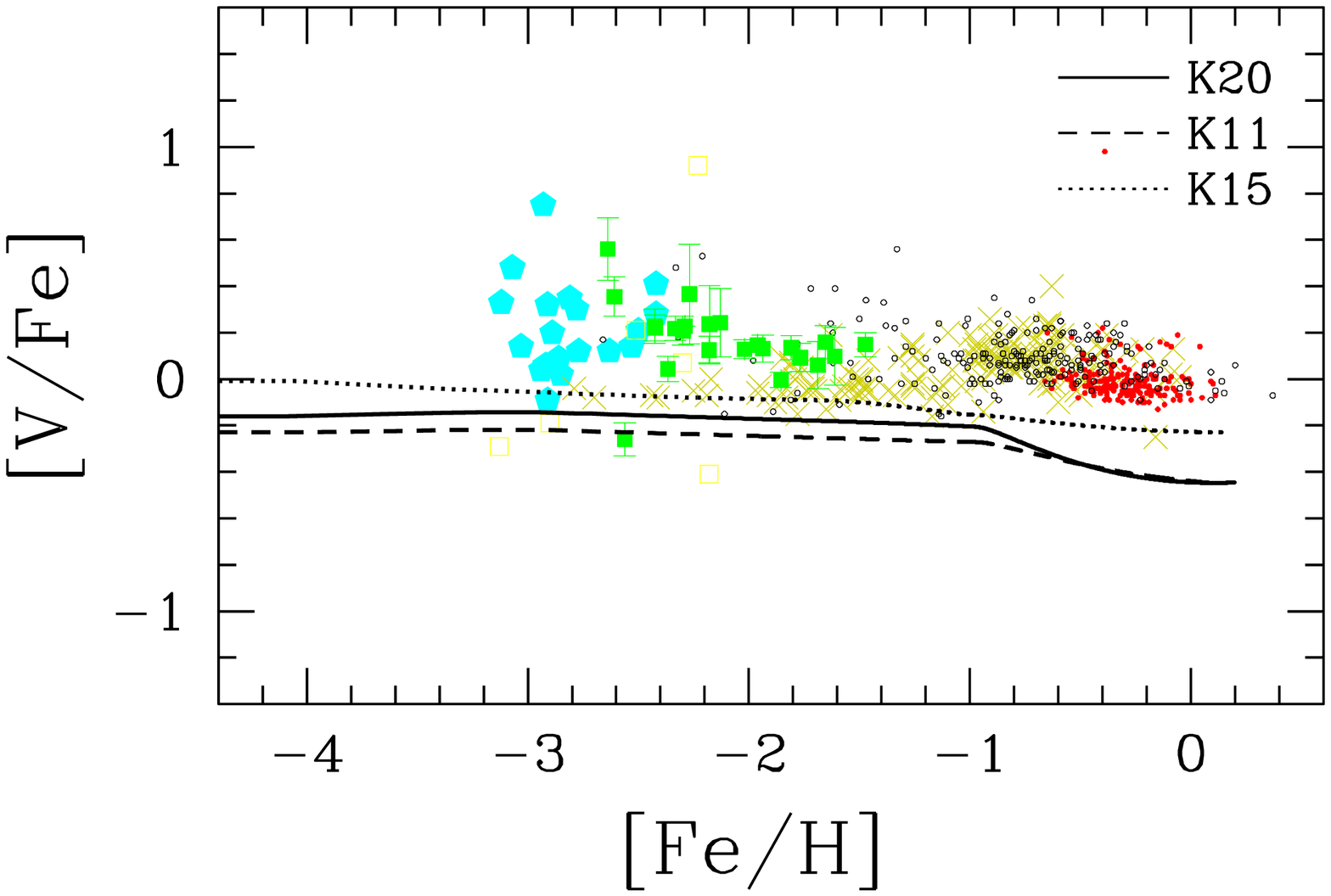}
\caption{\label{fig:v}
Same as Fig.~\ref{fig:sc}, but for the [V/Fe]--[Fe/H] relation.
Observational data sources are:
cyan filled pentagons, \citet{hon04};
yellow open squares, \citet{coh13};
green filled squares with error bars, \citet{reg17};
olive crosses, \citet{ful00};
small red filled and black open circles, \citet{red03,red06,red08} for thin and thick disk/halo stars, respectively.
}
\end{figure}

\begin{figure}\center
\includegraphics[width=8.5cm]{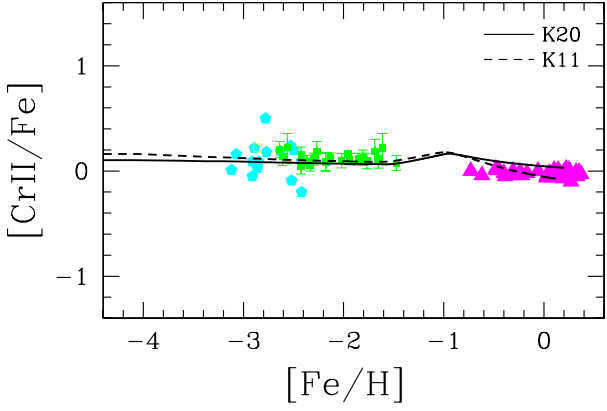}
\caption{\label{fig:cr}
Same as Fig.~\ref{fig:c}, but for the [Cr II/Fe]--[Fe/H] relation.
Observational data sources are:
%cyan filled pentagons, \citet{hon04};
%green filled squares with error bars, \citet{reg17};
magenta filled triangles, \citet{ben03};
and same as Fig.\ref{fig:v} for the other datapoints.
}
\end{figure}

\begin{figure}\center
\includegraphics[width=8.5cm]{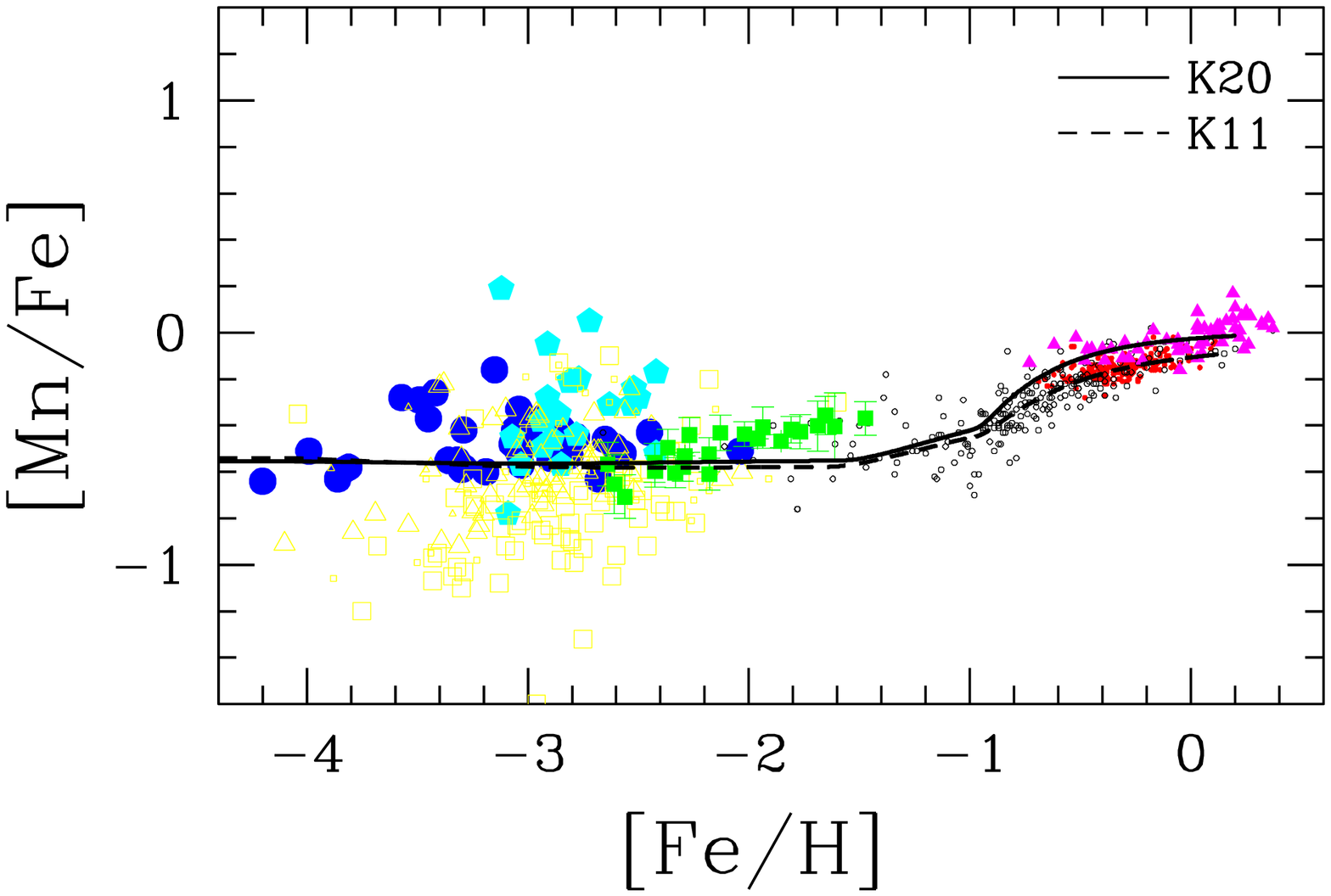}
\caption{\label{fig:mn}
Same as Fig.~\ref{fig:c}, but for the [Mn I/Fe]--[Fe/H] relation.
Observational data sources are:
blue filled circles, \citet{cay04};
cyan filled pentagons, \citet{hon04};
yellow open squares, \citet{coh13} with smaller symbols denoting CEMP stars;
yellow open triangles, \citet{yon13} with smaller symbols denoting CEMP stars;
green filled squares with error bars, \citet{reg17};
small red filled and black open circles, \citet{red03,red06,red08} for thin and thick disk/halo stars, respectively;
magenta filled triangles, \citet{fel07}.
}
\end{figure}

\begin{figure}\center
\includegraphics[width=8.5cm]{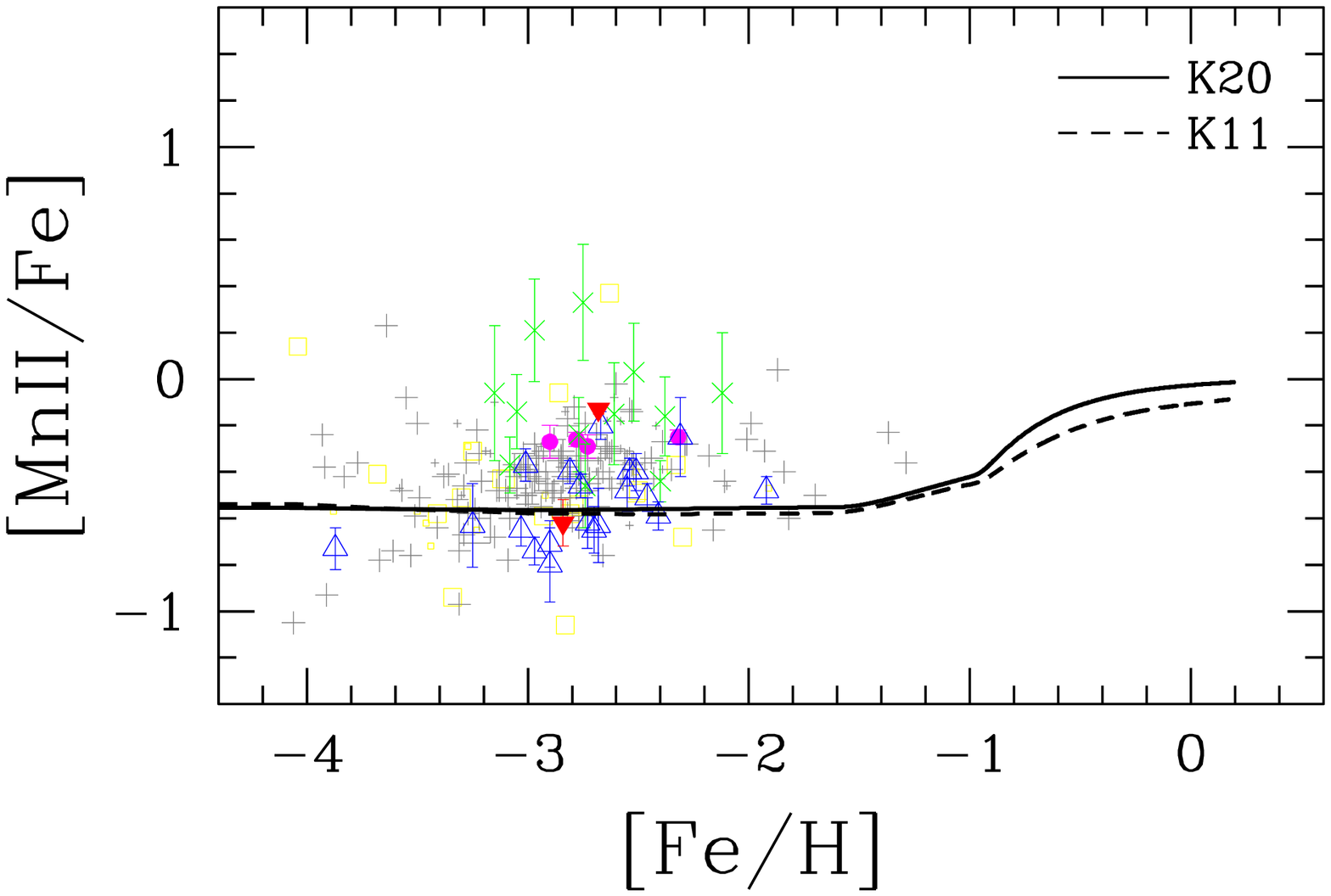}
\caption{\label{fig:mn2}
Same as Fig.~\ref{fig:c}, but for the [Mn II/Fe]--[Fe/H] relation.
Observational data sources are:
green crosses, \citet{joh02};
blue open triangles, \citet{lai08};
red filled upside-down triangles, \citet{mas10,mas14};
magenta filled circles, \citet{sne16,cow20};
yellow open squares, \citet{coh13} with smaller symbols denoting CEMP stars;
gray plus, \citet{roe14} with smaller symbols denoting CEMP stars.
}
\end{figure}

\begin{figure}\center
\includegraphics[width=8.5cm]{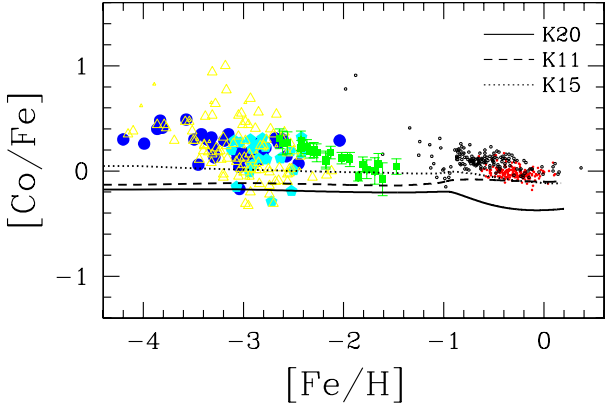}
\caption{\label{fig:co}
Same as Fig.~\ref{fig:sc}, but for the [Co I/Fe]--[Fe/H] relation.
See Fig.\ref{fig:mn} for the observational data sources.
Note that observed [Co II/Fe] ratios are $\sim 0$ \citep{cow20}.
%blue filled circles, \citet{cay04};
%cyan filled pentagons, \citet{hon04};
%yellow open squares, \citet{coh13};
%yellow open triangles, \citet{yon13};
%green filled squares with error bars, \citet{reg17};
%small red filled and black open circles, \citet{red03,red06,red08} for thin and thick disk/halo stars.
}
\end{figure}

\begin{figure}\center
\includegraphics[width=8.5cm]{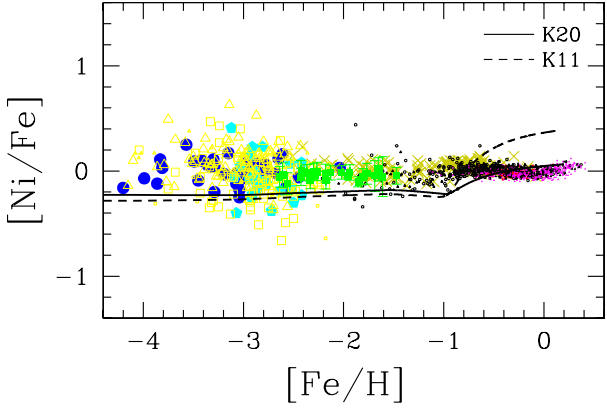}
\caption{\label{fig:ni}
Same as Fig.~\ref{fig:sc}, but for the [Ni/Fe]--[Fe/H] relation.
Observational data sources are:
%blue filled circles, \citet{cay04};
%cyan filled pentagons, \citet{hon04};
%yellow open squares, \citet{coh13};
%yellow open triangles, \citet{yon13};
%green filled squares with error bars, \citet{reg17};
olive crosses, \citet{ful00};
%small red filled and black open circles, \citet{red03,red06,red08} for thin and thick disk/halo stars.
magenta small filled triangles and black small open triangles, \citet{ben14} for thin and thick disk stars, respectively;
and same as Fig.~\ref{fig:mn} for the other datapoints.
}
\end{figure}

\begin{figure}\center
\includegraphics[width=8.5cm]{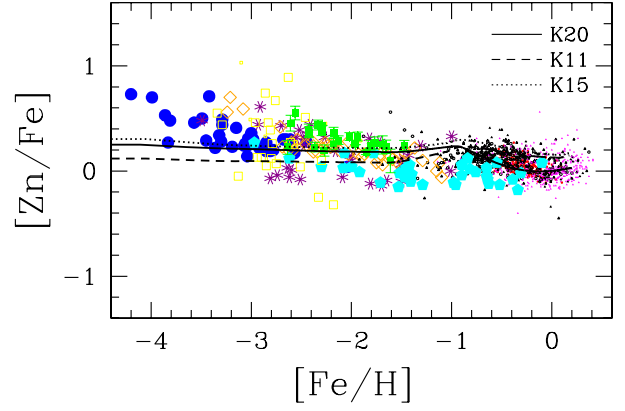}
\caption{\label{fig:zn}
Same as Fig.~\ref{fig:sc}, but for the [Zn/Fe]--[Fe/H] relation.
Observational data sources are:
%blue filled circles, \citet{cay04};
%yellow open squares, \citet{coh13};
purple asterisks, \citet{pri00};
orange open diamonds, \citet{nis07};
%green filled squares with error bars, \citet{ref17};
cyan filled pentagons, \citet{sai09};
%small red filled and black open circles, \citet{red03,red06,red08} for thin and thick disk/halo stars.
magenta small filled triangles and black small open triangles, \citet{ben14}, respectively for thin and thick disk stars;
and same as Fig.\ref{fig:mn} for the other datapoints.
}
\end{figure}

Based on our fiducial model (the red solid line in all the figures), which includes super-AGB stars, Figures \ref{fig:xfe} and \ref{fig:xfe2} show the evolution of elemental abundance ratios [X/Fe] against [Fe/H] from C to Zn in the solar neighborhood, compared to the other models.
In Figures \ref{fig:c}-\ref{fig:zn}, we compare the fiducial, K15, and K11 models with more observational data, not only from the NLTE analysis but also from other careful analysis.

The contribution to GCE from AGB stars (green dotted lines in Fig.\,\ref{fig:xfe}) can be seen mainly for C and N, and only slightly for Na, compared with the model that includes supernovae only (blue dashed lines).
Hence it seems not possible to explain the O--Na anti-correlation observed in globular cluster stars with a smooth star formation history as in the solar neighborhood.
Although AGB stars produce significant amounts of Mg isotopes (see \S \ref{sec:iso}), the inclusion of these do not affect the [Mg/Fe]--[Fe/H] relation.
The contribution from super-AGB stars (red solid lines) is very small; with super AGB stars, C abundances slightly decrease, while N abundances slightly increase.
It would be very difficult to put a constraint on super-AGB stars from the average evolutionary trends of elemental abundance ratios, but it might be possible to see some signatures of super-AGB stars in the scatters of elemental abundance ratios.
With ECSNe (magenta long-dashed lines), Ni, Cu and Zn are slightly increased. These yields are in reasonable agreement with the high Ni/Fe ratio in the Crab Nebula \citep{nom87,wan09}.
No difference is seen with/without SNe Iax (cyan dot-dashed lines) in the solar neighborhood because of the narrow mass range of hybrid WDs.
As noted before (\S \ref{sec:sn}), this mass range depends on convective overshooting, mass-loss, and reaction rates.
Even with a wider mass range in \citet[][$\Delta M\sim1M_\odot$]{kob15}, however, the SN Iax contribution is negligible in the solar neighborhood, but can be important at lower metallicities such as in dwarf spheroidal galaxies with stochastic chemical enrichment \citep{ces17}.

With failed SNe, our fiducial model (red solid lines in Fig.\,\ref{fig:xfe2}) is in good agreement with observations of most of the major elements. Strictly speaking, the predicted Mg, Si, and S abundances are slightly higher, Ca, Co, and Ni abundances are slightly lower than in the observations.
Compared with K11 model (cyan dot-dashed lines), the match is improved for most of elements, except for Ca at [Fe/H] $\ltsim-1$ and Co at all [Fe/H], which imply higher energies or a larger HN contribution.
The improvement is due to the inclusion of failed SNe (i.e., the exclusion of massive SNe II) and/or the updated solar abundances (see below for more details).
With the metal-dependent HN fraction (blue short-dashed line), Cu and Zn are also under-produced at [Fe/H] $\gtsim -1$ in the model, and this is why we use a constant HN fraction for our fiducial model (\S \ref{sec:sn}).
It is possible to keep the agreement without failed SNe (green dotted lines) if we change the upper limit of IMF (i.e. of both SNe II and HNe) as discussed above. This gives slightly better matching for Al and Cu, but not Na, at [Fe/H] $\gtsim-1$.
Finally, the VK18 model (magenta long-dashed lines) gives slightly too large ratios from O to S, relative to Fe, and the N/O ratios in \citet{vin18b} would be $\sim 0.1$ dex larger with the yields in our fiducial model.
To put a further constraint on supernova explosions, it is very important to measure elemental abundances with $\sim 0.1$ dex accuracy, not only at low metallicities but also at high metallicities.

{\bf Carbon} ---
Half of carbon in the Universe is produced by massive stars ($>10 M_\odot$), while the rest is mainly by low-mass AGB stars ($1-4 M_\odot$, K11).
However, the [C/Fe] ratio is enhanced efficiently by low-mass stars because these stars produce no Fe.
In Figure \ref{fig:c}, the fiducial model (solid line) reproduces the observed trend slightly better than the K11 model (dashed line).
When we include AGB yields (green dotted lines in Fig.\,\ref{fig:xfe}), [C/Fe] increases from [Fe/H] $\sim -1.5$, which corresponds to 
the lifetime of $\sim 4 M_\odot$ stars ($\sim 0.1$ Gyr).
At [Fe/H] $\sim -1$, [C/Fe] reaches $0.21$ ($0.16$ with s-process), which is $0.31$ dex larger than the case without AGB yields (blue short-dashed lines in Fig.\,\ref{fig:xfe}).
The inclusion of super-AGB stars (red solid line in Fig.\,\ref{fig:xfe}) increases [C/Fe] only by 0.004 dex.
The peak value of [C/Fe] is in excellent agreement with the measurements from \citet{zha16}, which are based on 1D NLTE analysis of CI lines, as well as 1D LTE analysis of molecular CH and C$_2$ lines.

At lower and higher metallicities than [Fe/H] $\sim -1$, however, the predicted [C/Fe] is $0.1-0.2$ dex lower than the observations.
This is at least partially due to the fact that the AGB contribution appears suddenly in the one-zone models. With inhomogeneous enrichment in chemodynamical simulations \citep{kob11mw}, the [C/Fe] variation would become weaker.
In particular, AGB stars can contribute at metallicities below [Fe/H] $\ltsim -1.5$ when inhomogeneous chemical enrichment is taken into account \citep{kob14mw,vin18a}.

At [Fe/H] $\gtsim -1$, [C/Fe] shows a decrease in the NLTE observation, which is consistent with the LTE observation from \citet{ben06} with the forbidden [CI] line at 872.7 nm.
Our models also show a decrease due to SNe Ia, but is steeper than shown by these observations.
C yields from AGB could be increased with overshoot \citep{pig16}, which could also increase s-process yields.
The K11 model (dashed line in Fig.\,\ref{fig:c}) gives lower [C/Fe] ratios at [Fe/H] $\ltsim -1$ than the fiducial model (solid line), which is due to the adopted higher solar abundance ($A_\odot({\rm C})=8.56$ in AG89, instead of $8.43$ in AGSS09).

At [Fe/H] $\ltsim -2.5$, the model [C/Fe] is in good agreement with the observations from \citet{spi06}, although the observational data show a significant scatter.
Note that \citet{spi06} flagged ``mixed'' stars, where C is likely to have been transformed into N. These are plotted with smaller symbols in the figures and should be excluded from the comparison.
It is known that a significant fraction of extremely metal-poor stars (the CEMP stars) show carbon enrichment, which are also plotted with smaller symbols for the data in \citet{coh13} and \citet{yon13} in the figures.
One of the scenarios for the CEMP stars is faint supernovae, which are not included in our models (\S \ref{sec:sn}).
The model [C/Fe] shows roughly $0$, with a very weak increase toward lower [Fe/H], which is consistent with the lower boundary of the plotted [C/Fe] ratios of the unmixed stars in \citet{spi06}.
The weak increase is also in good agreement with recent analysis of 3D/NLTE C and 3D/LTE Fe abundances from Fe II lines in \citet{ama19a}.

Figure \ref{fig:coo} shows the [C/O] ratio against [O/H] for the models in Fig.\,\ref{fig:ofe}.
At the low metallicity,
there is some variation in the plateau values among these models, but all models show a weak increase from [O/H] $\sim -1$ to $-3$.
Although \citet{ama19a} reported that [C/O] rather decreases toward lower [O/H] with their 3D/NLTE analysis, the slope of our models is in good agreement with the plotted observations including those from \citet{ama19a}.
At high metallicity, all models predict [C/O] ratios significantly lower than observed, although the model with metallicity-dependent HN fraction (blue short-dashed line) gives [C/O] ratios closest to the observational data.
C yields from AGB stars could be larger with more overshoot, but it is not clear if it would be enough to explain the $\sim 0.3$ dex offset.

{\bf Nitrogen} ---
Different from C, N is produced mainly by intermediate-mass AGB stars ($4-7 M_\odot$, K11).
Therefore the contribution from AGB stars is seen already from [Fe/H] $\sim -2.5$ (green dotted lines in Fig.\,\ref{fig:xfe}).
At [Fe/H] $\sim -1$, [N/Fe] reaches $0.37$ ($0.39$ with s-process), which is $0.94$ dex larger than the case without AGB yields (blue dashed lines in Fig.\,\ref{fig:xfe}).
With super-AGB stars, the peak [N/Fe] is slightly higher, $0.44$ (red solid lines in Fig.\,\ref{fig:xfe}), and the trend agrees very well with the plotted observational data.

At [Fe/H] $\gtsim -1$, the model [N/Fe] shows a decrease due to SNe Ia. In the observational data, such a decrease is not clearly seen, but at [Fe/H] $\sim 0$, the [N/Fe] ratio is $\sim0$, which is consistent with our new models with AGB and super-AGB stars.
The [N/Fe] ratio at [Fe/H] $=0$ is $-0.59$ without AGB (blue dashed line in Fig.\,\ref{fig:xfe}), and $-0.23$ in K11 model (dashed line in Fig.\,\ref{fig:n}).
Figure \ref{fig:n} shows that the fiducial model (solid line) gives a better match than the K11 model (dashed line) at all metallicity range.
This difference is caused mainly by the adopted solar abundance ($A_\odot({\rm N})=8.05$ in AG89, instead of $7.83$ in AGSS09).

Although no difference is seen at [Fe/H] $\ltsim -2.5$ with and without the AGB yields in these one-zone models (Fig.\,\ref{fig:xfe}),
AGB stars can also contribute to N production even at [Fe/H] $\ltsim -2.5$ when taking inhomogeneous chemical enrichment into account \citep{kob11mw,vin18a}, and \citet{vin18b} reproduced the observed N/O--O/H relations, not with rotating massive stars, but with the failed SN model of VK18.
Figure \ref{fig:noo} shows the N/O ratio against oxygen abundance for the solar neighborhood models represented in Fig.\,\ref{fig:ofe}.
All models shows a strong increase of N/O ratios toward higher metallicities, and the N/O increases the most in the VK18 model, which allowed to reproduce the observed N/O--O/H relation of galaxies \citep{vin18a,vin18b}.

{\bf $\alpha$ elements} ---
For all $\alpha$ elements (O, Ne, Mg, Si, S, Ar, and Ca), the same trend as O is present: the plateau caused by SNe II/HNe and the decrease from [Fe/H] $\sim -1$ by SNe Ia (Figs.\,\ref{fig:o}-\ref{fig:ca}).
The [O/Fe] plateau value of the NLTE observation is $\sim 0.6$, and the trend is surprisingly consistent with the LTE analysis in \citet{cle81}\footnote{\citet{sne79} first found the plateau at [O/Fe] $\sim 0.5$ for $-3 \ltsim$ [Fe/H] $\ltsim -0.5$.}, \citet{mel02}, \citet{ful03}, and \citet{ben03}.
On the other hand, the observed [Mg/Fe] plateau value may be $\sim 0.3$ dex lower.
This observed positive [O/Mg] ratio was discussed in Fig.\,9 of K06, although the results were not conclusive because of the uncertainty of the observational data.
Observationally, the [Mg/Fe] plateau value was reported to be $0.27$ \citep{cay04}, which was due to underestimated equivalent widths of the Mg lines.
This was updated by \citet{and10} to $0.31$ with LTE, and $0.61$ with NLTE analysis for Mg (but not for Fe), which would result in [O/Mg] $\sim 0$.
However, the differential analysis of \citet{reg17}, although with LTE, produces very similar results as the NLTE analysis of \citet{zha16}, with a low [Mg/Fe] plateau.
\citet{ber17} also obtained $\sim 0.3$ with 1D NLTE, and an even lower value with their $<$3D$>$ NLTE analysis.
The difference between these two NLTE Mg abundances should be investigated further.

In our fiducial model, [Mg/Fe] ratios are $0.45, 0.45, 0.43$ at [Fe/H] $=-3, -2, -1.1$ (Fig.\,\ref{fig:mg}), which are only $\sim 0.15$ dex lower than [O/Fe].
Figure \ref{fig:mgoo} shows the O/Mg ratio against oxygen abundance in the various GCE models for the solar neighborhood. The predicted [O/Mg] is never higher than $0.25$ and is lower than observed at $-1.5 \ltsim$ [O/H] $\ltsim -0.5$.

Since the majority of O and Mg are formed during hydrostatic burning of stellar evolution of massive stars, it is not possible to greatly modify the [O/Mg] ratios during supernova explosions.
As mentioned in \S \ref{sec:sn}, a different $^{12}$C($\alpha$,$\gamma$)$^{16}$O reaction rate could change the [O/Mg] ratio during stellar evolution and may explain the large [O/Mg] at the plateau ([O/H] $\ltsim -0.5$);
the core-collapse supernova yields used here were calculated with 1.3 times the value given in \citet{cau88}, which is up to a factor of 2 lower than that calculated by \citet{deBoer17}.
At high metallicities, however, it would be difficult to vary [O/Mg] as much as observed.
Some of the observational data (open circles and filled triangles) indicate that
[O/Mg] may decrease for higher metallicities, which might require a different new physical explanation (K06).

Similar to [Mg/Fe], the observed plateau values of [Si/Fe] and [Ca/Fe] are $\sim 0.3$.
In our fiducial model, [Si/Fe] ratios are $\sim 0.58, 0.51, 0.52$ at [Fe/H] $=-3, -2, -1.1$, which is $\sim 0.2$ dex higher, and [Ca/Fe] ratios are $0.28, 0.21, 0.25$ at [Fe/H] $-3, -2, -1.1$, which is $\sim 0.1$ dex lower than observed.
Si and Ca yields are affected by explosive burning, and it is unclear if the $^{12}$C($\alpha$,$\gamma$)$^{16}$O rate could solve these mismatches as well as for O and Mg.
Note that the differential analysis of \citet{reg17} leads to systematically lower [Si/Fe] ratios, compared to other studies. Their abundances are based on the Si I 390.5nm line that is blended with CH, and may suffer from NLTE effects in the metal-poor regime \citep[][up to $\sim +0.2$ dex]{ama17}.

S abundances are difficult to measure in stellar spectra with a significant NLTE effect depending on the lines.
The predicted [S/Fe] ratios are $0.52, 0.45, 0.47$ at [Fe/H] $-3, -2, -1.1$, which is $\sim 0.1$ dex higher than \citet{spi11} and \citet{nis07}, but is in good agreement with \citet{tak05} at low metallicities.
At high metallicities, K11 model gives slightly better match with \cite{che02} and also more recent observations by \citet{cos20}.
Note that for S, the solar abundance adopted in the K11 model was higher; $A_\odot({\rm S})=7.27$ in AG89, and 7.12 in AGS09.
Ne and Ar also show a similar trend, and the solar abundances are also decreased by 0.16 dex in AGS09.

{\bf Odd-Z elements} --
The production of odd-Z elements depends on the surplus of neutrons from $^{22}$Ne, which is made during He-burning from $^{14}$N produced in the CNO cycle, and hence the yields depends on the metallicity of the progenitors (see Fig.\,5 in K06).
In GCE models, at [Fe/H] $\ltsim -1$,
[(Na, Al, Cu)/Fe] show a decrease toward lower metallicities (Figs.\,\ref{fig:na}-\ref{fig:cu}).
The observed Na and Al abundances are largely affected by NLTE effects, and our models are in good agreement with the NLTE observations \citep{and07,and08,zha16,mas17} as well as the LTE differential analysis \citep{reg17}.

For Cu, the LTE data shows a decrease with a large scatter \citep{pri00}, and our models reproduce the average trend very well, giving [Cu/Fe] $=-0.63$ at [Fe/H] $=-2.5$.
Our [Cu/Fe] trend is also quantitatively consistent with that first found by \citet{sne88}.
The recent NLTE analysis by \citet{and18}, however, found no such Cu decrease at $-4 \ltsim$ [Fe/H] $\ltsim -1.5$.
Another NLTE analysis by \citet{shi18} found a decrease very similar to our model.
This could be tested with Cu II lines; \citet{roe18} found a very similar decrease at $-2.5 \ltsim$ [Fe/H] $\ltsim -1$ with LTE, while \citet{kro18} found a shallower decrease.
It is important to obtain NLTE abundances using Cu I and Cu II lines for a larger sample of metal-poor stars.

At [Fe/H] $\gtsim -1$, Na and Al show a decrease toward higher metallicities owing to the contribution from SNe Ia, which is shallower than the trend for the $\alpha$ elements.
With the updated reaction rates\footnote{Updated rates of the $^{22}{\rm Ne}(p,\gamma)^{23}{\rm Na}$, $^{23}{\rm Na}(p, \gamma)^{24}{\rm Mg}$, and $^{23}{\rm Na}(p, \alpha)^{20}{\rm Ne}$ reactions result in $\sim 6$ to 30 times less Na is produced by intermediate-mass models with HBB.}, Na yields from AGB stars were reduced in K11.
Nonetheless, in the K11 model (dashed line in Fig.\,\ref{fig:na}), Na and Al were over-produced, and this problem is solved in the fiducial model (solid line) of this paper.
Although the predicted [Na/Fe] is in excellent agreement, [Al/Fe] may be decreased slightly too much, compared with the NLTE abundances in \citet{zha16}.

Similar to Al, Cu was over-produced at [Fe/H] $\gtsim -1$ in the K11 model (dashed line in Fig.\,\ref{fig:cu}), and this problem is also solved in the fiducial model (solid line).
[Cu/Fe] may be slightly too much decreased, compared with the NLTE abundances in \citet{zha16}.
With the s-process (dot-dashed line), AGB stars produce some Cu but this gives only a small contribution from [Fe/H] $\sim -2$, and [Cu/Fe] is increased only by $0.03$ dex at [Fe/H] $=0$ with s-process (\S \ref{sec:sr}).

Recently, P abundances became available with near UV or IR spectra.
The predicted [P/Fe] shows a weak decrease toward lower metallicities due to the same metallicity dependence of the yields, which is in reasonably good agreement with the observations (Fig.\,\ref{fig:p}).
It would be better if P yields increase toward higher metallicity to reach a peak [P/Fe] $\sim 0.4$ at [Fe/H] $\sim -1$ and then sharply decrease due to SNe Ia.
Note that the solar P abundance is decreased by 0.16 dex, compared with K11.

{\bf Cl, K, Sc, V, and Ti} ---
K, Sc, V and Ti are known to be under-produced at all metallicity ranges in theoretical models with respect to the observations (K06; Figs.\,\ref{fig:k}-\ref{fig:ti}), and it has been shown that some multi-dimensional effects can increase Sc, V, and Ti abundances, as in the K15 model (dotted lines).
Namely, Sc and Ti yields are greatly increased in the nucleosynthesis calculation of 2D jet-induced supernovae (\S \ref{sec:sn}).
K, Sc, and V yields can also be affected by the neutrino process \citep{kob11f}, whose effects are not included in any of models in this paper.
Stellar rotation can enhance Cl, K, and Sc abundances, but not V \citep{pra18}.
Cl, K, and Sc may also be enhanced by the O-C mergers during hydrostatic burning that is seen in one of the 1D stellar evolution calculations \citep[$15M_\odot$, ][]{rit18b}.
In our models, [(Cl, K, Sc)/Fe] show a weak increase from [Fe/H] $\sim -3$ to $\sim -1$, which is due to the metallicity dependence of SN II/HN yields.
Note that the solar Cl abundance is increased by 0.37 dex, and the predicted [Cl/Fe] is negative overall, giving $-0.8$ at [Fe/H] $=-2$, which is as low as for K and Sc.
Ti and V yields do not depend very much on the progenitor metallicity, and thus [(Ti, V)/Fe] show a plateau at [Fe/H] $\ltsim -1$.
At [Fe/H] $\gtsim -1$, all of these elemental abundances show a weak decrease toward higher metallicities because of SNe Ia.
K is also known for strong NLTE effects \citep[][up to $-0.7$ dex]{takeda02,reg19}, and the plotted data in Figure \ref{fig:k} are from NLTE analysis.

{\bf Iron-peak elements} ---
Iron peak elements are synthesized in thermonuclear explosions of supernovae, as well as in incomplete or complete Si-burning during explosive burning of core-collapse supernovae (K06), and therefore it is very important to obtain the exact abundances for constraining the explosion mechanism.
Observationally, NLTE effects of iron-peak elements other than iron have not been well studied yet, except for a few cases \citep[e.g.,][]{ber08,ber10co,ber10cr}.
However, \citet{sne16} and \citet{cow20} obtained consistent abundances between neutral and ionized lines using updated atomic data, except for Cr and Co, which implies that the NLTE effects may not be so large.
Given these previous studies, it is a matter of urgency to check the NLTE effects for iron-peak elements with updated atomic data.
In Figures \ref{fig:cr}-\ref{fig:zn}, we compare our models to LTE observations.
For Cr, as shown in Figs.\,20 and 21 of K06, the difference between Cr I and Cr II abundances are significant and we use only the Cr II observations in this paper.
The difference in the adopted solar abundances is up to 0.1 dex for iron-peak elements, and it is $\sim 0.1$ dex decrease for Mn, Cu, Zn, while it is $\sim 0.1$ dex increase for Co.

At $-2.5 \ltsim$ [Fe/H] $\ltsim -1$, [(Cr, Mn, Zn)/Fe] are consistent with the observed mean values ($0.07, -0.56, 0.18$ at [Fe/H] $=-2$, respectively).
[Co I/Fe] is $-0.20$ at [Fe/H] $=-2$, which is $\sim 0.3$ dex lower than the observations (Fig.\,\ref{fig:co}) and is slightly lower than in the K11 model (dashed line), but can be increased by the HN jet effects (dotted line).
However, \citet{cow20} showed a large difference between Co I and Co II abundances, showing [Co II/Fe] $\sim 0$ at $-3\ltsim$ [Fe/H] $\ltsim-2.2$, in contrast to the very high NLTE abundances in \citet{ber10co}. Our models are in good agreement with the Co II observations, and it is necessary to increase the sample to discuss which of our models is the best.
There is no such a difference between Ni I and Ni II abundances \citep{cow20}.
The predicted [Ni/Fe] is $-0.19$ at [Fe/H] $=-2$, which is $\sim 0.2$ dex lower than the observations (Fig.\,\ref{fig:ni}).
Both Ni and Fe are produced in the complete Si-burning region (K06) and it is very difficult to change the ratio; a deeper mass cut could slightly increase $^{58}$Ni yields but not $^{60}$Ni yields, and the latter isotope is the majority at low metallicities (\S \ref{sec:iso}).

At [Fe/H] $\gtsim -1$, these abundance ratios stay roughly constant, except for [Mn/Fe] (see the next paragraph), because iron-peak elements are also produced by SNe Ia.
There was a Ni over-production problem by SNe Ia (dashed line in Fig.\,\ref{fig:ni}), which is eased in our models because of the new metallicity-dependent yields of SNe Ia (\S \ref{sec:ia}; \citealt{kob19ia}).
At [Fe/H] $\ltsim -2.5$, observational data (namely, \citealt{reg17}) show an increase of [(Co I, Zn)/Fe] toward lower metallicities, 
which is not discussed here since inhomogeneous chemical enrichment is becoming increasingly important there.

{\bf Manganese} ---
Mn is the most important element for constraining the physics of SNe Ia, since it is produced more by SNe Ia than SNe II/HNe relative to iron \citep{kob09}.
Mn yields depend on the explosion model of SNe Ia, and thus indirectly depend on the progenitor model of SNe Ia.
In this paper we use the 2D delayed detonation model for the SN Ia yields given as a function of metallicity (\S 2.1).

At [Fe/H] $\ltsim -1$, [Mn/Fe] shows a plateau with $\sim -0.56, -0.55, -0.43$ at [Fe/H] = $-3,-2,-1.1$, which is determined by the IMF-weighted SN II/HN yields and is consistent with the plotted observations in Figure \ref{fig:mn}.
The small difference in the adopted solar abundances is cancelled out with a small reduction of Mn yields by failed supernovae.
Recently, \citet{eit20} suggested a large NLTE correction for Mn I lines ($\sim -0.4$ dex, up to 0.6 dex) and [Mn/Fe] is $\sim 0$ at $-4 \ltsim$ [Fe/H] $\ltsim 0$.
However, using the same Mn atomic model, \citet{ama20} found a steeper NLTE [Mn/Fe] relation at [Fe/H] $\gtsim -2$. \citet{ber19} also showed a positive or negative 3D effect depending on the spectral lines used and model parameters. Consequently, 3D-NLTE analysis for a large sample of stars is needed.
Theoretically, Mn is synthesized in the incomplete Si-burning regions together with Cr, and thus higher [Mn/Fe] would result in higher [Cr/Fe], which is inconsistent with the Cr II observations and even more inconsistent with Cr I observations.
In Figure \ref{fig:mn2}, we also show Mn II observations. Since it is harder to obtain the Mn II lines, there is not much literature on this. [Mn/Fe] plateau value may be $0.1-0.2$ dex higher than in the model but should not be as high as [Mn/Fe] $\sim 0$.

Above [Fe/H] $\sim -1$, [Mn/Fe] shows an increase toward higher metallicities, which is caused by the 
delayed enrichment of SNe Ia and has been used to constrain the progenitors of SNe Ia \citep{sei13,ces17,kob19ia,eit20}.
Similar evolutionary trend with a plateau and an increase were first found by \citet{gra89}\footnote{\citet{bey78} found an increase but their metallicity range was not low enough to find the plateau.}.
\citet{fel07} showed a steep slope at [Fe/H] $>0$ than in \citet{red03}, which is better reproduced with the fiducial model than with the K11 model (dashed line).

\begin{figure*}\center
\includegraphics[width=17.5cm]{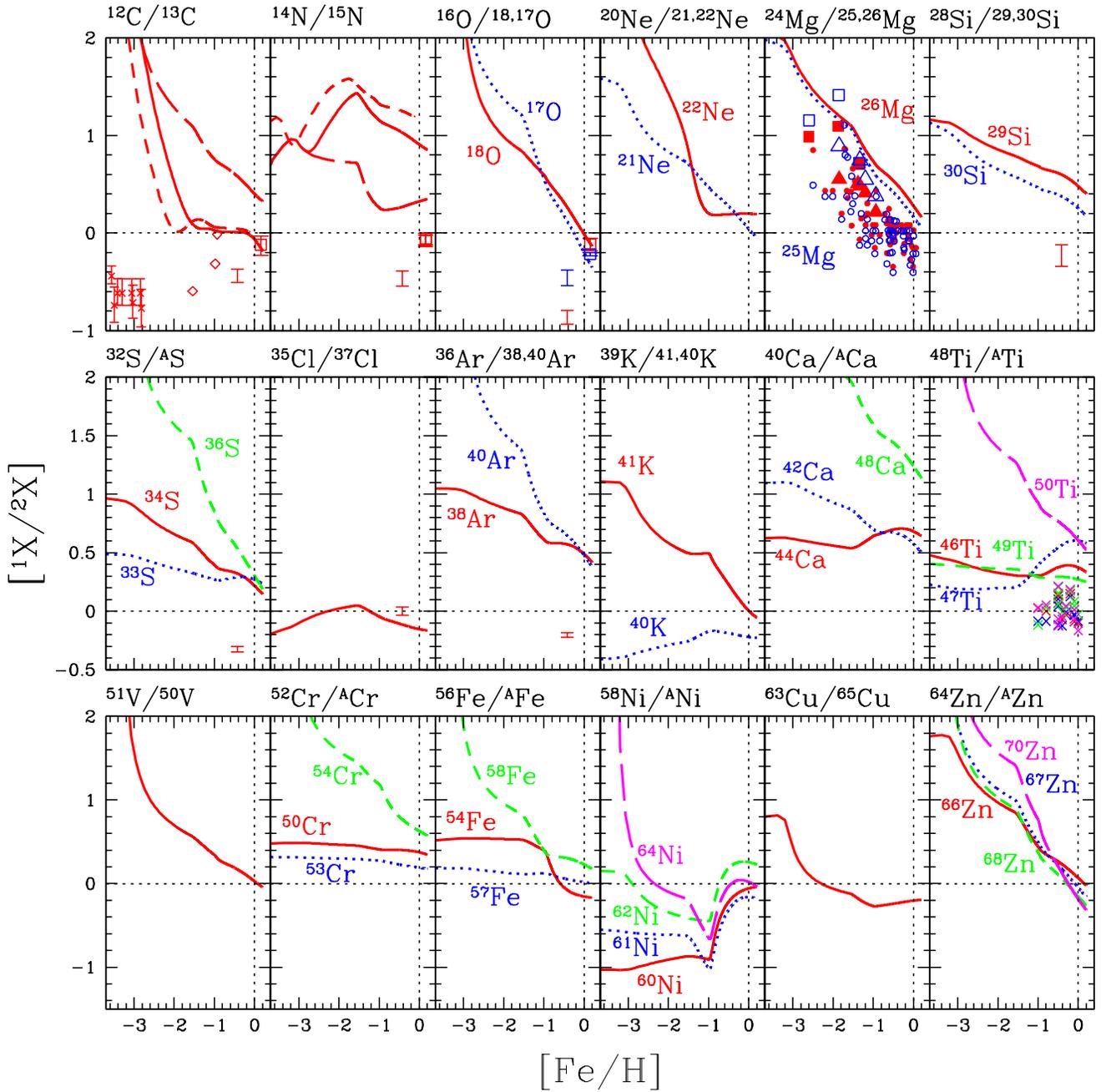}
\caption{\label{fig:iso}
Evolution of isotope ratios relative to the solar ratios, 
$\log ({^1}{\rm X}/{^2}{\rm X}/({^1}{\rm X}_\odot/{^2}{\rm X}_\odot))$, against [Fe/H]
for the solar neighborhood model.
For C and N, the models
without super-AGB stars (red short-dashed lines), without AGB/super-AGB stars (red long-dashed lines) are also shown comparing to the fiducial model (red solid lines).
Observational data sources for stars include:
for C,
\citet{car00}, diamonds;
\citet{spi06}, crosses;
for $^{25}$Mg and $^{26}$Mg, 
\citet{yon03}, small open and filled circles;
\citet{mel07}, open and filled squares;
\citet{car18}, open and filled triangles, respectively;
for Ti,
\citet{cha09}, crosses.
The squares with error bars at [Fe/H] $\sim +0.2$ are for the local ISM:
\citet{milam05} for C;
\citet{colzi18} for N;
\citet{poleh05} for $^{18}$O; and
\citet{woute08} for $^{17}$O.
The small error bars at [Fe/H] $\sim -0.4$ are for a spiral galaxy at $z=0.89$ from \citet{mul11,mul14,mue15}.
}
\end{figure*}

{\bf Zinc} ---
Zn is one of the most important elements for the physics of core-collapse supernovae as $^{64}_{30}$Zn is enhanced in the deepest region of HNe with higher explosion energy and entropy (K06), and thus [Zn/Fe] is increased with multi-dimensional effects (dotted line in Fig.\,\ref{fig:zn}).
Zn yields potentially depend on the neutrino processes and the resultant $Y_{\rm e}$ as well.
However, it is worth noting that Zn can be enhanced at the stellar layer with $Y_{\rm e} \sim 0.5$, but $Y_{\rm e}$ too close to $0.5$ gives too small $^{55}_{25}$Mn yields. From these dependencies, $Y_{\rm e}=0.4997$ is chosen for the incomplete Si-burning regions in K06.
Neutron-rich isotopes of zinc ($^{66-70}$Zn) can also be produced by neutron-capture processes, of which yields are larger for higher metallicity massive SNe II (\S \ref{sec:iso}).
The contribution from the s-process in AGB stars is very small, increasing [Zn/Fe] only by $0.004$ dex at [Fe/H] $=0$.
As noted before, ECSNe can enhance Zn as well as Ni and Cu, but the contribution to GCE is small, up to $0.04$ dex (magenta long-dashed lines in Fig.\,\ref{fig:xfe}).

The predicted [Zn/Fe] is about $\sim 0.2$ for a wide range of metallicities if we apply such a large fraction of HNe (\S 2.1).
This is $\sim 0.1$ dex larger than in the K11 model, where the different is mostly due to the adopted solar abundance.
Our [Zn/Fe] ratios are in good agreement with the observational data, and also the results from \citet{roe18}, who obtained Zn II lines and found that the NLTE effect should be less than 0.1 dex, at [Fe/H] $\gtsim -2.5$.
At [Fe/H] $\ltsim$ $-2.5$, \citet{sne88} first suggested an increase of [Zn/Fe], more recent observations show a linear increase from [Fe/H] $\sim -1$ to lower metallicities \citep{pri00,nis07,sai09}, and the increase may be steeper with the NLTE correction \citep{takeda05}.
This may imply that the HN fraction is larger in the earlier stages of galaxy formation (K06).

[Zn/Fe] observations seem to decrease from [Fe/H] $\sim -1$ to $\sim -0.5$, and then slightly increase to [Fe/H] $\sim 0$ \citep{sai09}. The NLTE correction there is found to be mostly $\ltsim 0.1$ dex \citep{takeda16}.
This up-turn trend is well reproduced with our fiducial model.
With the metallicity dependent HN fraction, the [Zn/Fe] ratio shows a continuous decrease from [Fe/H] $\sim -1$ to $\sim 0$, which gives lower values than some of the observational data at high metallicities (Fig.\,\ref{fig:xfe2}).
A similar problem also arises for Co and Cu with the metallicity dependent HN fraction.
However, this problem is not seen in the chemodynamical simulations of \citet{kob11mw} (\S \ref{sec:sn}).

\subsection{Isotopic ratios from C to Zn}
\label{sec:iso}

Figure \ref{fig:iso} shows the evolution of isotopic ratios against [Fe/H] for the solar neighborhood models.
Comparing to the fiducial model including AGB and super-AGB stars (red solid lines), the models without super-AGB stars (red short-dashed lines) and with neither AGB nor super-AGB stars (red long-dashed lines) are also shown for C and N.
The predicted $^{12}$C/$^{13}$C ratio is 77.0 at [Fe/H] $=0$, which is slightly lower than the solar ratio (89.4, AGS09), but is 62.7 at 13.8 Gyr, which is in good agreement with the local ISM observations \citep[$68\pm15$][squares]{milam05}.
Without AGB stars, the $^{12}$C/$^{13}$C ratio would be too high. Super-AGB stars further decrease $^{12}$C/$^{13}$C, which is consistent with the finding in \citet{rom19}.
At low metallicities, as discussed for N, the effect of inhomogeneous enrichment is important, which could explain the low $^{12}$C/$^{13}$C ratios at low metallicities in stellar observations.

The under-production problem of $^{15}$N is known, and may require other sources such as novae (\citealt{rom03,rom19}; see K11 for more details) and/or H-ingestions in massive stars \citep{pig15}.
Super-AGB stars produce more $^{14}$N than $^{15}$N, which results in an even larger $^{14}$N/$^{15}$N ratios.
N isotopic ratios are also observed in carbon stars, showing $^{14}$N/$^{15}$N ratios of $\sim 1000$ for N-type carbon stars \citep{hed13}, which might require $^{15}$N production in the He-shell.

The predicted $^{16}$O/$^{18}$O ratio is 484 at [Fe/H] $=0$, which is in good agreement with the solar ratio (499, AGS09), and is 389 at 13.8 Gyr, which is also in excellent agreement with the local ISM observations \citep[$385\pm56$][]{poleh05}.
$^{17}$O is, however, over-produced by AGB and super-AGB stars, giving $^{16}$O/$^{17}$O by a factor of $\sim 1.5$ lower than the solar ratio. However, new sets of yields need to be calculated and tested using the new rate of the $^{17}$O(p,$\alpha$)$^{14}$N measured underground by the LUNA collaboration \citep{bruno16}. The new rate is 2 to 2.5 times higher than the previous rate used in the models considered here and will result in a decrease of the contribution to $^{17}$O from AGB and super-AGB stars potentially of the same magnitude needed to reach the solar value.

The solar $^{21,22}$Ne/$^{20}$Ne ratios are reasonably well reproduced with AGB stars. The predicted $^{24}$Mg/$^{25,26}$Mg ratios are higher than the stellar observations, even with the inclusion of super-AGB stars.
There is no significant difference among these models for the other ratios, and these are determined by the SN II/HN yields (K11); small mismatches for the solar ratios still remain.
Because minor isotopes are enhanced by higher metallicity SNe II/HNe, the major-to-minor isotopic ratios decrease as a function of metallicity.
The model without failed SNe but with $M_{\rm u}=40M_\odot$ (green dotted lines in Fig.\,\ref{fig:xfe2}) gives almost the same trends of isotopic ratios, while the model with the metal-dependent HN fraction (blue short-dashed lines in Fig.\,\ref{fig:xfe2}) gives slightly shallower evolution of these isotopic ratios.
The solar Zn isotopic ratios can be better explained with the metal-dependent HN fraction, although the Zn abundance is better explained without it.

Isotopic ratios are also available with the detailed analysis of molecular lines in radio observations, and an IMF variation is suggested from the low values at high redshifts \citep[e.g.,][]{zhang18}.
This figure also shows some data for a high-redshift galaxy (small error bars), which are far from any of these theoretical predictions at $t=6.3$ Gyr (corresponding to the observed redshift). Since these data are for a spiral galaxy, it is unlikely that the SFRs and/or the IMF are very different from our solar-neighborhood models. These mismatches should be noted when the IMF is constrained from radio observations.

\begin{figure*}\center
\includegraphics[height=15.5cm,angle=90]{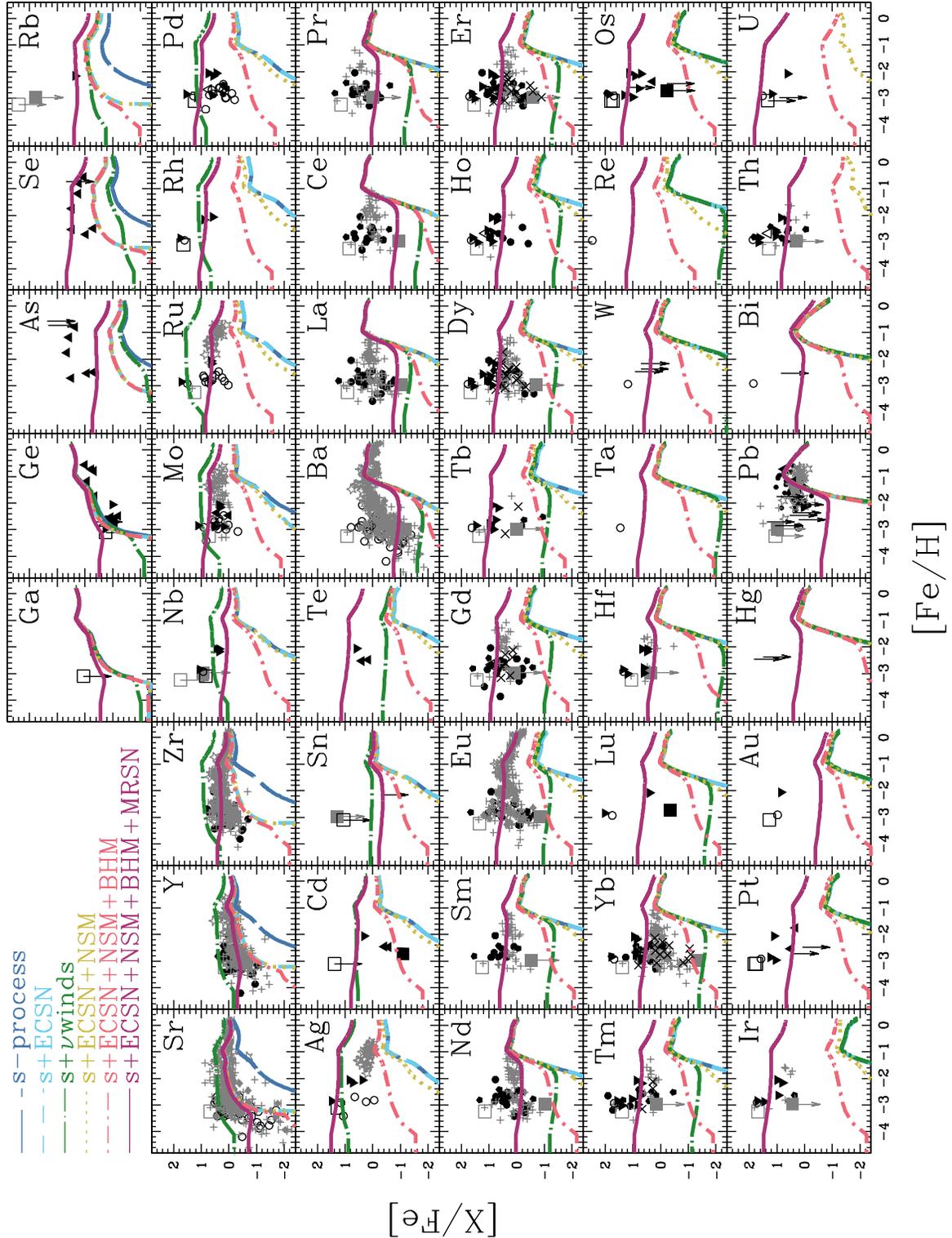}
\caption{\label{fig:xfesr}
Evolution of the neutron-capture elemental abundances [X/Fe] against [Fe/H]
for the models in the solar neighborhood,
with s-process from AGB stars only (blue long-dashed lines),
with s-process and ECSNe (light-blue short-dashed lines),
with s-process, ECSNe, and $\nu$-driven winds (green dot-long-dashed lines),
with s-process, ECSNe, and NS-NS mergers (olive dotted lines),
with s-process, ECSNe, and NS-NS/NS-BH mergers (orange dot-short-dashed lines),
with s-process, ECSNe, NS-NS/NS-BH mergers, and MRSNe (red solid lines).
Observational data sources are:
filled diamonds with error bars, \citet{zha16,mas17};
open circles, \citet{and09,and11,bar11,spi18};
filled circles, \citet{fra07};
filled pentagons, \citet{hon04};
stars, \citet{han12,han14};
gray plus, \citet{roe14}.
For the elements except for Sr, Y, Zr, Ba, La, Ce, Pr, Nd, Sm, and Eu,
crosses, \citet{joh02};
open triangles, \citet{lai08};
filled triangles, \citet{roe12,roe12te,roe14n}.
The large gray filled and open squares indicate the Sneden and Honda stars, respectively, and the filled upside-down triangles
%from \citep{wes00,sne03,roe10,roe14}.
and upper limits are for r-II stars (see the text for the observational data sources).
% from \citet{wes00,cow02,iva06,sne09,hay09,roe10,roe12te,mas10,mas14}.
}
\end{figure*}

\subsection{Neutron-capture element abundances}
\label{sec:sr}

\begin{figure}\center
\includegraphics[width=8.5cm]{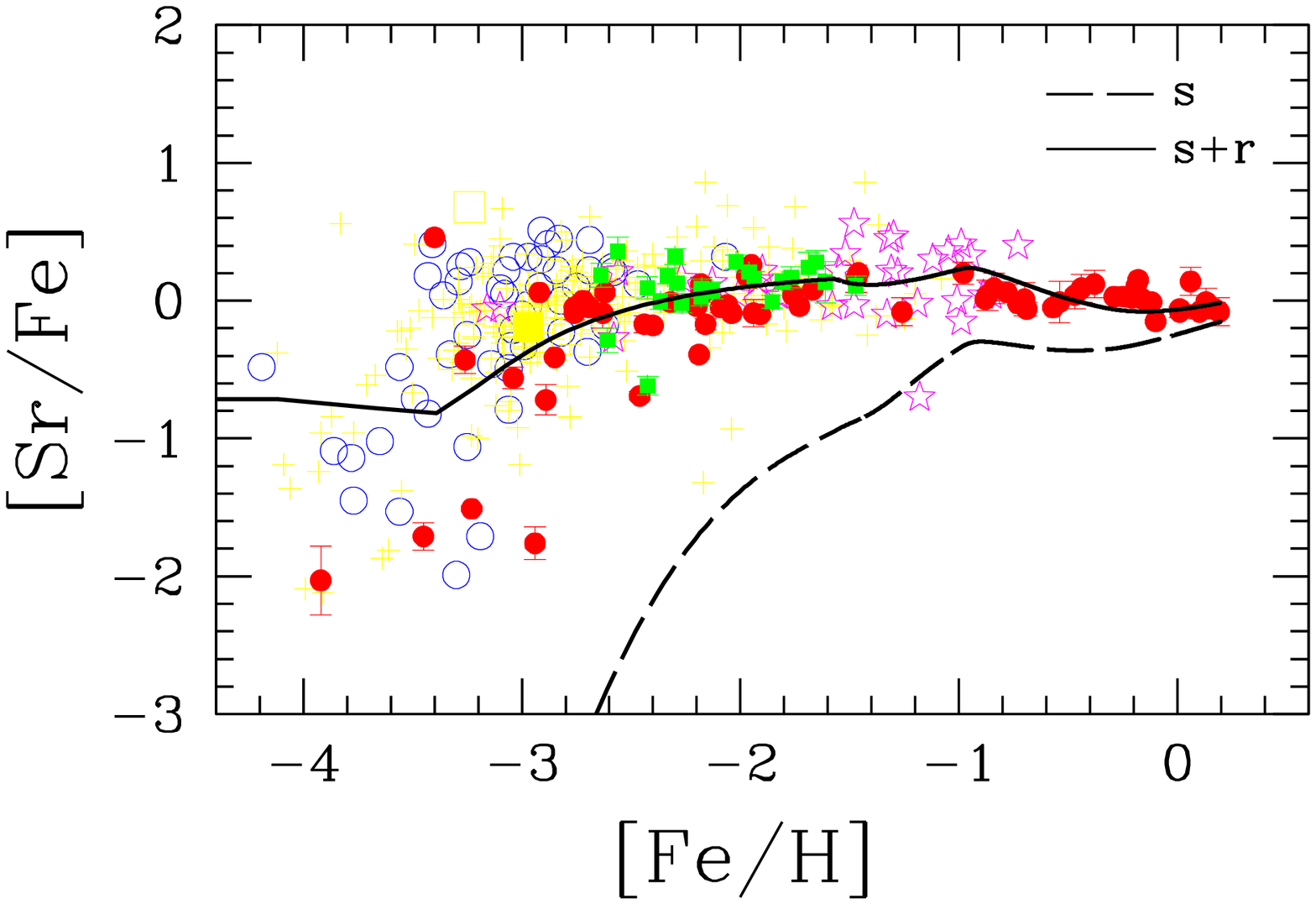}
\caption{\label{fig:sr}
The [Sr/Fe]--[Fe/H] relation for the solar neighborhood models in our model with s-process only (dashed line) and with s- and r-processes (solid line).
Observational data sources are:
red filled circles with error bars, \citet{zha16,mas17};
green filled squares with error bars, \citet{reg17};
blue open circles, \citet{and11};
magenta stars, \citet{han12,han14};
yellow plus, \citet{roe14} for C-normal stars.
The large yellow filled and open squares indicate the Sneden and Honda stars, respectively.
}
\end{figure}

\begin{figure}\center
\includegraphics[width=8.5cm]{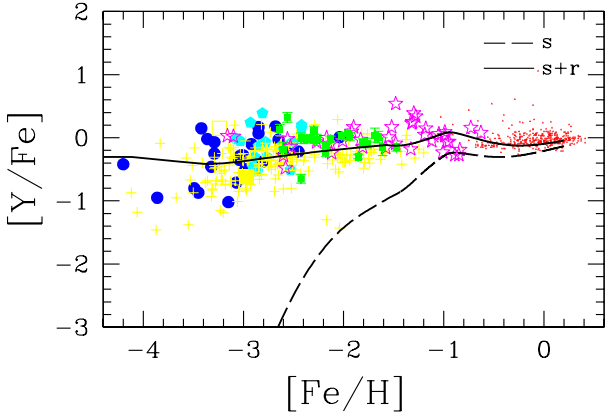}
\caption{\label{fig:y}
Same as Fig.~\ref{fig:sr}, but for the [Y/Fe]--[Fe/H] relation.
Observational data sources are:
blue filled circles, \citet{cay04};
cyan filled pentagons, \citet{hon04};
filled triangles, \citet{ben14} for thin disk stars;
and same as Fig.\ref{fig:sr} for the other datapoints.
}
\end{figure}

\begin{figure}\center
\includegraphics[width=8.5cm]{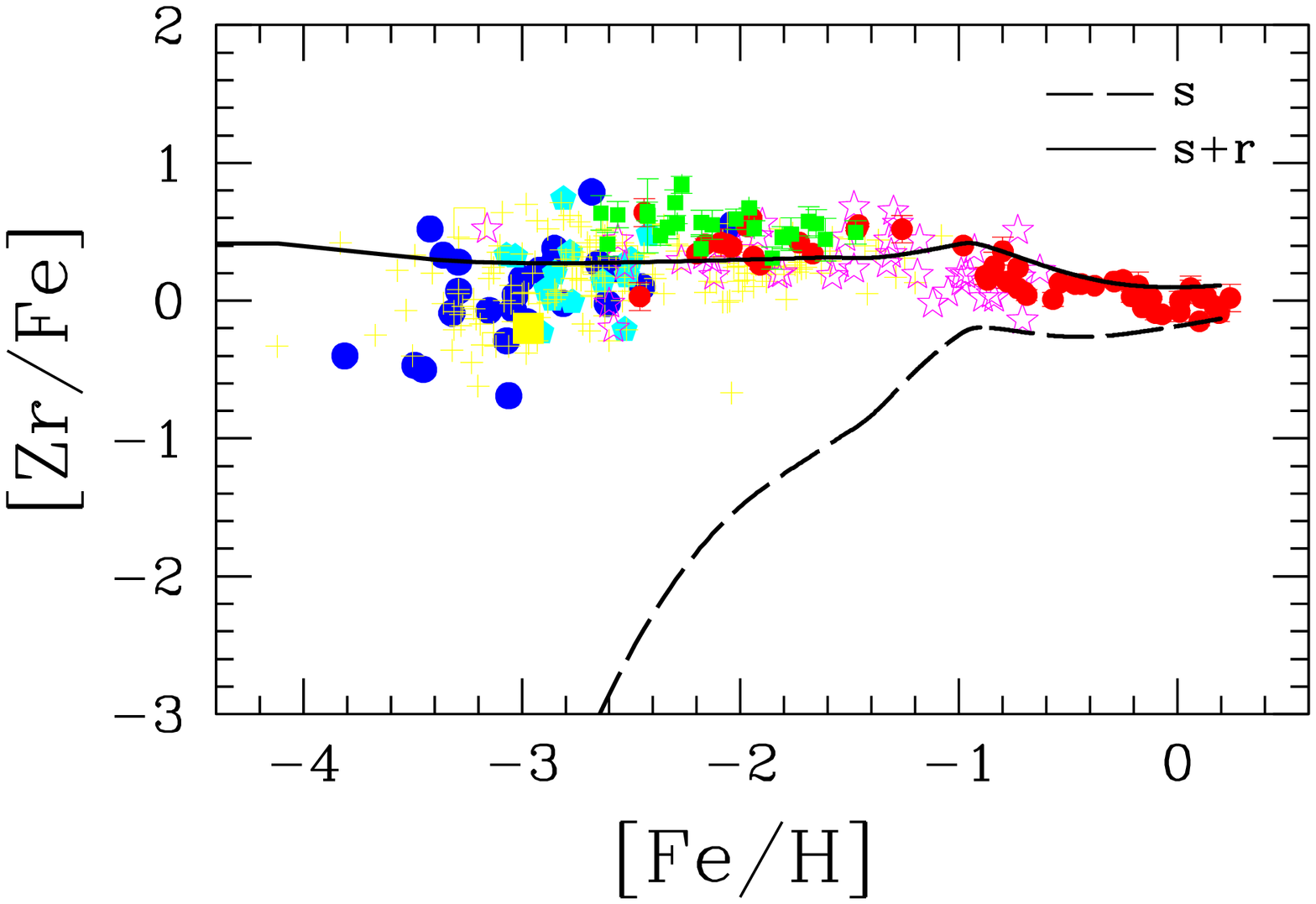}
\caption{\label{fig:zr}
Same as Fig.~\ref{fig:y}, but for the [Zr/Fe]--[Fe/H] relation.
}
\end{figure}

Figure \ref{fig:xfesr} shows the evolution of neutron-capture elements in the solar neighborhood, for all neutron-capture elements that have stellar abundance estimates or upper limits in the literature; some of the elements including Au are measurable only in UV spectra from the Hubble Space Telescope, and thus the number of measurements is very limited. It is necessary to increase the sample to obtain a comprehensive picture on the origin and evolution of these heavy elements.
The number of candidates is being increased by strategic surveys such as the R-process Alliance \citep{han18}.
Two characteristic stars that show different r-process enrichment\footnote{CS 22892-052 is a CEMP star, and its large r-process enhancement was reported by \citet{sne94}. The observational data are primarily taken from \citet{roe14} with [Fe I/H] $= -3.24$ and [Fe II/H] $=-3.16$, and additionally from \citet{sne03} with [Fe I/H] $= -3.10$ and [Fe II/H] $=-3.09$ for Ga, Ge, Rh, Pd, Ag, Cd, Sn, Lu, Os, Pt, Au, and U.
For HD 122563, relatively low r-process abundances were reported by \citet{sne83}. The observational data are primarily taken from \citet{roe14} with [Fe I/H] $= -2.97$ and [Fe II/H] $=-2.93$, and additionally from \citet{wes00} for Ge and from \citet{roe10} for Cd, Lu, Os II, which use [(Fe I, Fe II)/H] $=-2.72$. Note that the [Fe/H] of this star was $-2.77$ in \citet{hon04} and $-2.63$ in \citet[][NLTE]{mas17}.}
are highlighted with the large gray filled and open squares, respectively.
For the elements except for Sr--Zr and Ba--Gd, so-called r-II stars\footnote{Our sample of r-II stars are: HD 115444 \citep{wes00,sne09,roe10}, BD +17$^\circ$3248 \citep{cow02,sne09,roe10,roe12te}, HD 221170 \citep{iva06,sne09}, CS 29491-069 and HE 1219-0312 \citep{hay09}, HE 2327-5642 \citep{mas10}, HE 2252-4225 \citep{mas14}, and CS 29497-004 \citep{hil17}, in addition to CS 31078-018 \citep{lai08} and CS 31082-001 \citep{bar11,siq13,spi18} that are plotted also in the other figures.} 
are also plotted but this may cause a selection bias; some elements are detected because of the large r-process enhancement.
This figure shows our six GCE models switching the neutron-capture enrichment sources one by one.
There are no differences in the evolution of the elements up to Ni when considering the different models shown in this figure.
Ga and Ge are produced mainly from core-collapse supernovae, and the predicted trends are consistent with the observations \citep{sne03,roe14n} although the sample is very limited.

The s-process from AGB stars (blue long-dashed lines) can produce elements up to Pb, and the contribution appears from [Fe/H] $\sim -2$ for light s-process elements (e.g., Sr, Y, Zr) and only from [Fe/H] $\sim -1.5$ for heavy s-process elements (e.g., Pb).
This is because the light s-process elements are produced also from intermediate-mass AGB stars, while the heavy s-process elements are mainly produced by low-mass AGB stars.
At [Fe/H] $=0$, $\sim 70$\% of the solar abundances of the elements belonging to the first s-process peak, Sr, Y, and Zr, are produced from AGB stars, while only $\sim50$\% or less of the solar abundances of Mo, Ru, and Ag are of s-process origin, as well known \citep{arlandini99}. The elements Ba, La, Ce, belonging to the second s-process peak, 
are $\sim50$\%  overproduced, while the solar abundances of Pr and Nd are reproduced.
The elements from Eu to Tm as well as Ir are typical r-process elements, with less than 30\% of their solar abundances contributed by the s-process. The elements Yb and Hf have a significant contribution from the s-process.
Finally, Pb belongs to the third s-process peak, and it is also overproduced by $\sim 30$\%.

The over-production of the second (Ba) and third (Pb) s-process peak elements at [Fe/H] $=0$ suggests that the contribution from the s-process with the adopted yield set is too large. This contribution can be reduced by decreasing the parameter $M_{\rm mix}$ in the AGB models, which controls the extent in mass of the region where the s-process elements are produced (\S \ref{sec:sn}). 
As noted in \S 2.1, $M_{\rm mix}$ values are already reduced from those in \citet{kar16} (e.g., from $2\times10^{-3}$ to $1\times10^{-3}$ in the $2M_\odot$ models with $Z=0.0028$ and $0.014$).
A decrease in $M_{\rm mix}$ brings the model predictions for the second peak s-process elements qualitatively closer to the direct observations of carbon-rich AGB stars at solar metallicity from \citet{abia02}, although it is difficult to perform a detailed analysis given the large observational error bars ($\pm$0.4 dex). Smaller $M_{\rm mix}$ values than the standard considered here also provide a better coverage of the spread in isotopic ratios observed for Sr, Zr, and Ba in meteoritic stardust SiC grains \citep{lugaro18}.

With ECSNe (light-blue short-dashed lines), the enhancement is as small as $\sim 0.1$ dex for [(Cu, Zn)/Fe] (Fig.\,\ref{fig:xfe}), but a larger enhancement is seen for As, Se, Rb, Sr, Y, Zr starting from [Fe/H] $\sim -3$.
Rb is produced more by ECSNe than by AGB stars.
The ECSN contribution appears before AGB stars start to contribute because the progenitors of ECSNe are more massive than those of AGB stars at a given metallicity (\S \ref{sec:sn}).
There is also a slightly earlier increase in Mo, while for heavier elements the contribution from ECSNe is negligible.
This result is different from \citet{ces14} mainly because they assumed ECSNe from all $8-10M_\odot$ stars.

With $\nu$-driven winds (green dot-long-dashed lines), the elements from Sr to Ag are largely over-produced, which is a crucial problem.
Therefore, the $\nu$-driven winds are not included in the other models in this figure.
Since $\nu$-driven winds are associated with core-collapse supernovae, the contribution can appear at [Fe/H] $<<-3$. This rapid contribution may explain some of the observations at [Fe/H] $\ltsim -3$.

With NS-NS mergers (olive dotted lines), the elements heavier than Zr show an increase starting from [Fe/H] $\sim -3$ due to the r-process. Although the progenitor masses of NSs are larger than those of ECSNe at a given metallicity, there is a time-delay for the merging of two NSs (\S \ref{sec:r}).
The time-delay is shorter for NS-BH mergers, and the first increase in the abundances is seen already at [Fe/H] $\sim -4$ if a similar r-process occurs in NS-BH mergers (orange short-dot-dashed lines). 
However, this time-delay is still not short enough to explain the observations at [Fe/H] $\ltsim -3$, although our one-zone models cannot give a strong conclusion at [Fe/H] $\ltsim -2.5$ where chemical enrichment takes place inhomogeneously.
NS-NS/NS-BH mergers do not produce Pb, but Th and U.

\begin{figure}\center
\includegraphics[width=8.5cm]{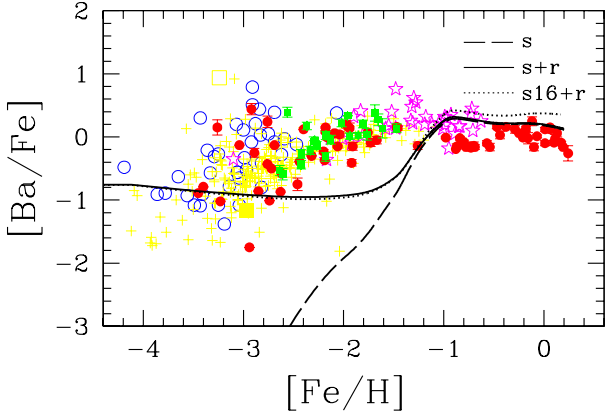}
\caption{\label{fig:ba}
Same as Fig.~\ref{fig:sr}, but for the [Ba/Fe]--[Fe/H] relation.
The dotted line shows a model with the s-process yields calculated using a twice larger $M_{\rm mix}$ at $Z\ge0.0028$ (see the text for the details).
Observational data sources are:
blue open circles, \citet{and09};
and same as Fig.\ref{fig:sr} for the other datapoints.
}
\end{figure}

Observations of metal-poor stars show that by [Fe/H] $\ltsim$ \\\noindent
$-3$ there is already an enhancement of all neutron-capture elements, which requires production from a different site with a shorter time-delay than NS-NS/NS-BH mergers \citep[e.g.,][]{weh15,hay19,cot19}.
MRSNe are very good candidates for the rapid enrichment as they are massive core-collapse supernovae with a very short time-delay ($10^6$ yrs).
In fact, the model with MRSNe (red solid lines) show a plateau at [Fe/H] $\ltsim -3$, which is similar to $\nu$-driven winds, but the elemental abundance ratios are much more consistent with observations than with $\nu$-driven winds.
From Sr to Ru, the plateau values are lower than with $\nu$-driven winds, and are $\sim -1$ and $\sim 0$ for [(Sr, Y)/Fe] and [(Zr, Mo, Ru)/Fe], respectively, which are in good agreement with the observations (see also Figs.\,\ref{fig:sr}-\ref{fig:zr} for more detailed comparison).
For Ag, both $\nu$-driven winds and MRSNe seem to over-produce its abundance, although observational data have been provided by only a few studies \citep[e.g.,][]{han12,spi18}. Similar over-production may also be seen for Pd and Cd.
For Ba, the models with MRSNe or $\nu$-driven winds can explain the low [Ba/Fe] of some metal-poor stars. However, AGB stars should contribute more at $-3\ltsim$ [Fe/H] $\ltsim -2$ (Fig.\,\ref{fig:ba}), which is possible with inhomogeneous chemical enrichment \citep[see also][]{raiteri99}.

\begin{figure}\center
\includegraphics[width=8.5cm]{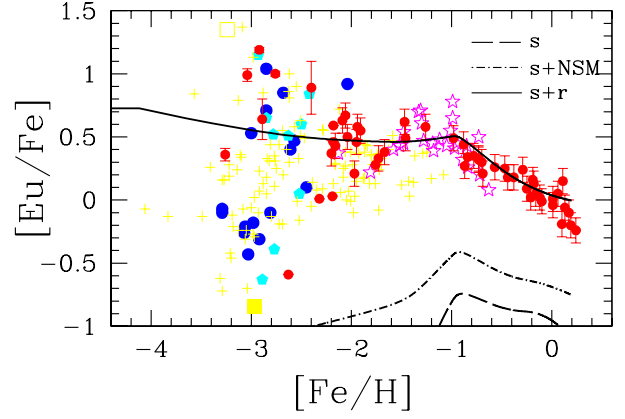}
\caption{\label{fig:eu}
Same as Fig.~\ref{fig:y}, but for the [Eu/Fe]--[Fe/H] relation.
The dot-dashed line shows a model without MRSNe.
}
\end{figure}

For the elements heavier than Sn, the plateau values are higher with MRSNe than with $\nu$-driven winds, which is more consistent with the observations.
The Te data are provided by only one study \citep{roe12te}, which seem to favour MRSNe than $\nu$-driven winds.
Although the model with MRSNe is in good agreement with the observations for Pr, Nd, Sm, Eu, Gd, Tb, Dy, Ho, Er, Tm, and Yb, the model prediction is lower than observed for La, Ce, and Pr at [Fe/H] $\ltsim -2$.
Note that the atomic data for Pr, Dy, Tm, Yb, and Lu have been updated \citep{sne09}, which may affect some of the observational data plotted here.
Similar to Ba (Fig.\,\ref{fig:ba}), these elements can be enhanced at $-3\ltsim$ [Fe/H] $\ltsim -2$ possibly by AGB stars with inhomogeneous chemical enrichment.
In contrast, the MRSN model can reproduce the average trend of [Eu/Fe] very well (Fig.\,\ref{fig:eu}). Note that the inhomogeneous enrichment could slightly increase the contribution from NSMs at $-3\ltsim$ [Fe/H] $\ltsim -2$.
This model is also acceptable for Os, Ir, and Pt, while Au is underproduced in the model. Au measurements are available only for three well-known r-process enhanced stars: BD +17$^\circ$3248 \citep[][filled upside-down triangle]{cow02}, CS 22892-052 \citep[][open square]{sne03}, and CS 31082-001 \citep[][open circle]{bar11}.
Finally, the MRSN model can also explain the observed Th abundances of metal-poor stars and the Sun.
Note that our GCE model predictions are after the long-term decay at each time and are normalized by the proto-solar abundance (\S \ref{sec:gce}), and thus the lines are expected to go through [X/Fe] $=$ [Fe/H] $=0$. The observational data, however, are normalized by the present-day solar abundance from AGS09, assuming that the observed stars are as old as the Sun.
Compared with Th, however, U may be overproduced in the nucleosynthesis yields; this is more serious for MRSNe, which give $M({\rm Th})/M({\rm U}) = 0.18$, while NSMs give $M({\rm Th})/M({\rm U}) = 0.58$. In our adopted solar abundances, $M_\odot({\rm Th})/M_\odot({\rm U}) = 1.7$.

\begin{figure}\center
\includegraphics[width=8.5cm]{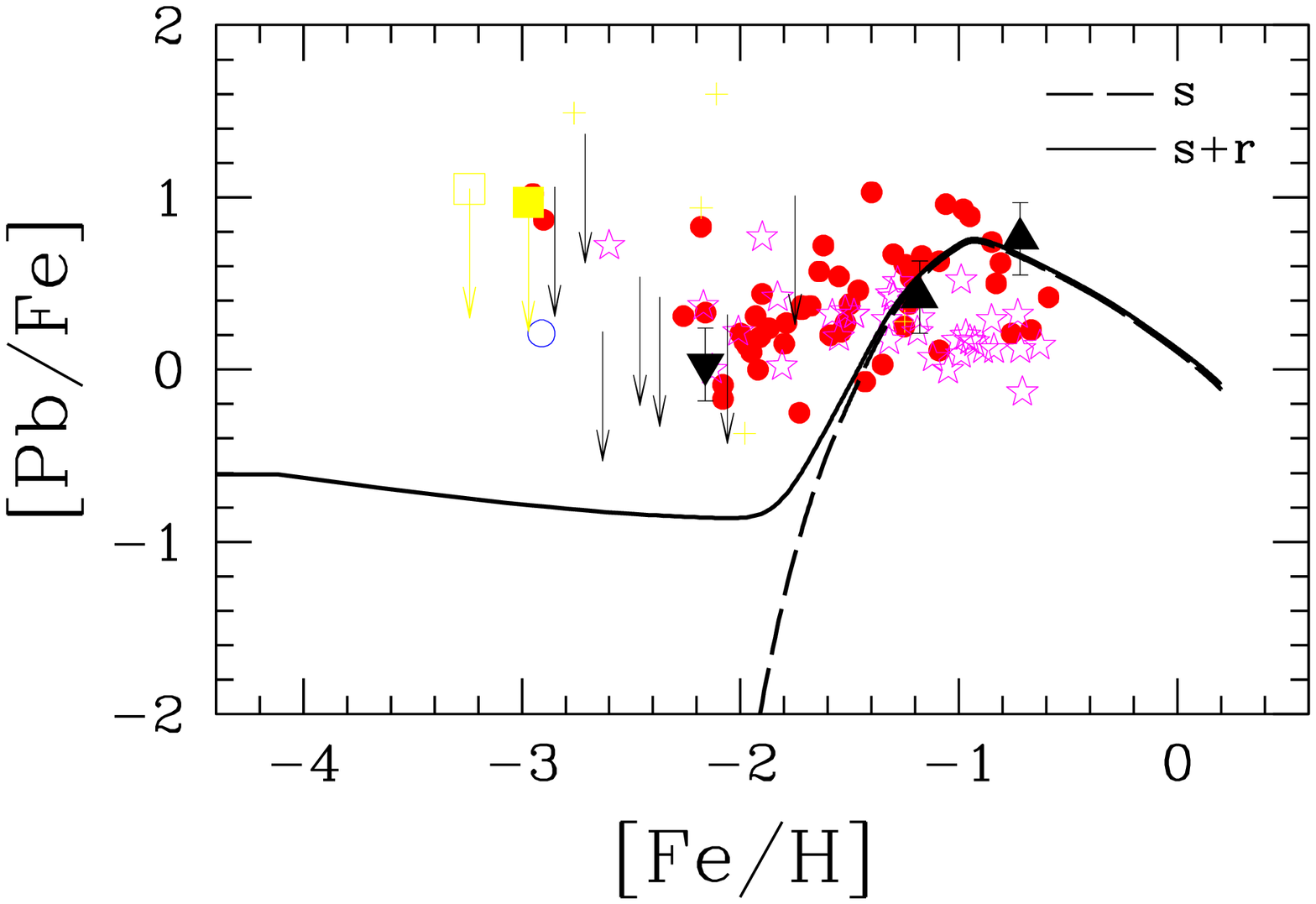}
\caption{\label{fig:pb}
Same as Fig.~\ref{fig:y}, but for the [Pb/Fe]--[Fe/H] relation.
Observational data sources are:
red circles, \citet[][NLTE]{mas12};
blue open circle, \citet{bar11};
magenta stars, \citet{han12};
yellow plus, \citet{roe14} for C-normal stars;
black filled triangles \citet{roe12};
black filled upside-down triangle \citep{iva06};
and upper limits \citep{roe10,roe12,roe14n,mas14};
The large yellow filled and open squares indicate the Sneden \citep{sne03,roe14} and Honda stars \citep{roe14}, respectively.
}
\end{figure}

{\bf Sr, Y, and Zr ---}
Figures \ref{fig:sr}-\ref{fig:eu} compare more observational data to our elemental abundance tracks of the model with s-process only (dashed lines) and with s-process, ECSNe, NS-NS/NS-BH mergers, and MRSNe (the s+r model, solid lines).
In Figure \ref{fig:sr} the base level of [Sr/Fe] $\sim -0.8$ at [Fe/H] $\ltsim -3.5$ is caused by MRSNe. The average [Sr/Fe] increases from [Fe/H] $\sim -3.5$ due to ECSNe, from [Fe/H] $\sim -2.5$ due to AGB stars before decreasing at [Fe/H] $\gtsim -1$ because of SNe Ia, and becomes [Sr/Fe] $=-0.064$ at [Fe/H] $=0$ in the s+r model (solid line).
The slope of the decrease becomes flat at [Fe/H] $\sim -0.3$ due to the increase of the AGB contribution (dashed line).
This trend is in excellent agreement with the observational data, except for one star with low [Sr/Fe]. Differential analysis by \citet{reg17} gives slightly higher ratios than \citet{zha16}'s NLTE abundances on the average.
At [Fe/H] $\ltsim -2.5$ there is a large scatter, where LTE abundances by \citet{roe14} agree well with \citet{and11}'s and \citet{zha16}'s NLTE abundances.

In Figure \ref{fig:y}, starting from [Y/Fe] $\sim -0.4$ at [Fe/H] $\sim -3.5$, the average [Y/Fe] also increases gradually from [Fe/H] $\sim -3$ to $\sim -1$ before decreasing at [Fe/H] $\gtsim -1$ due to SNe Ia, and becomes [Y/Fe] $=-0.096$ at [Fe/H] $=0$ in the s+r model (solid line).
The predicted [Y/Fe] trend is also in excellent agreement with the observations, although the differential analysis by \citet{reg17} gives slightly higher ratios. Note that there are no NLTE abundances available at any metallicities for Y.

In Figure \ref{fig:zr}, the average [Zr/Fe] is rapidly enhanced to $\sim +0.4$ by MRSNe, stays roughly constant until [Fe/H] $\sim -1$, and then decreases at [Fe/H] $\gtsim -1$ due to SNe Ia reaching [Zr/Fe] $=0.098$ at [Fe/H] $=0$ in the s+r model (solid line).
Similar to Sr and Y, the predicted [Y/Fe] is slightly lower than the differential analysis by \citet{reg17} but in excellent agreement with the other observations including the NLTE abundances from \citet{zha16}.

These three elements are similar in the sense that they are mainly produced by ECSNe and AGB stars, but the relative contribution is different.
Nonetheless, our s+r model can reproduce the observations of the three elements consistently without introducing a free parameter, and without adding extra LEPP (\S \ref{sec:sn}).
The large scatter at [Fe/H] $\ltsim -2.5$ may be caused by the rareness of ECSNe under the inhomogeneous enrichment; the stars with higher [(Sr, Y, Zn)/Fe] ratios may be locally enriched by ECSNe. It might be possible to constrain the mass ranges of ECSNe and the fate of super-AGB stars from the scatters in chemodynamical simulations.

{\bf Barium ---}
Ba is the characteristic element for the s-process in AGB stars.
In Figure \ref{fig:ba}, the average [Ba/Fe] is already enhanced up to the $\sim -1$ base level by the r-process\footnote{The base [Ba/Fe] is $-1.3$ if only 1\% of HNe are MRSNe.}, which is consistent with the NLTE observations.
Then, [Ba/Fe] shows an increase from [Fe/H] $\sim -2$ in the s+r model, but from [Fe/H] $\sim -3$ both in the NLTE and LTE observations.
Finally, [Ba/Fe] shows a plateau at $\sim 0$ from [Fe/H] $\sim -1$ in the s+r model, but from [Fe/H] $\sim -2$ in both NLTE and LTE observations.
These mean that Ba enrichment is slower in the s+r model than in the observations; this mismatch should at least partially be caused by the lack of inhomogeneous enrichment in our GCE models, and should be tested together with its effect on the other s-process elements with chemodynamical simulations such as in \citet{hay19}.
The dotted line shows a model with the s-process yield set recommended in \citet{kar16}, where a larger $M_{\rm mix}$ at $Z\ge0.0028$ was adopted.
In this model, [Ba/Fe] is already overproduced at [Fe/H] $\sim -1$, and the model [Ba/Fe] is $0.37$ at [Fe/H] $=0$, which is reduced to be $0.20$ with the lower $M_{\rm mix}$ in this paper.
The new yields with a smaller $M_{\rm mix}$ predict [(Zr,Ba)/Fe] ratios within 0.15~dex of the FRUITY yields \citep{cri15}.

{\bf Europium ---}
Different from Ba, Eu is mostly enhanced by the r-processes.
In Figure \ref{fig:eu}, the average [Eu/Fe] ratio is already super-solar at [Fe/H] $<<-3$ in the s+r model.
This plateau value, $\sim +0.5$, depends on the MRSN rates, and 3\% of HNe (\S \ref{sec:r}) is chosen to match these observations\footnote{The plateau [Eu/Fe] is $\sim 0$ if only 1\% of HNe are MRSNe.}.
From [Fe/H] $\sim -1$, the model [Eu/Fe] decreases by SNe Ia, reaching [Eu/Fe] $=0.038$ at [Fe/H] $=0$, consistent with the solar ratio.
This trend is in excellent agreement with the observations. 
At very high metallicities, although with a scatter, the NLTE abundances from \citet{zha16} may be slightly lower than in the s+r model.
The dot-dashed line shows a model without MRSNe; clearly it is not possible to explain the observed [Eu/Fe] ratios with NSMs alone and the contribution from MRSNe is necessary.

{\bf Lead ---}
Pb is also a characteristic element for AGB stars and belongs to the third-peak of the s-process, in contrast to Ba. The observational data are very limited, and can be $0.2-0.4$ dex underestimated in the LTE analysis \citep{mas12}.
This is why we adopt the meteoritic solar abundance for Pb (\S \ref{sec:gce}).
In Figure \ref{fig:pb}, the evolutionary trend is similar to that of Ba, and the model prediction is lower than the observations at $-2\ltsim$ [Fe/H] $\ltsim -1.5$, which could also be improved with inhomogeneous enrichment in chemodynamical simulations.
At [Fe/H] $\ltsim -2$, a small amount of Pb is produced by MRSNe, but the predicted plateau value is lower than the four measurements at [Fe/H] $\sim -3$.

\begin{figure*}\center
\includegraphics[width=17.5cm]{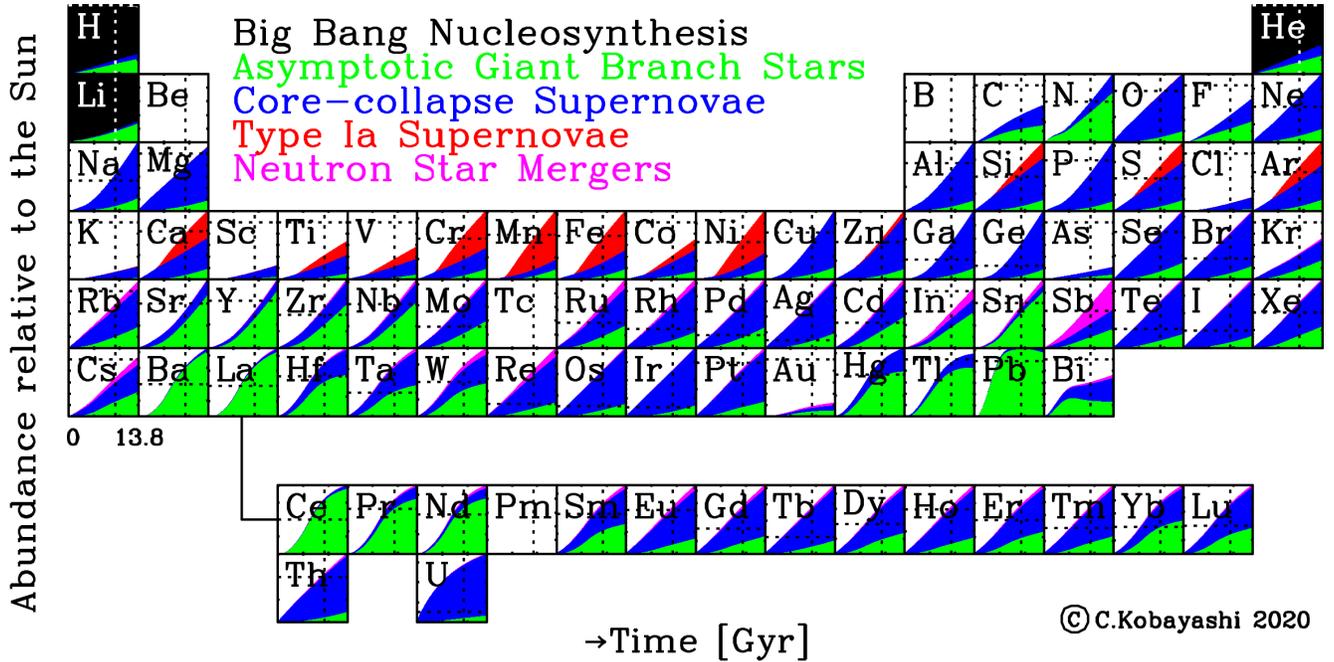}
\caption{\label{fig:origin}
The time evolution (in Gyr) of the origin of elements in the periodic table: Big Bang nucleosynthesis (black), AGB stars (green), core-collapse supernovae including SNe II, HNe, ECSNe, and MRSNe (blue), SNe Ia (red), and NSMs (magenta).
The amounts returned via stellar mass loss are also included for AGB stars and core-collapse supernovae depending on the progenitor mass.
The dotted lines indicate the observed solar values.
}
\end{figure*}

\subsection{The Origin of Elements in the Solar System}
\label{sec:origin}

Using our GCE model that includes neutron capture processes, we summarize the origin of elements in the form of a periodic table. In each box of Figure \ref{fig:origin}, the contribution from each chemical enrichment source is plotted as a function of time: Big Bang nucleosynthesis (black), AGB stars (green), core-collapse supernovae including SNe II, HNe, ECSNe, and MRSNe (blue), SNe Ia (red), and NSMs (magenta).
It is important to note that the amounts returned via stellar mass loss are also included for AGB stars and core-collapse supernovae depending on the progenitor star mass.
The x-axis of each box shows time from $t=0$ (Big Bang) to $13.8$ Gyrs, while the y-axis shows the linear abundance relative to the Sun, $X/X_\odot$.
The dotted lines indicate the observed solar values, i.e., $X/X_\odot=1$ and $4.6$ Gyr for the age of the Sun.
Since the Sun is slightly more metal-rich than the other stars in the solar neighborhood (Fig.\,\ref{fig:sfr}), the fiducial model goes through [O/Fe]$=$[Fe/H]$=0$ slightly later compared with the Sun's age.
Thus, a slightly faster star formation timescale ($\tau_{\rm s}=4$ Gyr instead of 4.7 Gyr) is adopted in this model.
The evolutionary tracks of [X/Fe] are almost identical.
The adopted star formation history is similar to the observed cosmic star formation rate history, and thus this figure can also be interpreted as the origin of elements in the Universe.
Note that Tc and Pm are radioactive.
The origin of stable elements can be summarized as follows:
\begin{itemize}
\item H and most of He are produced in Big Bang nucleosynthesis. The small green and blue areas also includes the amounts returned to the ISM via stellar mass loss and some He newly synthesized in stars.
The Li model is very uncertain because the initial abundance and nucleosynthesis yields are uncertain \citep[see also][]{gri19}.
Be and B are supposed to be produced by cosmic rays \citep{pra93}, which is not included in our model either.
\item 49\% of C, 51\% of F, and 74\% of N are produced by AGB stars (at $t=9.2$ Gyr). Note that extra production from Wolf-Rayet stars is not included and may be important for F \citep{jonsson14,spito18}.
For the elements from Ne to Ge, the newly synthesized amounts are very small for AGB stars, and the small green areas are mostly for mass loss.
\item $\alpha$ elements (O, Ne, Mg, Si, S, Ar, and Ca) are mainly produced by core-collapse supernovae, but 22\% of Si, 29\% of S, 34\% of Ar, and 39\% of Ca are come from SNe Ia. 
These fractions would become higher with sub-Ch-mass SNe Ia instead of Ch-mass SNe Ia in this model \citep{kob19ia}.
Therefore, in the [X/Fe]--[Fe/H] relations, the slopes of the decrease from [Fe/H] $\sim -1$ to $\sim 0$ are shallower for these elements than those for O and Mg.
\item A large fraction of Cr, Mn, Fe, and Ni are produced by SNe Ia. In classical works, most of Fe was thought to be produced by SNe Ia, but the fraction is only 60\% in our model, and the rest is mainly produced by HNe. The inclusion of HNe is very important as it changes the cooling and star formation histories of the Universe significantly.
Co, Cu, Zn, Ga, Ge are largely produced by HNe.
\item Among neutron-capture elements, as predicted from nucleosynthesis yields, AGB stars are the main enrichment source for the s-process elements at the second (Ba) and third (Pb) peaks. 
\item 32\% of Sr, 22\% of Y, and 44\% of Zr can be produced from ECSNe even with our conservative mass ranges, which is included in the blue areas. Combined with the contributions from AGB stars, it is possible to perfectly reproduce the observed trends (Figs.\,\ref{fig:sr}-\ref{fig:zr}), and no extra LEPP is needed. The inclusion of $\nu$-driven winds in GCE simulation results in a strong overproduction of the abundances of the elements from Sr to Sn with respect to the observations.
\item For the heavier neutron-capture elements contributions from both NS-NS/NS-BH mergers and MRSNe are necessary, and the latter is included in the blue areas.
\end{itemize}

In this model, the O and Fe abundances go though the cross of the dotted lines, meaning [O/Fe] $=$ [Fe/H] $=0$ at 4.6 Gyr ago.
This is also the case for some important elements including N, Ca, Cr, Mn, Ni, Zn, Eu, and Th.
Mg is slightly under-produced in the model.
The under-production of the elements around Ti is a long-standing problem.
The s-process elements are slightly overproduced even with the updated s-process yields in this paper.
Notably, Ag is over-produced by a factor of $6$, while Au is under-produced by a factor of $5$. U is also over-produced. These problems may require revising nuclear reaction rates.

\subsection{Uncertainties}

As discussed in previous sections, these GCE predictions mainly depend on the input nucleosynthesis yields, and hence it is difficult to quantify the uncertainties, i.e., theoretical error bars.
Very roughly speaking, $\alpha$-element abundances can vary $\sim \pm0.2$ dex depending on the detailed physics during hydrostatic stellar evolution, e.g., mass loss, convection, rotation, and magnetic fields. However, the ratios among $\alpha$-elements such as O/Mg and Si/S do not so much depend on these, and but on the $^{12}$C($\alpha$,$\gamma$)$^{16}$O reaction rate (K06).
C, N and odd-Z element abundances largely depend on rotation ($\sim +0.5$ dex), which can be found in \citet{lim18}, as well as any mixing of hydrogen into the He-burning layer (K11).

Uncertainties in the treatment of mass loss and convection also affect AGB yields (e.g., \citealt{ventura05a,ventura05b,herwig05,kar14}; for C, N, F and s-process elements \citealt{kar16}). Observations of s-process elements in AGB stars can be used to constrain for example the efficiency of third dredge-up mixing but selection biases in observations means that the whole stellar parameter space is not well sampled. Other observations including white dwarfs in clusters \citep{marigo20} or the number of carbon stars in a stellar population \citep{boyer19} provide additional constraints but are also subject to their own biases.  The uncertainties on AGB yields are difficult to quantify for the following reasons. If we vary one parameter, say slow down the rate of mass-loss in an intermediate-mass star with HBB then the yields of C can change by a factor of $\sim 2$, the yields of N by almost a factor of 10, while the yields of heavy elements can vary by almost a 100 \citep[using models from][]{kar12}. Hence varying one input parameter can lead to different variations in yields for different elements, depending upon their production mechanism.

The uncertainties of the yields from core-collapse supernovae come from the lack of physical explosion models (\S \ref{sec:sn}) including neutrino heating and black hole formation.
This could largely change the iron-peak element abundances relative to $\alpha$ elements, and thus we use the iron yields constrained from another independent observation -- supernova light-curves and spectra, which is not the case in the other supernova yields such as \citet{woo95} and \citet{lim18}.
Once the iron yields are fixed, the ratios among iron-peak elements are not so flexible, in particular, the ratios among the elements formed in the same layer, i.e., Cr/Mn and Ni/Fe, are fixed (see more discussion on Mn and Ni for Figs.\,\ref{fig:mn} and \ref{fig:ni}).
The effect of jet-like explosions can be quantified by the difference between the K11 and K15 models in Figs.\,\ref{fig:sc}-\ref{fig:v}, but an additional difference can be caused by the neutrino process. These two can cause a non-linear effect, and it is necessary to use 3D simulations to quantify the error bars.
SN Ia yields also depend on an explosion mechanism, but the main uncertainties are caused by the modelling of progenitor systems including the WD masses \citep[\S \ref{sec:ia}; see][for more discussion]{kob19ia}.

Among s-process elements, relative discrepancies between different abundances, such as the overproduction of Ba and Pb with respect to Sr, may be related to the physics of the nucleosynthetic models \citep{cri15}, for example, the extent of the partial mixing zone $M_{\rm mix}$ in AGB stars discussed above and/or the occurrence of relatively slow, diffusive mixing within the pocket \citep{battino19}.  The contribution of ECSNs and/or $\nu$-driven winds to the abundances of Sr, Y, and Zr should also be considered, as well as the possible effects of the intermediate neutron-capture process \citep[the $i$ process]{hampel16,hampel19}, see, e.g. \citet{cot18}, whose resulting abundance pattern and stellar site is however currently unknown and debated \citep{banerjee18,clarkson18,denissenkov19}. Future analysis of the production of the isotopic rather than elemental abundances in the solar system will be a powerful tool to constrain the models in more details.

The r-process nucleosynthesis is the cutting edge of nuclear astrophysics, and there are only a small number of yields available (\S \ref{sec:r}). The dependence on initial conditions, e.g., the progenitor mass for ECSNe/MRSNe and the compact object masses for NSMs, have not been studied yet. The result also depends on the detailed modelling, e.g., numerical resolution, dimensionality, inclusion of general relativity, and modelling of neutrino physics, as well as nuclear reactions.
% ML
It should also be kept in mind that major uncertainties are present in the nuclear physics inputs of the r-process model calculations, including the mostly theoretical predictions for nuclear masses and of fission fragments \citep{kajino19}.
 We note that all previous GCE models of the neutron-capture elements in the Milky Way Galaxy \citep{travaglio04,pra18} included the r-process yields as calculated using the r-residuals method, i.e., by subtracting the s-process contribution from the solar system abundances, and assumed the universal r-process to occur in SNe II. 
Here, instead we implemented a first-principle approach by including the r-process nucleosynthesis yields calculated for various r-process astrophysical sites, so that the mismatches between the model predictions and observations can be used to deepen our understanding of nuclear astrophysics.

\begin{figure*}\center
\includegraphics[width=17.5cm]{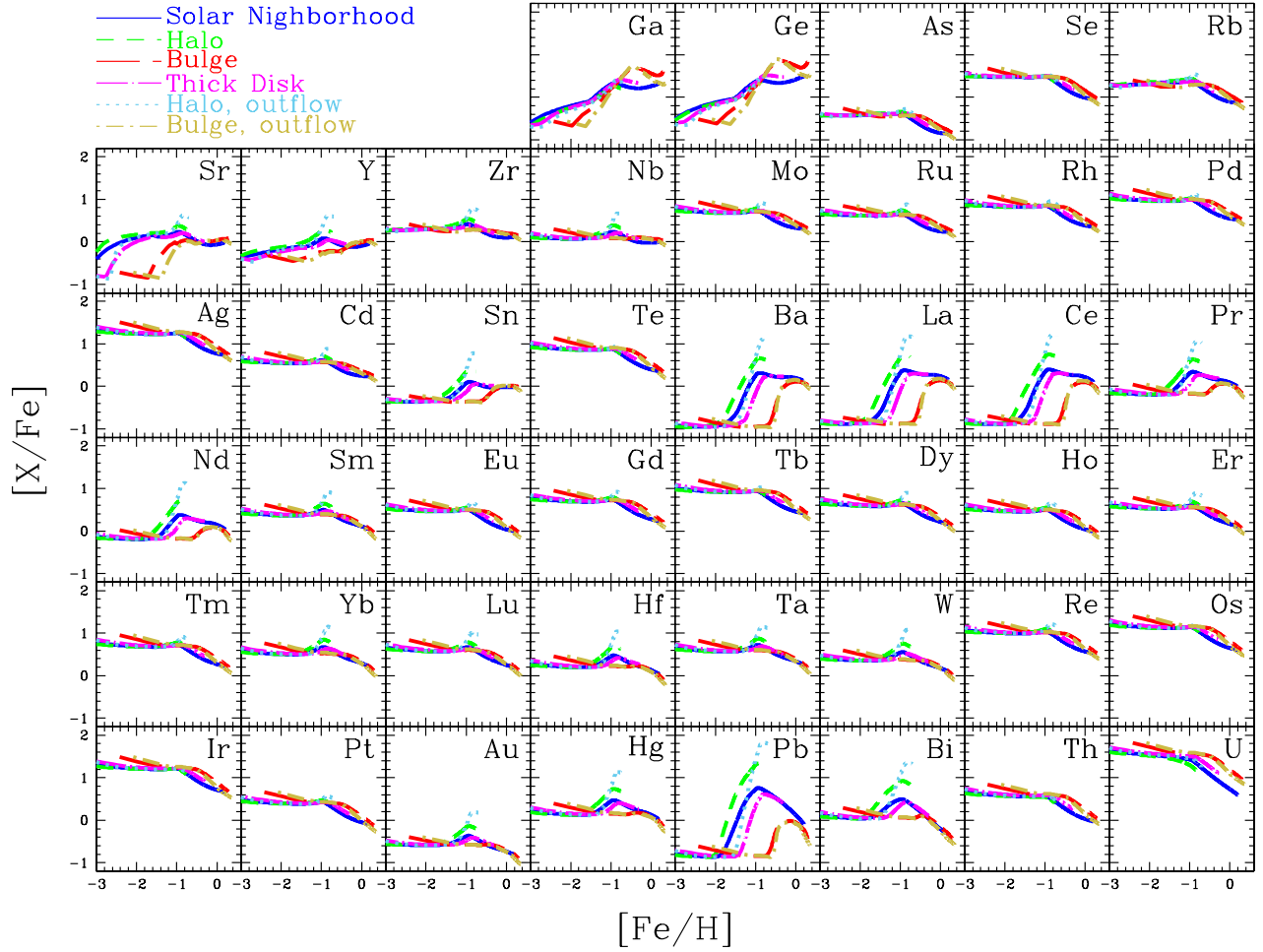}
\caption{\label{fig:xfesr2}
Evolution of the neutron-capture elemental abundances [X/Fe] against [Fe/H]
for the solar neighborhood (blue solid lines),
halo (green short-dashed lines),
halo with stronger outflow (light-blue dotted lines),
bulge (red long-dashed lines),
bulge with outflow (olive dot-short-dashed lines),
and thick disk (magenta dot-long-dashed lines).
}
\end{figure*}

\subsection{Halo, Bulge, and Thick Disk Models}
\label{sec:bulge}

Enrichment sources produce different elements on different timescales, and thus the time evolution of the elements varies as a function of location in a galaxy, depending on the star formation history.
In Figure \ref{fig:xfesr2} we show the evolution of elemental abundance ratios [X/Fe] against [Fe/H]
for the solar neighborhood (blue solid lines),
halo (green short-dashed and cyan dotted lines),
bulge (red long-dashed and olive dot-short-dashed lines),
and thick disk (magenta dot-long-dashed lines),
where the contributions from AGB stars, ECSNe, NSMs, and MRSNe are included (and those from the $\nu$-driven winds are not).

\begin{figure}\center
\includegraphics[width=8.5cm]{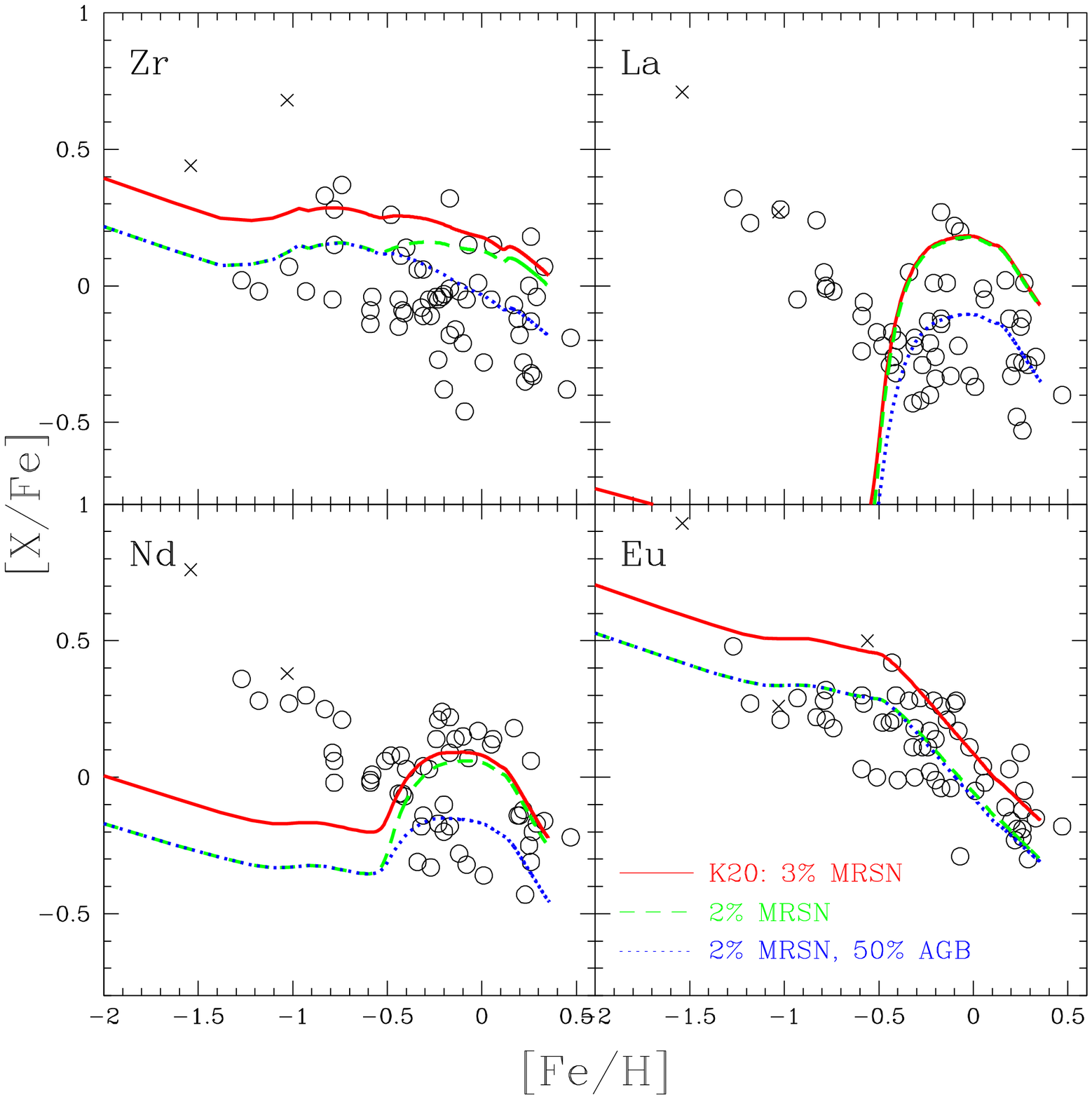}
\caption{\label{fig:bulge}
Evolution of [(Zr, La, Nd, Eu)/Fe] against [Fe/H] in the Galactic bulge
for the bulge with outflow model in Fig.~\ref{fig:xfesr2} with the same r-process model as for the solar neighborhood (red solid lines), with lower MRSN rates  (2\% of HNe, green dashed lines), and with a half contribution from AGB stars (blue dotted lines).
The observational data are taken from \citet{joh12}, respectively for the stars with normal (open circles) and anomalously high n-capture element abundances (crosses).
}
\end{figure}

{\bf Bulge and thick-disk ---}
If the star formation timescale is shorter than in the solar neighborhood (blue solid lines) as in our bulge 
(red long-dashed and olive dot-short-dashed lines) and thick-disk (magenta dot-long-dashed lines) models, the contribution from stars of a given 
lifetime appear at a higher metallicity than in the solar neighborhood.
Intermediate-mass AGB stars, low-mass AGB stars, and SNe Ia start to contribute at [Fe/H] $\sim -2.5, -1.5$, and $-1$, 
respectively in the solar neighborhood, but at a higher [Fe/H] in the bulge and thick disk models.
The [(C, N, F)/Fe] ratios peak at higher metallicities (see K11 for more discussion).
At [Fe/H] $\gtsim -1$, [$\alpha$/Fe] is higher and [Mn/Fe] is lower than in the solar neighborhood, which is consistent with observations \citep[e.g.,][]{ben04,joh14}.

Similar behaviour is also expected for neutron-capture elements. With our s+r models, [s/Fe] are lower in the bulge and thick disk than in the solar neighborhood, at a given metallicity with [Fe/H] $\gtsim -1.5$.
If the contribution from NSMs is larger, then [r/Fe] would also be lower.
The difference between the two bulge models is seen only at the metal-rich end, where the outflow model shows slightly lower [$\alpha$/Fe] and higher [r/Fe] than the wind model where star formation is totally quenched.
These trends seem very different from observations. \citet{joh12} showed that [(La, Nd, Eu)/Fe] ratios are positive at [Fe/H] $\ltsim -1$ and decrease from [Fe/H] $\sim -1$ to higher metallicities, which is similar to the [$\alpha$/Fe]--[Fe/H] relations.
This is not produced in our s+r models, and may suggest a different origin of neutron-capture elements.
In Figure \ref{fig:bulge}, the green dashed lines show the outflow bulge model with a different contribution of r-processes;
the fraction of MRSNE to HNe is decreases from 3\% to 2\%, so that the plateau value of [Eu/Fe] is consistent with observations.
In the blue dotted lines, the contribution from AGB stars is halved as an experiment. [Zr/Fe] matches with the observations, but not [La/Fe] and [Nd/Fe].
However, it is still  not possible to reproduce the abundances of Zr, La, and Nd, simultaneously.
Similar observational results are also shown by \citep{luc19}.

{\bf Halo ---}
If the chemical enrichment efficiency is lower than in the solar neighborhood as in our standard halo model (green short-dashed lines), the contribution from AGB stars and NSMs becomes larger compared with that from core-collapse supernovae, and thus [(s, r)/Fe] ratios are higher than in the solar neighborhood at [Fe/H] $\gtsim -2$.
If the chemical enrichment timescale is also shorter than in the solar neighborhood as in our second halo model with stronger outflow (light-blue dotted lines), [(s, r)/Fe] ratios becomes even higher toward higher metallicities.
The number of these relatively metal-rich halo stars should be very small, but by finding those stars with a large survey and measuring their various elemental abundances, it may be possible to re-construct the star formation history of the Galactic halo.

\section{Conclusions and Discussion}

We quantify of the origin of elements in the periodic table (Fig.\,\ref{fig:origin}) by constructing GCE models for all stable elements from C ($A\!=\!12$) to U ($A\!=\!238$) from the first principles, i.e., with using theoretical nucleosynthesis yields and event rates of all chemical enrichment sources.
Compared with the model in \citet{kob11agb},
we update our GCE models including i) new solar abundances, ii) failed supernovae, iii) super-AGB stars, iv) the s-process from AGB stars, and v) various r-process sites, i.e., ECSNe, $\nu$-driven winds, NSMs, and MRSNe.
We then compare the evolutionary trends of elemental abundance ratios to the most reliable observational abundances such as the NLTE analysis of \citet{zha16} and the LTE differential analysis of \citet{reg17}.
This has enabled us to understand the origin of the elements as a function of time and environment and to draw the following conclusions.

\begin{itemize}
\item
As required from recent observational and theoretical studies of core-collapse supernovae, we find that stars with initial masses of $M>30M_\odot$ can become failed supernovae if there is a significant contribution from hypernovae at $M\sim 20-50 M_\odot$, with a fraction of $\ge 1$\% at the solar metallicity and $\sim50$\% below one-tenth of the solar metallicity.
Observationally, this rate is comparable to the observed rate of broad-line SNe Ibc at the present. The cosmic supernova rates will give more constraints on the contribution of hypernovae.
Theoretically, it is a matter of urgency to understand the explosion mechanism of hypernovae, which requires GR-MHD simulations with detailed micro-physics.

\item
Although the fate of super-AGB stars (with $M \sim 8-10M_\odot$ at solar metallicity) is crucial for supernova rates, their contribution to GCE is negligible, unless hybrid WDs from the low-mass end of super-AGB stars explode as so-called Type Iax supernovae, or the high-mass end of super-AGB stars explode as ECSNe. 
Because the mass ranges are shifted toward lower masses for lower metallicities, the rates of these supernovae will be higher in the low metallicity environment such as in dwarf spheroidal galaxies.

\item
The observed abundances of the second (e.g., Ba) and third (Pb) s-process peaks are well reproduced with a smaller mass extent of $^{13}$C pockets in this paper. The standard $^{13}$C pockets assumed in the models of \citet{kar16} can explain the observed s-process abundances up to Pb, but the elements belonging to the second and third s-process peaks are overproduced relative to the solar abundances. This depends on the choice of the mass extent of the $^{13}$C pocket in the models and needs to be tested further together with the contribution of ECSNe and $\nu$-driven winds to the first (e.g., Sr) s-process peak.

\item
Although the enhancement due to ECSNe is small, $\sim 0.1$ dex for [(Cu, Zn)/Fe] ratios, ECSNe can provide enough light neutron-capture elements such as Sr, Y, and Zr, together with AGB stars to reproduce their observed trends and solar abundances. No extra LEPP is needed.
Adding the yields from $\nu$-driven winds result in a strong over-production of these light s-process elements. The yields we use are calculated separately from core-collapse SNe yields, whereas they should be consistently calculated together. For this reason it is better to exclude their contribution from GCE until self-consistent yields become available.

\item
NSMs can produce r-process elements up to Th and U, but it is not possible to explain the evolution of r-process elements with NSMs alone because i) the rates are too low and ii) the timescales are too long to explain the observations at low metallicities.
Note that, however, we adopt a metallicity-dependent delay-time distribution from binary population synthesis, which involves a few unknown physics of binaries.
Also, we apply 3D nucleosynthesis yields of a NS-NS merger to both NS-NS and NS-BH mergers, however, the yields of asymmetric NS-NS mergers or NS-BH mergers may be different. It is necessary to calculate nucleosynthesis yields of NSMs with varying parameters.

\item
The observed evolutionary trends such as for Eu can be well explained if $\sim 3$\% of $25-50M_\odot$ hypernovae produce the r-process elements, as in magneto-rotational supernovae.
In this paper we apply 2D nucleosynthesis yields of rotating iron-core collapse with magnetic fields.
It is unclear if the envelope totally collapses to the central BH or not, and it is necessary to simulate a long-time evolution of a whole star instead of iron core.
If the ejecta does not totally collapse, then C, N, and $\alpha$ elements might be ejected as well as the r-process elements.

\item
Our purely-theoretical models allow us to discover consistencies, and inconsistencies, that arise only by considering all the elements together. For example, we find that silver is overproduced by a factor of $6$, while gold is underproduced a factor of 5 in the model (Fig.\,\ref{fig:xfesr}). It would be worth revisiting the nuclear reaction rates relevant to these elements. It is also necessary to increase the samples of observational data, in particular with Hubble Space Telescope.

\item
The chemical evolutionary tracks depend on the location within the Galaxy. In general, at a given metallicity, rapid star formation such as in the bulge and thick disk results in lower s-process elemental abundances relative to iron, while inefficient star formation such as in the halo gives higher neutron-capture elemental abundances ratios than in the solar neighborhood (Fig.\,\ref{fig:xfesr2}).
This is because the contributions from the long time-delay sources, i.e., AGB stars, ECSNe, and NSMs, are lower in the case of rapid star formation. Thus, this difference depending on the locations is larger for the elements that are mainly produced from low-mass AGB stars, e.g., Ba, La, Ce, and Pb, and smaller for r-process elements such as Eu.
Available observational data in the bulge may suggest that the origins of neutron-capture elements are more complicated and  may depend on the location (Fig.\,\ref{fig:bulge}).
\end{itemize}

We stress that these one-zone chemical evolution models do not include the inhomogeneous enrichment that is particularly important at [Fe/H] $\ltsim -2.5$. However regarding the origins of neutron-capture elements, similar conclusions are obtained with more realistic, chemo-hydrodynamical simulations of a Milky Way-type galaxy \citep[e.g.,][]{hay19}.
In the case of inhomogeneous enrichment, the contribution from AGB stars also appears at low metallicities, and fast rotating stars for N abundances are not needed \citep{vin18a,vin18b}. This should also be tested for the s-process together with the contribution from ECSNe (Haynes \& Kobayashi 2020, in prep.).
Finally, in our GCE models, the effects of binary evolution are only partially included as SNe Ia and NSMs.
The AGB nucleosynthesis yields may also be affected by binary interactions \citep{izz06}, and future work should study this effect also in relation to the s-process.

\acknowledgments
We would like to thank the late M. Burbidge for directly/indirectly encouraging women in nuclear and astrophysics.
We also thank C. Doherty, N. Tominaga, S. Wanajo, and N. Nishimura for providing nucleosynthesis data, D. Van Beveren for binary population synthesis data, and C. Hansen and L. Mashonkina for observational data.
We are grateful to S. Smartt, K. Nomoto, F. Vincenzo, L. Spina, K. Lind, A. Amarsi, I. Roederer, and J. Vink\'o for fruitful discussions.
CK acknowledges funding from the UK Science and Technology Facility Council (STFC) through grant ST/M000958/1 \& ST/R000905/1.
This research was partially supported by the Australian Government through the Australian Research Council's Discovery Projects funding scheme (project DP170100521); as well through a Lend\"{u}let grant (LP17-2014) from the Hungarian Academy of Sciences to M.L. and the NKFIH KH\_18 (130405) project. 
Parts of this research were supported by the Australian Research Council Centre of Excellence for All Sky Astrophysics in 3 Dimensions (ASTRO 3D), through project number CE170100013.
This work used the DiRAC Data Centric system at Durham University, operated by the Institute for Computational Cosmology on behalf of the STFC DiRAC HPC Facility (www.dirac.ac.uk). This equipment was funded by a BIS National E-infrastructure capital grant ST/K00042X/1, STFC capital grant ST/K00087X/1, DiRAC Operations grant ST/K003267/1 and Durham University. DiRAC is part of the National E-Infrastructure.
CK and ML acknowledge support from the ``ChETEC'' COST Action (CA16117), supported by COST (European Cooperation in Science and Technology).
CK and AIK also acknowledge Kavli Institute for the Physics and Mathematics of the Universe funded by World Premier Research Center Initiative (WPI) in Japan.

\end{document}